\documentclass[showpacs,twocolumn,superscriptaddress]{revtex4-1}
\usepackage{doi}%<----------
\usepackage{hyperref}
\hypersetup{
  colorlinks=true,        % false: boxed links; true: colored links
  linkcolor=blue,         % color of internal links
  citecolor=cyan,         % color of links to bibliography
}

\usepackage{graphicx}% Include figure files
\usepackage{dcolumn}% Align table columns on decimal point
\usepackage{bm}% bold math
\usepackage{color}
\usepackage{enumitem}
\usepackage{amsmath}
\usepackage{amssymb}
\usepackage{orcidlink}
\usepackage{subcaption}
\usepackage{graphicx} % Required for inserting images
\begin{document}
\title{Periodic orbits and observational accretion disk around a Schwarzschild-like black hole surrounded by dark matter halo}

\author{Javokhir Sharipov}
\email{javohirsh100@gmail.com}
\affiliation{Institute of Fundamental and Applied Research, National Research University TIIAME, Kori Niyoziy 39, Tashkent 100000, Uzbekistan}

\author{Tursunali Xamidov}
\email{xamidovtursunali@gmail.com}
\affiliation{Institute of Fundamental and Applied Research, National Research University TIIAME, Kori Niyoziy 39, Tashkent 100000, Uzbekistan}
\affiliation{Institute for Theoretical Physics and Cosmology, Zhejiang University of Technology, Hangzhou 310023, China}

\author{Qiang Wu}
\email{wuq@zjut.edu.cn}
\affiliation{Institute for Theoretical Physics and Cosmology, Zhejiang University of Technology, Hangzhou 310023, China}
\affiliation{United Center for Gravitational Wave Physics (UCGWP), Zhejiang University of Technology, Hangzhou 310023, China}

\author{Sanjar Shaymatov}
\email{sanjar@astrin.uz}
\affiliation{Institute of Fundamental and Applied Research, National Research University TIIAME, Kori Niyoziy 39, Tashkent 100000, Uzbekistan}
\affiliation{Institute for Theoretical Physics and Cosmology,
Zhejiang University of Technology, Hangzhou 310023, China}
\affiliation{University of Tashkent for Applied Sciences, Str. Gavhar 1, Tashkent 100149, Uzbekistan}

\author{Tao Zhu}
\email{zhut05@zjut.edu.cn}
\affiliation{Institute for Theoretical Physics and Cosmology, Zhejiang University of Technology, Hangzhou 310023, China}
\affiliation{United Center for Gravitational Wave Physics (UCGWP), Zhejiang University of Technology, Hangzhou 310023, China}

\date{\today}

\begin{abstract}

In this work, we investigate the dynamics of periodic orbits and the properties of accretion disks around a Schwarzschild-like black hole (BH) immersed in a King-type dark matter (DM) halo. Our analysis focuses on how the presence of the King DM halo influences both the behavior of periodic orbits and the radiative characteristics of the accretion disk. We begin by examining time-like periodic geodesic orbits for various configurations characterized by different energy and angular momentum values, represented by the integers $(z, w, v)$. Furthermore, we explore the effects of the King DM halo on time-like periodic geodesics, marginally bound orbits, and innermost stable circular orbits, thereby providing a deeper understanding of how the DM halo environment modifies the behavior of these stable orbits and timelike particle geodesics. Finally, we analyze the null geodesics and the accretion disk properties by studying their direct and secondary images, redshift distributions, and radiation fluxes as observed at infinity for a range of inclination angles. This approach allows us to gain valuable insights into the spacetime geometry of a Schwarzschild-like BH within the King-type DM halo, its physical and radiative properties in the accretion disk, and the corresponding observational implications.

\end{abstract}

\maketitle

\section{Introduction}

In General Relativity (GR), black holes (BHs) not only arise naturally as exact solutions to Einstein's field equations but also remain as one of the most fascinating objects in the universe. Their existence was first predicted theoretically nearly a century ago. However, modern breakthroughs, such as the detection of gravitational waves by LIGO \cite{Abbott16a, Abbott16b} and the first images of BH shadows captured by the Event Horizon Telescope (EHT) \cite{Akiyama19L1, Akiyama19L6, Akiyama22L12}, have not only provided compelling evidence for the existence of BHs in nature, but have also offered powerful tools for probing their unknown properties. Despite these advances, many open questions persist regarding BH behavior and their interactions with the surrounding environment. Consequently, identifying observational signatures that reveal the interplay between BH gravity and its surroundings remains a crucial goal in modern astrophysics. In this context, dark matter (DM) represents a particularly intriguing candidate capable of influencing or exhibiting such interactions within the framework of GR. 

In an astrophysical context, BHs are expected to be surrounded by DM halos \cite{Iocco15NatPhy, Bertone18Nature} and are characterized by complex and dynamic environments due to the evidence indicating that supermassive BHs located at the centers of most galaxies serve as the primary engines powering active galactic nuclei (AGN) \cite{Rees84ARAA, Kormendy95ARAA}. This is consistent with the composition of the universe, which measurements of the cosmic microwave background show is approximately 27\% dark matter and 68\% dark energy. Consequently, the ordinary matter familiar to us constitutes only about 5\% of the universe. It can thus be inferred that the distribution of DM throughout the universe can have a significant impact on galactic rotation curves and on large-scale astrophysical phenomena such as the bullet cluster collision (see, e.g., \cite{Rubin70ApJ, Davis85ApJ, Bertone18Nature, Corbelli00MNRAS, Clowe06ApJL}). These phenomena are widely regarded as key evidence confirming the presence of DM halos surrounding galaxies and clusters \cite{Bertone05,deSwart17Nat, Wechsler18}. Furthermore, the presence of a DM halo around a supermassive BH can influence the dynamics of extreme and intermediate mass-ratio inspirals (EMRIs and IMRIs) \cite{Amaro-Seoane18LRR, Babak17PRD}, offering a potential pathway to probe its properties.

These findings have confirmed the occurrence of star formation near galactic centers, supporting the formation of DM halos around galaxies and clusters, and suggesting that spiral and giant elliptical galaxies evolve within these surrounding DM halos \cite{Valluri04ApJ, Akiyama19L1, Akiyama19L6, Akiyama22L12}. This implies that the rotational velocities of stars orbiting spiral and giant elliptical galaxies can be accounted for by the presence of DM halos \cite{Persic96}. Motivated by the fundamental role of DM halos, it is therefore crucial to investigate various theoretical models to deepen our understanding of their nature and influence on the surrounding environment through both astrophysical observations and numerical simulations. To this end, several analytical BH solutions incorporating DM halos have been proposed within various theoretical models, such as the Einasto profile \cite{Merritt_2006,Dutton_2014}, the Navarro–Frenk–White (NFW) model \cite{Navarro_1996}, the Burkert profile \cite{Burkert_1995}, the Dehnen-type DM model \cite{Dehnen93,Xu18JCAP,Shukirgaliyev21A&A,Gohain_2024,Pantig_2022,Al-Badawi25JCAP}, universal rotation curve DM profile \cite{Jusufi:2019nrn}, superfluid DM halo \cite{Jusufi:2020cpn}, and DM distributions associated with phantom scalar fields \cite{Li-Yang12,Shaymatov21d,Shaymatov21pdu,Shaymatov22a}. Following the Dehnen-type DM halo, there have been several investigations on these lines \cite{Al-Badawi25CPC,Al-Badawi25CTP_DM,Uktamov25DM,Alloqulov25GW1,Arpan2025PhRvD,Xamidov:25epjc} addressing the influence of the DM halo on quasinormal modes, the BH shadow, and gravitational waveforms. In addition, analytical BH solutions describing supermassive BHs embedded within DM halos have also been developed and explored in detail (see, e.g., \cite{Cardoso22DM,Hou18-dm,Shen24PLB,Shen25PLB}). Analyzing EMRI signals was also used for probing the DM halo distribution around supermassive black holes (SMBHs) \cite{Navarro96ApJ,Gondolo99PRL, Zhang:2021bdr, Zhang:2022roh}.

To gain a deeper insight into the properties of BHs embedded within DM halos, it is essential to examine the geodesic motion of test particles and photons, since their trajectories serve as powerful probes of BH geometry and dynamics. In this context, the study of geodesic motion plays a decisive role in revealing the effects of external fields that influence their trajectories and in elucidating the behavior of the background spacetime in astrophysical environments \cite{Joshi19,Bini12,Shaymatov15,Dadhich18,Shaymatov19b,Shaymatov20egb}. In particular, bound orbits around BHs provide valuable clues to possible deviations from the standard geometry of astrophysical BHs. Recently, periodic orbits have attracted growing attention, especially since the detection of gravitational waves \cite{Abbott16a, Abbott16b} due to their significance in understanding the sources of these waves, including compact binary systems and extreme mass-ratio inspirals (EMRIs). Periodic bound trajectories frequently display zoom-whirl dynamics, where a compact object performs an integer number of radial and angular oscillations per orbit. During the whirl phase, the object makes several rapid revolutions near the BH before zooming outward again. These orbits are commonly classified by three integers, such as zoom ($z$), whirl ($w$), and vertex ($v$) and characterized by the ratio of their angular to radial frequencies \cite{Levin_2008,Levin_2009}. Numerous studies have examined periodic bound orbits near BHs across various gravity frameworks~\cite{Levin2010,Babar17PRD,Liu_2019,Deng20,Wei2019PRD, Jiang2024PDU,Tu23PRD,Chen:2025aqh, Lu:2025cxx,Yang2025JCAP,Haroon25PRD,Alloqulov25GW1, Ahmed:2025azu, Yang:2024cnd, Wang:2025hla, Wang:2025wob, Zare:2025aek}, revealing their strong dependence on the background spacetime geometry.

An important point is that astrophysical processes provide a critical testing ground for gravitational theories, both within GR and beyond. Observations of accretion disks, particularly their structures~\cite{Bambi17e,Chandrasekhar98} and X-ray spectra \cite{Bambi12a,Bambi16b}, and their dynamical properties~\cite{Abramowicz13,Fender04mnrs,Auchettl17ApJ}, offer a powerful method for probing the strong-field regime of gravity. Consequently, accretion disks around BHs are regarded as key probes of both spacetime geometry and the high-energy processes in their immediate vicinity. As matter spirals inward, losing angular momentum and converting gravitational potential energy into radiation, accretion disks become natural laboratories for studying background gravity and high-energy astrophysical phenomena. This makes them essential for understanding phenomena such as X-ray binaries \cite{Yao01AIP,Schultz05BOOK,Bu23BOOK} and quasars \cite{Czerny23ApSS,Morgan10ApJ}. A prime example is Sagittarius A$^{\star}$ (Sgr A$^{\star}$) at our Galactic center, which provides a unique setting for exploring BH accretion physics and testing gravity in the strong-field regime \cite{you2024,Nucita07,Ghez05,Ghez00,Imdiker23CQG,Vagnozzi23EHT,Afrin23ApJ}. Motivated by these findings, numerous recent studies \cite{Gyulchev20,Boshkayev:2020kle,Gyulchev2021E,Shaymatov2023,Collodel_2021,Alloqulov24CPC,Boshkayev21PRD,Hu2023,Cui24,Alloqulov24EPJP,Liu:2020ola, Chen2025,Ziqiang25,Nozari2025,Liu:2020vkh, Liu:2021yev, Igata2025,Nozari2025b,Zhu:2019ura, Jiang:2023img, Yan:2025mlg} have investigated the properties of accretion disks within various theories of gravity. Considering the astrophysical significance of DM halos in astrophysical environments, analyzing BHs within a King-type DM halo is particularly relevant. It can offer a compelling framework for studying how DM influences the dynamics of bound and null geodesics for modeling accretion disk structures, thus affecting key observables such as the innermost stable circular orbits (ISCOs) radius and the properties of the accretion disk. 

From an astrophysical perspective, a BH solution embedded in a DM halo offers an alternative avenue for understanding BH–DM systems, providing valuable insights not only into the fundamental properties of DM halos but also into their potential interactions within the framework of GR. In this work, we investigate periodic orbits and accretion disk properties around a BH in a King-type DM halo \cite{King1962AJ}, focusing on the influence of the halo on both the dynamics of periodic orbits and the radiative characteristics of the accretion disk. Specifically, we first examine time-like periodic geodesic orbits with their plots classified by integers $(z,w,v)$ and then explore how the King DM halo affects marginally bound orbits (MBOs) and innermost stable circular orbits (ISCOs) under the relevant stability conditions \cite{Dadhich22a}. Subsequently, we analyze the structure of accretion disks around the BH within the DM halo by investigating their observable features, such as images, redshift distributions, and radiation fluxes as detected by far away observers. Through this analysis, we aim to shed light on the spacetime structure of BH–DM systems, their physical nature, and possible observational implications.

The structure of this paper is organized as follows. In Section~\ref{Sec:II}, we briefly introduce a Schwarzschild-like BH spacetime within a King-type DM halo and examine time-like periodic geodesics, determining marginally bound and innermost stable circular orbits influenced by the DM halo profile. Section~\ref{Sec:III} investigates periodic geodesic orbits for various configurations characterized by integers $(z, w, v)$, considering different values of energy and angular momentum of time-like geodesic particles. In Section~\ref{Sec:IV}, we focus on null geodesics and investigate how the King DM halo profile modifies the photon trajectories around the BH. Section~\ref{Sec:V} is devoted to analyzing the properties of the accretion disk around the Schwarzschild-like BH in the presence of the King-type DM halo, focusing on the direct and secondary images, redshift distributions, and radiation fluxes as observed at infinity. Finally, Section~\ref{Sec:conclusion} summarizes our main findings and conclusions.

\section{Spacetime metric and timelike geodesics}\label{Sec:II}

\begin{figure*}[!htb]
    \centering
    \includegraphics[scale=0.5]{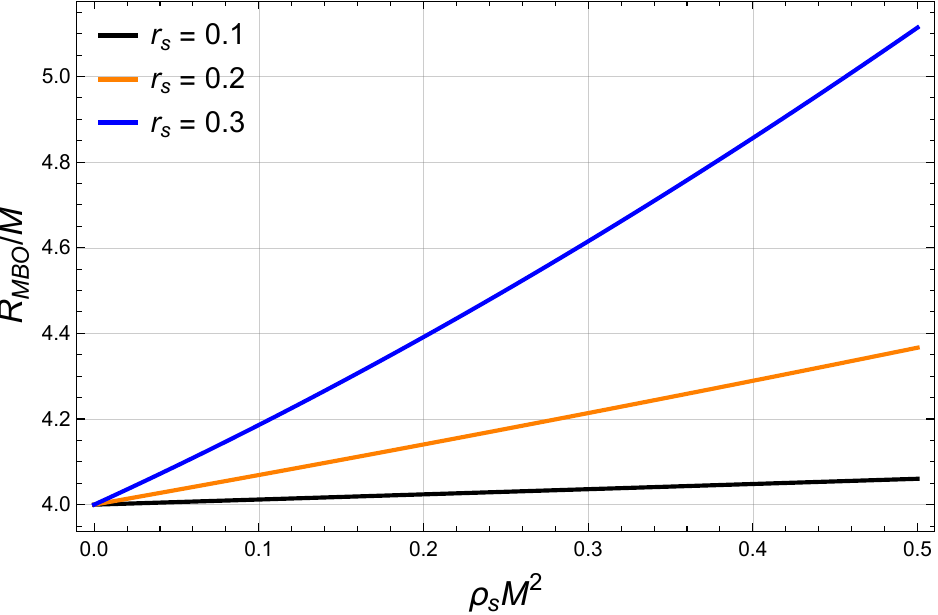}
    \includegraphics[scale=0.5]{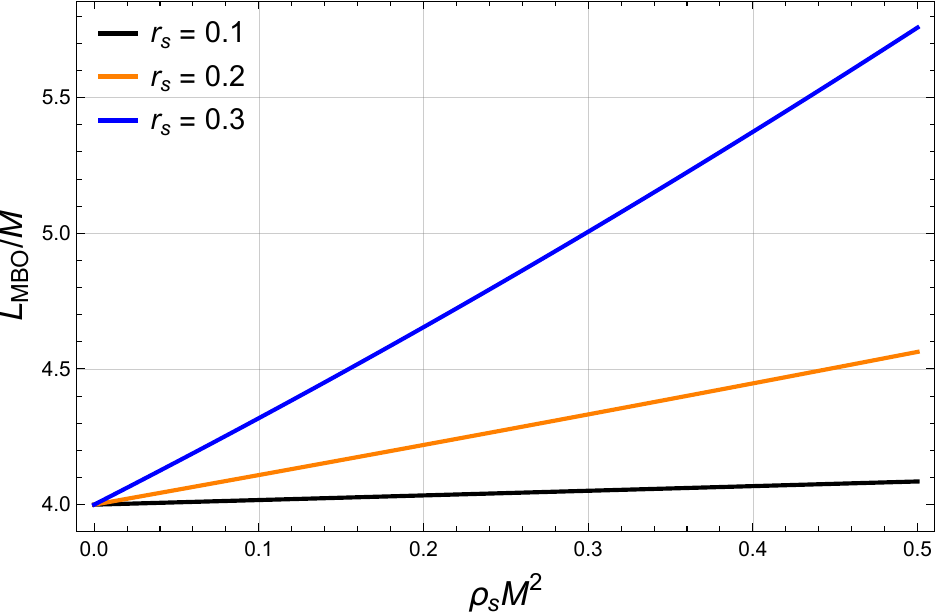}
    \caption{Left: MBO radius as a function of the DM halo density $\rho_s$ for various scale radii $r_s$. Right: Variation of the MBO angular momentum with the DM halo density $\rho_s$ for different $r_s$ values.}
    \label{fig:mbo}
\end{figure*}
\begin{figure*}[htbp]
    \centering
    \includegraphics[scale=0.32]{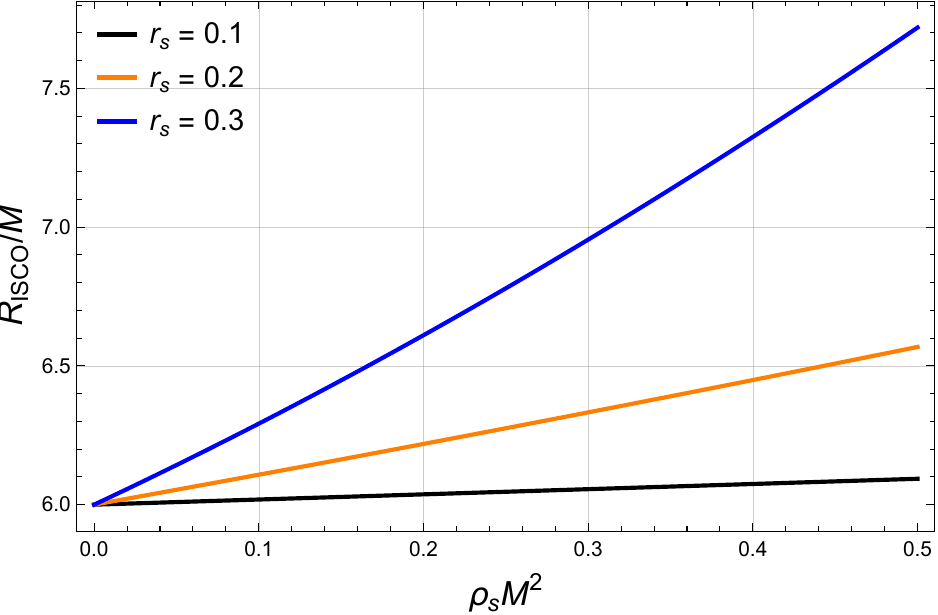}
    \includegraphics[scale=0.32]{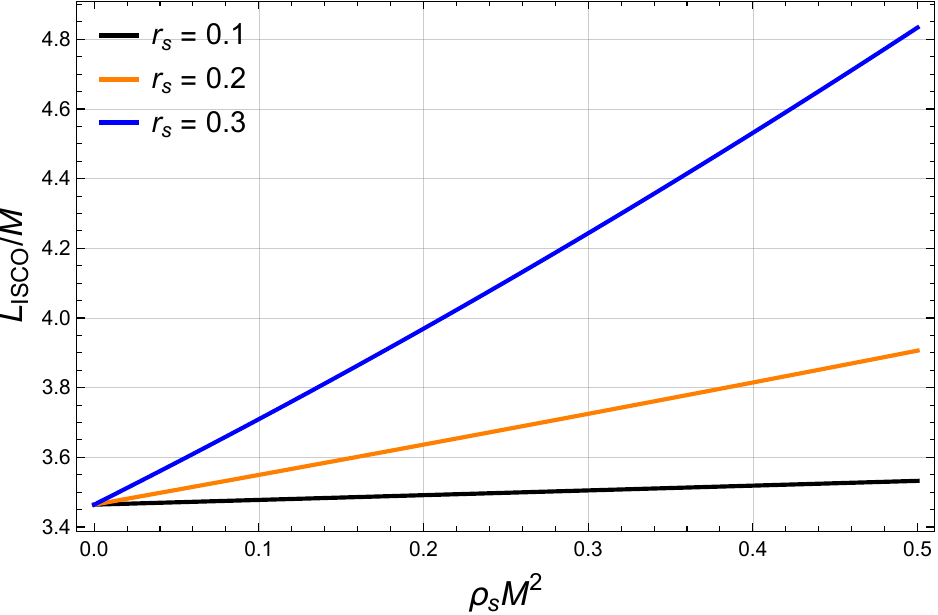}
    \includegraphics[scale=0.32]{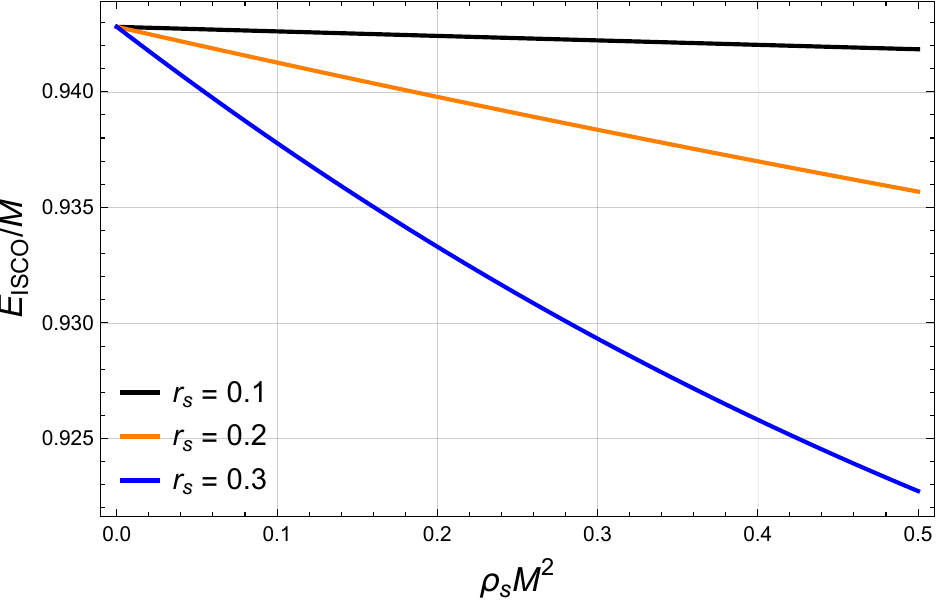}
    \caption{The ISCO parameters $R_{\mathrm{ISCO}}$, $L_{\mathrm{ISCO}}$, and $E_{\mathrm{ISCO}}$ as functions of the DM halo density $\rho_s$ for different values of the scale radius $r_s$.}
    \label{fig:isco}
\end{figure*}
\begin{figure*}[htbp]
\centering
 \includegraphics[scale=0.5]{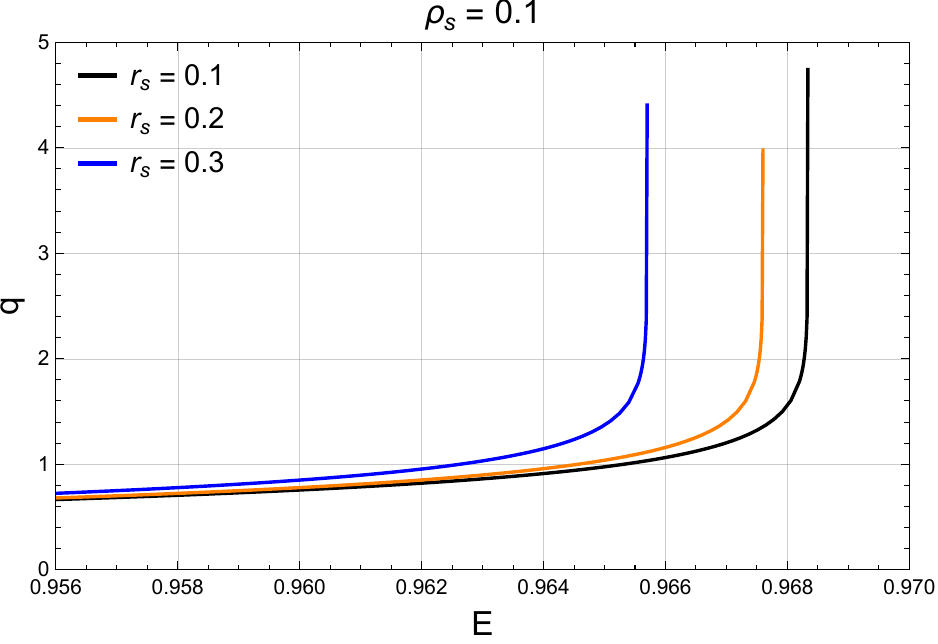}
 \includegraphics[scale=0.5]{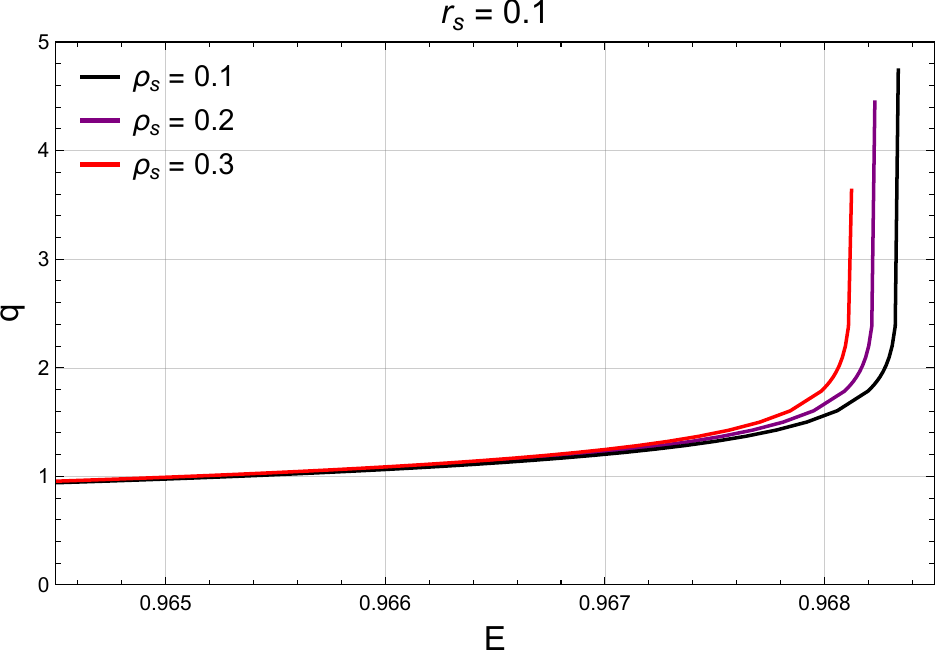}
  \includegraphics[scale=0.5]{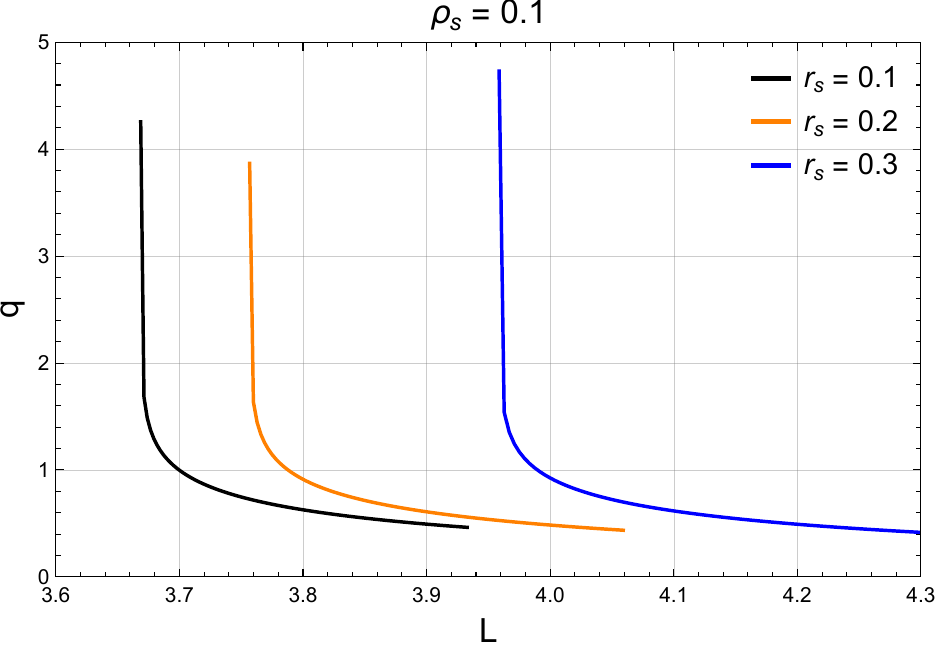}
 \includegraphics[scale=0.5]{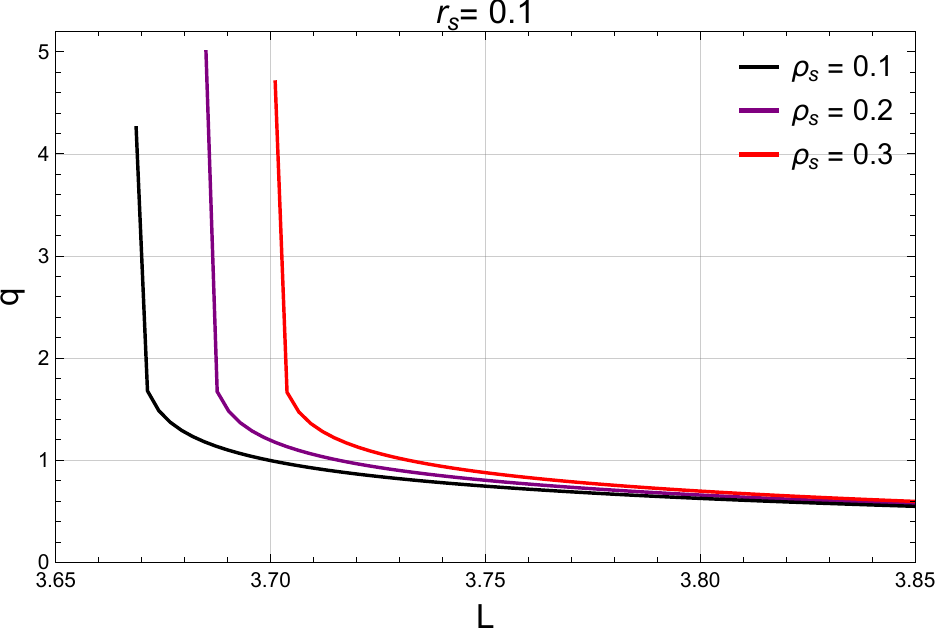}
 \caption{Top: The rational number $q$ as a function of the energy $E$ for orbits around a Schwarzschild BH surrounded by a King DM halo, shown for different values of the parameters $r_s$ (top-left panel) and $\rho_s$ (top-right panel). The orbital angular momentum is fixed at $L = \tfrac{1}{2}(L_{\mathrm{MBO}} + L_{\mathrm{ISCO}})$.
Bottom: The rational number $q$ as a function of the orbital angular momentum $L$ for orbits around a Schwarzschild BH in a King DM halo, plotted for different values of $r_s$ and $\rho_s$. The energy is fixed at $E = 0.96$.}
 \label{fig:q}
\end{figure*}

In this section, we study the time-like geodesics around the Schwarzschild BH embedded in a King DM halo ~\cite{King1962AJ}. The corresponding spacetime metric can be written as follows:
\begin{equation}
ds^{2}=-f(r)dt^{2}+f(r)^{-1}dr^{2}+r^{2}d\theta^{2}+r^{2}\sin^{2}\theta d\phi^{2},
\end{equation}
with~\cite{Kar2025}
\begin{align}
    f(r) =& 1 - \frac{2M}{r} 
+
\frac{8\pi \rho_{s} r_s^{3}}{\sqrt{r^{2} + r_s^{2}}}\nonumber\\
&+ \frac{8\pi \rho_{s} r_s^{3}}{r}
\ln\!\left(
\frac{\sqrt{r^{2} + r_s^{2}} - r}{r_s}
\right)\ ,
\label{metric}
\end{align}
where $r_s$ represents the scale radius and $\rho_s$ is the halo density.

Suppose a test particle moves around a Schwarzschild BH surrounded by a King DM halo. The Lagrangian governing the motion of the test particle takes the form ~\cite{1983mtbh.book.....C}
\begin{equation}
\mathcal{L}=\frac{1}{2}m\,g_{\mu\nu}\,\frac{dx^{\mu}}{d\tau}\,\frac{dx^{\nu}}{d\tau},
\end{equation}
with $\tau$ and $m$ denote the proper time and rest mass of the particle, respectively. By choosing $m=1$ for simplicity, the generalized momentum of the particle can be written as 
\begin{equation}
    p_{\mu}=\frac{\partial {\cal L}}{\partial \dot{x}^{\mu}}=g_{\mu \nu}\dot{x}^{\nu}\, .
\end{equation}
Using the above relations, the equations governing the particle’s motion take the form
\begin{eqnarray}\label{eq:eqmotion}
p_{t} &=& g_{tt}\dot{t} = -f(r) \dot{t}=-E \, , \nonumber\\
p_{\phi} &=&g_{\phi\phi}\dot{\phi} =  r^{2}\sin^{2}\theta\dot{\phi}=L \, , \nonumber \\
p_{r} &=&g_{rr}\dot{r} =f(r)^{-1} \dot{r} \, , \nonumber\\
p_{\theta} &=&g_{\theta\theta}\dot{\theta} = r^{2}\dot{\theta} \, ,
\end{eqnarray}
where $E$ and $L$ denote the conserved energy and angular momentum of the particle, respectively.

By restricting the motion to the equatorial plane ($\theta = \pi/2$) and applying the normalization condition $g_{\mu\nu}\dot{x}^{\mu}\dot{x}^\nu = -1$, the effective potential can be expressed as follows:
\begin{equation}
V_{\rm eff}=f(r)\left(1+\frac{L^{2}}{r^{2}}\right)\, .
\end{equation}

\renewcommand{\arraystretch}{1.2}
\begin{table*}[]
\centering
\resizebox{1.0\textwidth}{!}{
\begin{tabular}{|c|c|c|c|c|c|c|c|c|c|}
\hline
$\rho_s$ & $L$ & $E_{(1,1,0)}$ & $E_{(1,2,0)}$ & $E_{(2,1,1)}$ & $E_{(2,2,1)}$ & $E_{(3,1,2)}$ & $E_{(3,2,2)}$ & $E_{(4,1,3)}$ & $E_{(4,2,3)}$ \\ \hline
0.0    & 3.732050 & 0.965425   & 0.968383   & 0.968026   & 0.968435   & 0.968225   & 0.968438   & 0.968285   & 0.968440   \\ \hline
0.1    & 3.829283 & 0.964568   & 0.967539   & 0.967181   & 0.967590   & 0.967380   & 0.967594   & 0.967440   & 0.967596   \\ \hline
0.2    & 3.928165 & 0.963743   & 0.966729   & 0.966369   & 0.966781   & 0.966569   & 0.966785   & 0.966630   & 0.966786   \\ \hline
0.3    & 4.028690 & 0.962949   & 0.965952   & 0.965591   & 0.966005   & 0.965792   & 0.966009   & 0.965853   & 0.966010     \\ \hline
\end{tabular}
}
\caption{The energy $E$ of periodic orbits is tabulated for different orbit configurations $(z, w, v)$. Here, we set $L = \tfrac{1}{2}(L_{\mathrm{MBO}} + L_{\mathrm{ISCO}})$ and $r_s = 0.2$.}
\label{table1}
\end{table*}
\begin{table*}[]
\resizebox{1.0\textwidth}{!}{
\begin{tabular}{|c|c|c|c|c|c|c|c|c|}
\hline
$\rho_s$ & $L_{(1,1,0)}$ & $L_{(1,2,0)}$ & $L_{(2,1,1)}$ & $L_{(2,2,1)}$ & $L_{(3,1,2)}$ & $L_{(3,2,2)}$ & $L_{(4,1,3)}$ & $L_{(4,2,3)}$ \\ \hline
0.0    & 3.68358    & 3.65344    & 3.65759    & 3.65270    & 3.65533    & 3.65263    & 3.65462    & 3.65261    \\ \hline
0.1    & 3.78789    & 3.75765    & 3.76178    & 3.75697    & 3.75954    &  3.75691    & 3.75884    & 3.75689    \\ \hline
0.2    & 3.89373    & 3.86336    & 3.86745    & 3.86270    & 3.86522    & 3.86264    & 3.86453    & 3.86262    \\ \hline
0.3    & 4.00114    & 3.97055    & 3.97462    & 3.96990    & 3.97240    & 3.96984    & 3.97170    & 3.96983    \\ \hline
\end{tabular}
}
\caption{The values of the orbital angular momentum $L$ are tabulated for different periodic orbits characterized by $(z, w, v)$. Here, we set $E = 0.96$ and $r_s = 0.2$.
}
\label{table2}
\end{table*}
\begin{figure*}[!htb]
    \centering
    \includegraphics[scale=0.5]{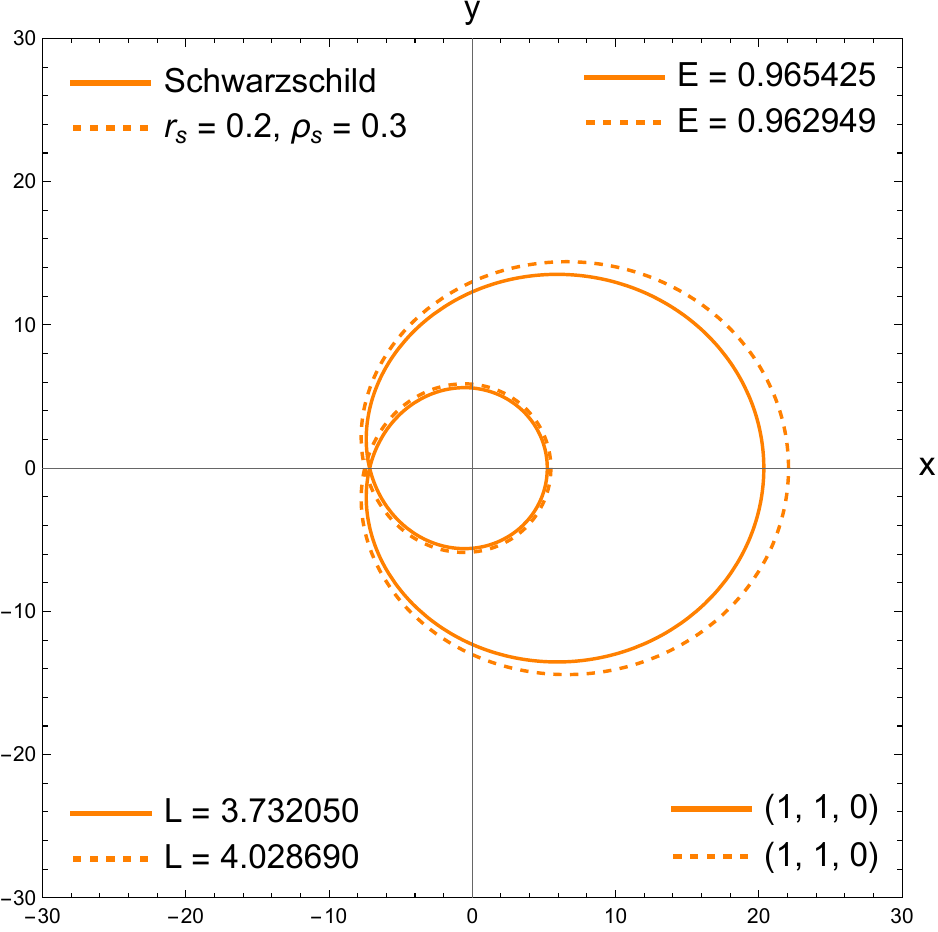}
    \includegraphics[scale=0.5]{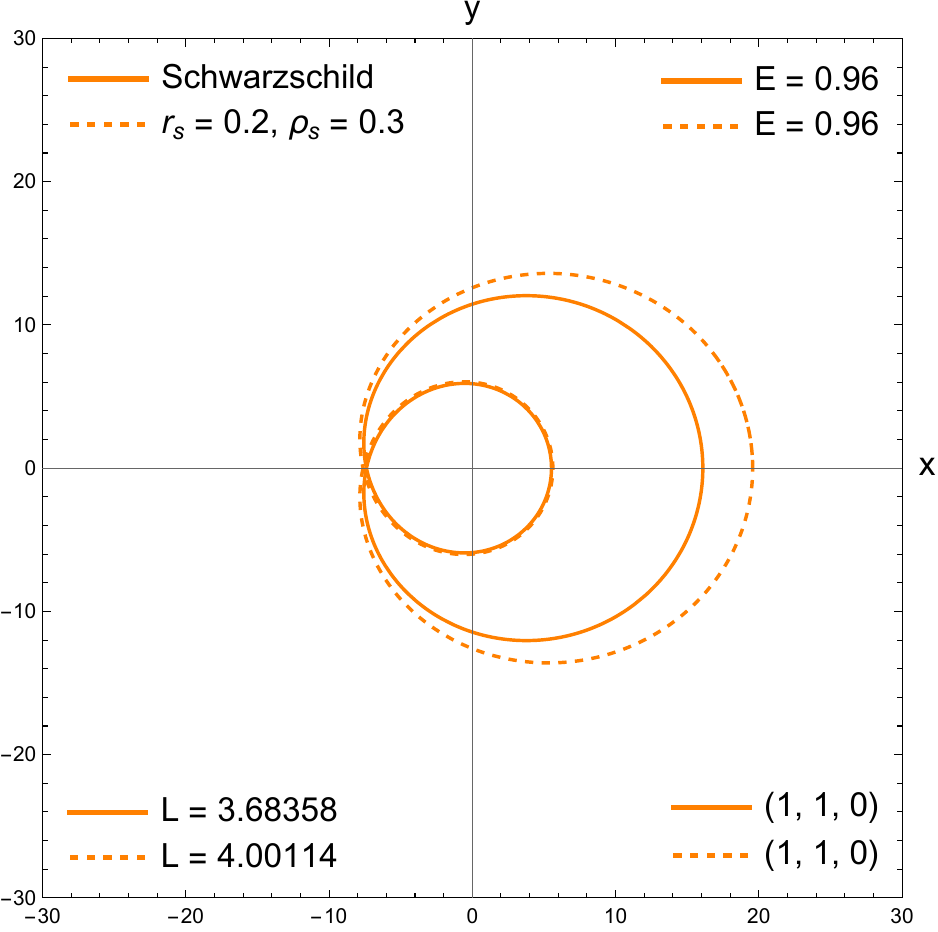}
    \caption{Comparison of periodic orbits with parameters $(1, 1, 0)$ for the Schwarzschild BH with and without a King DM halo.}
    \label{fig:comparison}
\end{figure*}

For a particle to move along a general bound orbit, its angular momentum and energy must lie within the following ranges:
\begin{equation}
    L_{ISCO}\leq L \quad\mbox{and}\quad E_{ISCO}\leq E \leq E_{MBO}=1\, ,
\end{equation}
where $E_{\mathrm{ISCO}}$ and $L_{\mathrm{ISCO}}$ represent the energy and angular momentum associated with the innermost stable circular orbit (ISCO), whereas $E_{\mathrm{MBO}}$ corresponds to the energy at the marginally bound orbit (MBO). The marginally bound orbit (MBO) is defined by the following conditions:
\begin{equation}
    V_{eff}=1\quad\mbox{and}\quad \frac{d V_{eff}}{dr}=0.
\end{equation}
Based on the above relations, we analyze numerically how the presence of the King DM halo influences the radius and orbital angular momentum of the MBO. Figure~\ref{fig:mbo} illustrates the dependence of $R_{\mathrm{MBO}}$ and $L_{\mathrm{MBO}}$ on the King DM halo density for different values of the halo radius. As shown in Fig.~\ref{fig:mbo}, an increase in the parameters $r_s$ and $\rho_s$ leads to a corresponding increase in $R_{\mathrm{MBO}}$ and $L_{\mathrm{MBO}}$. This implies that the presence of the King DM halo enables bound motion to occur at larger radii and higher angular momenta. Therefore, the halo exhibits a gravitational influence analogous to that of a BH, strengthening the effective gravitational field.

We now proceed to analyze another class of bound orbits, the ISCO, determined by the following conditions:
\begin{equation}
    \dot{r}=0, \quad \frac{d V_{eff}}{dr}=0 \quad\mbox{and}\quad \frac{d^2V_{eff}}{dr^2}=0.
\end{equation}
Figure~\ref{fig:isco} shows the ISCO radius, orbital angular momentum, and energy as functions of the DM halo density for various scale radii $r_s$. In this figure, $R_{\mathrm{ISCO}}$ and $L_{\mathrm{ISCO}}$ increase with increasing $r_s$ and $\rho_s$, indicating a strengthening of the gravitational field. The corresponding decrease in energy with increasing $r_s$ and $\rho_s$ is attributed to the outward shift of the ISCO radius from the BH as the gravitational influence of the DM halo becomes stronger.

\begin{figure*}[htbp]
   \centering
     \includegraphics[width=0.32\textwidth]{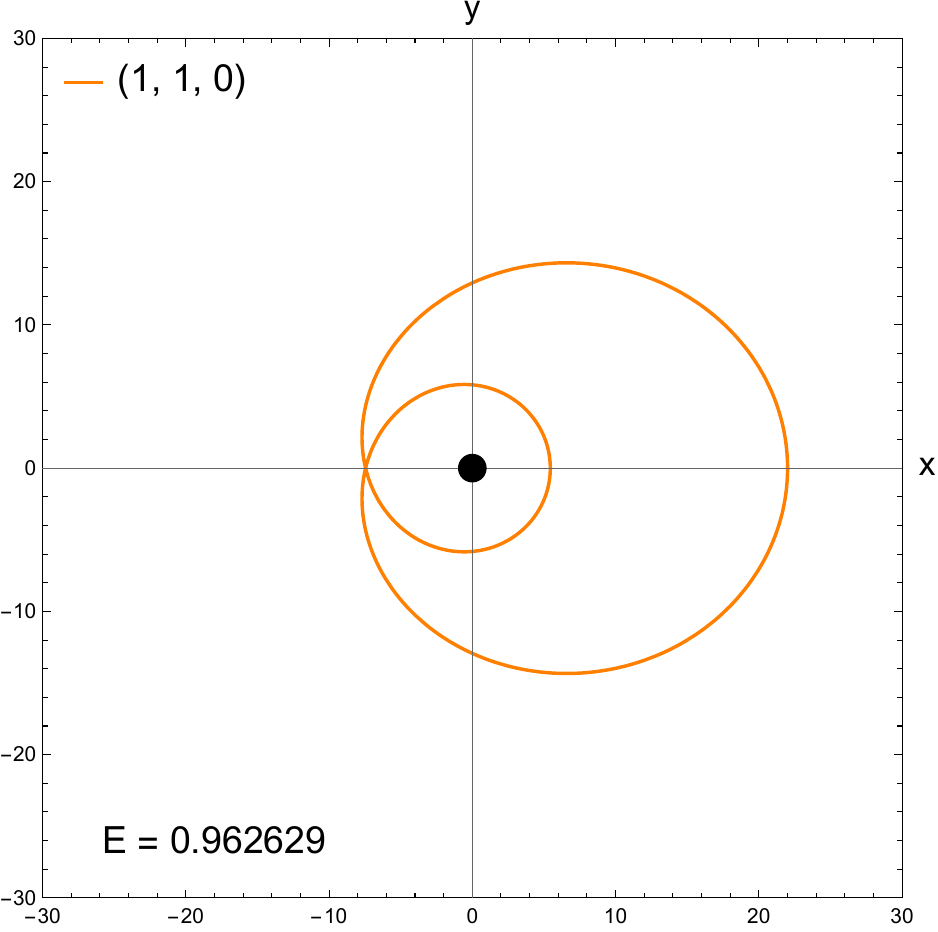} \hfill
    \includegraphics[width=0.32\textwidth]{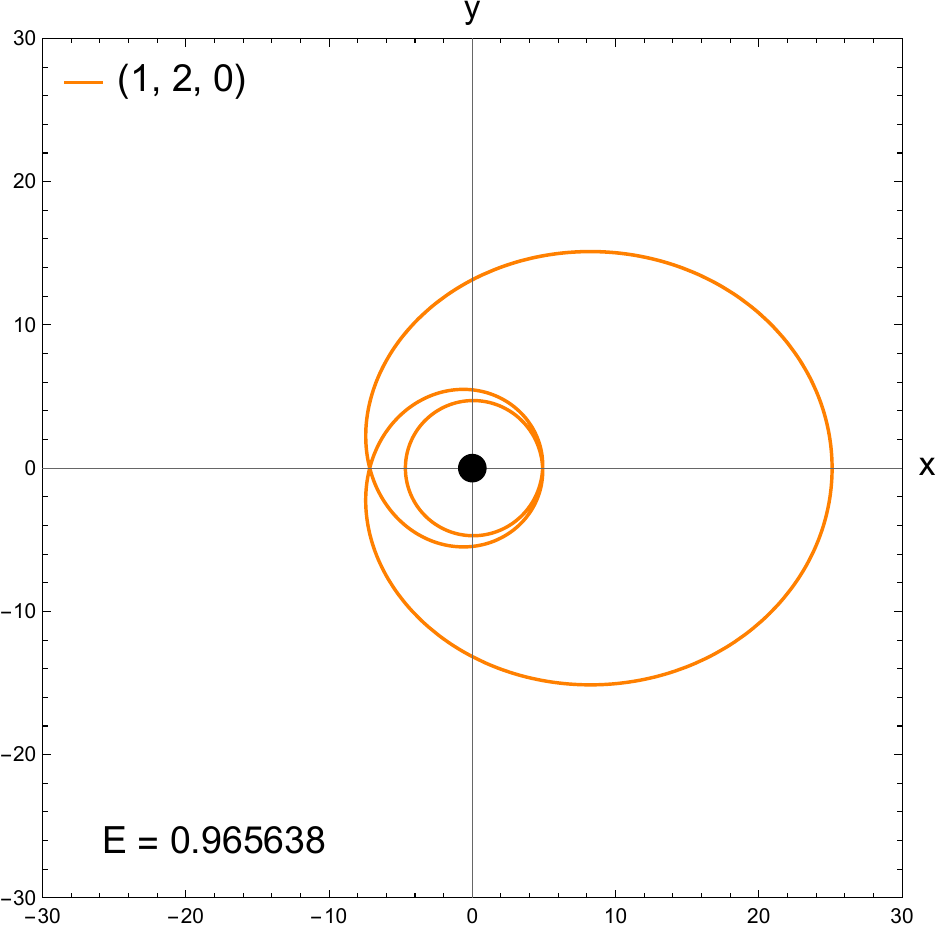} \hfill
    \includegraphics[width=0.32\textwidth]{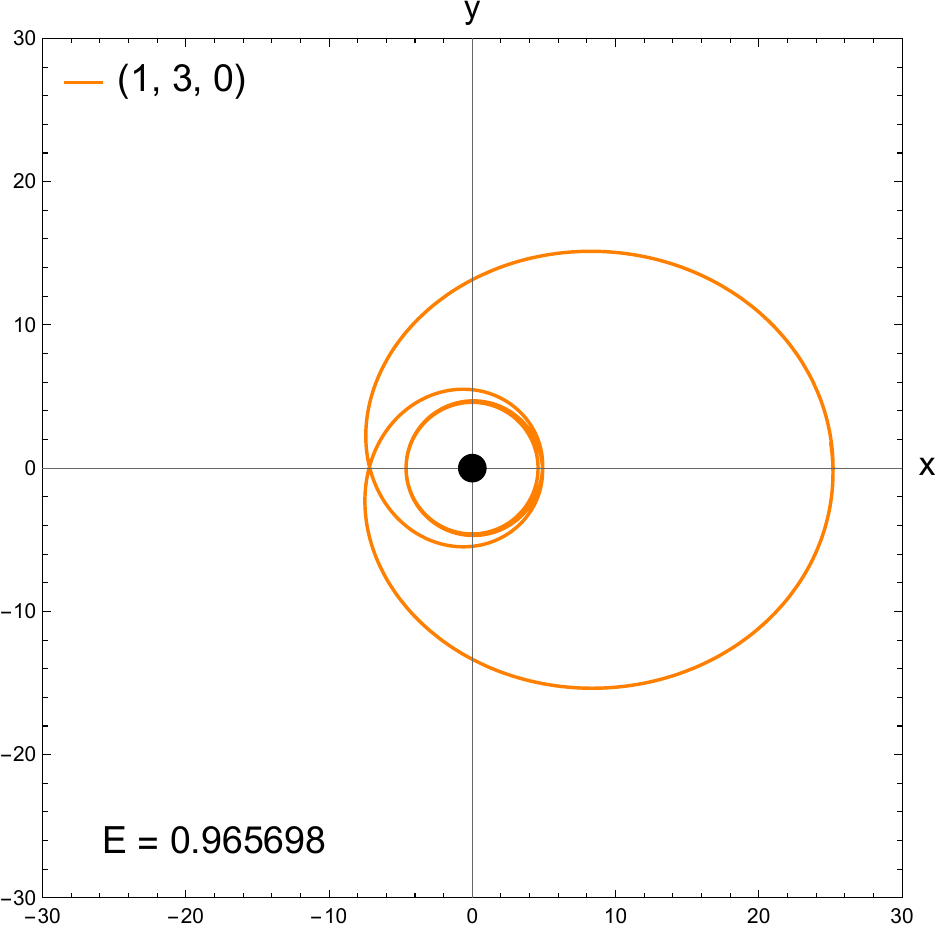} \\
    
    \vspace{0.2cm} 
    \includegraphics[width=0.32\textwidth]{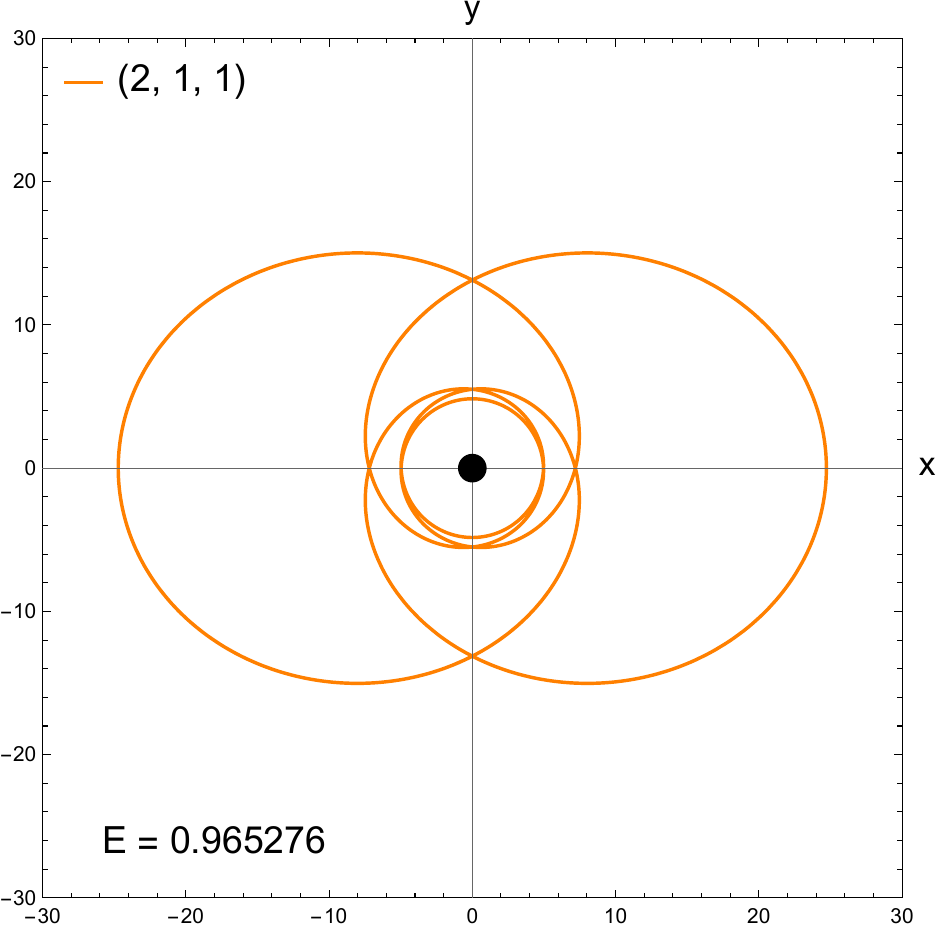} \hfill
    \includegraphics[width=0.32\textwidth]{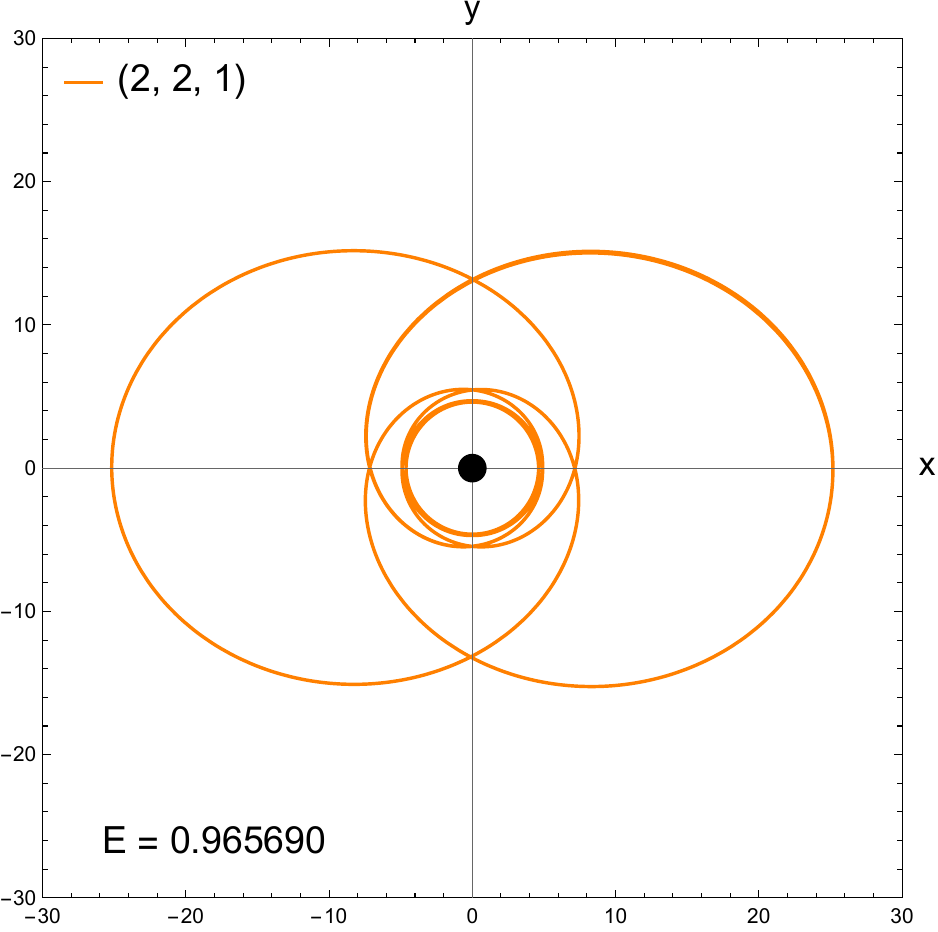} \hfill
    \includegraphics[width=0.32\textwidth]{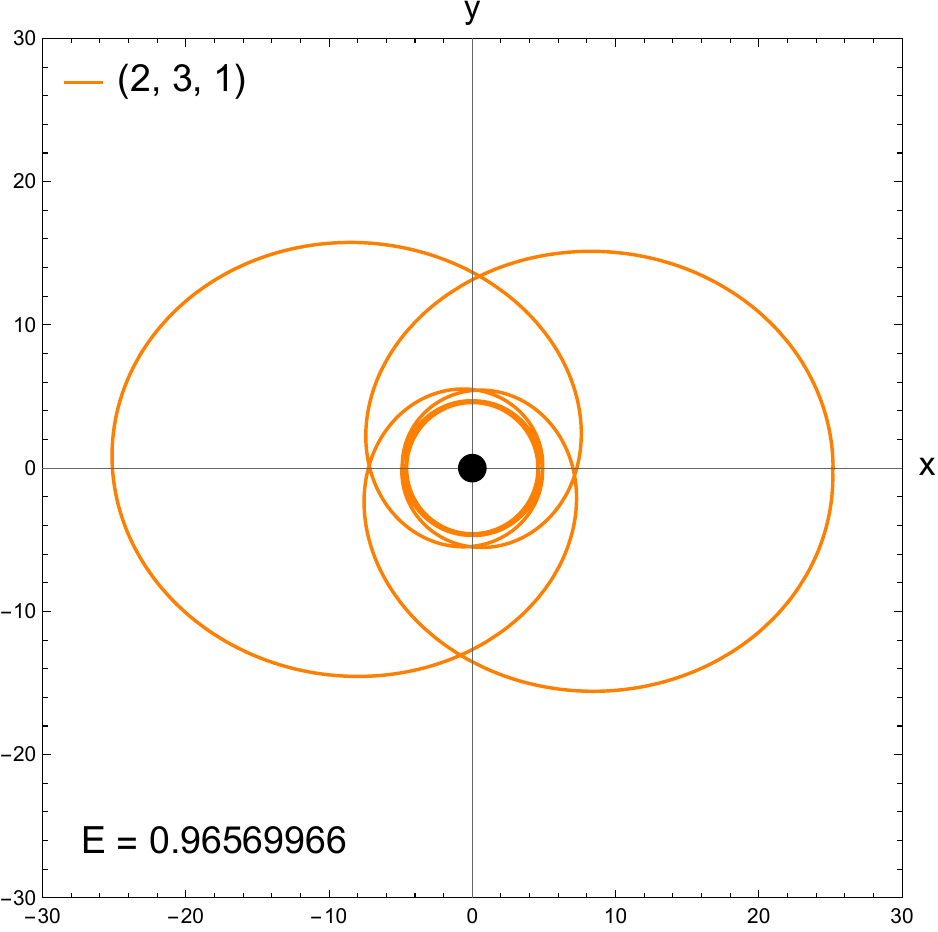} \\
    
    \vspace{0.2cm} 
    \includegraphics[width=0.32\textwidth]{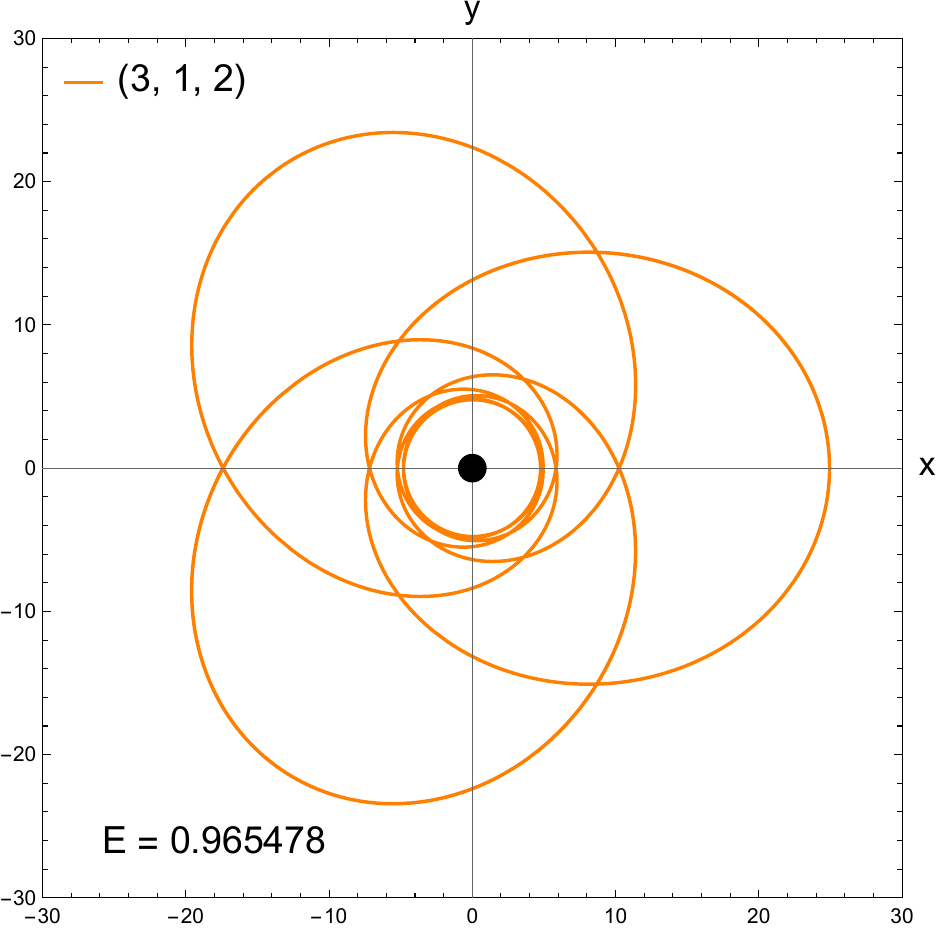} \hfill
    \includegraphics[width=0.32\textwidth]{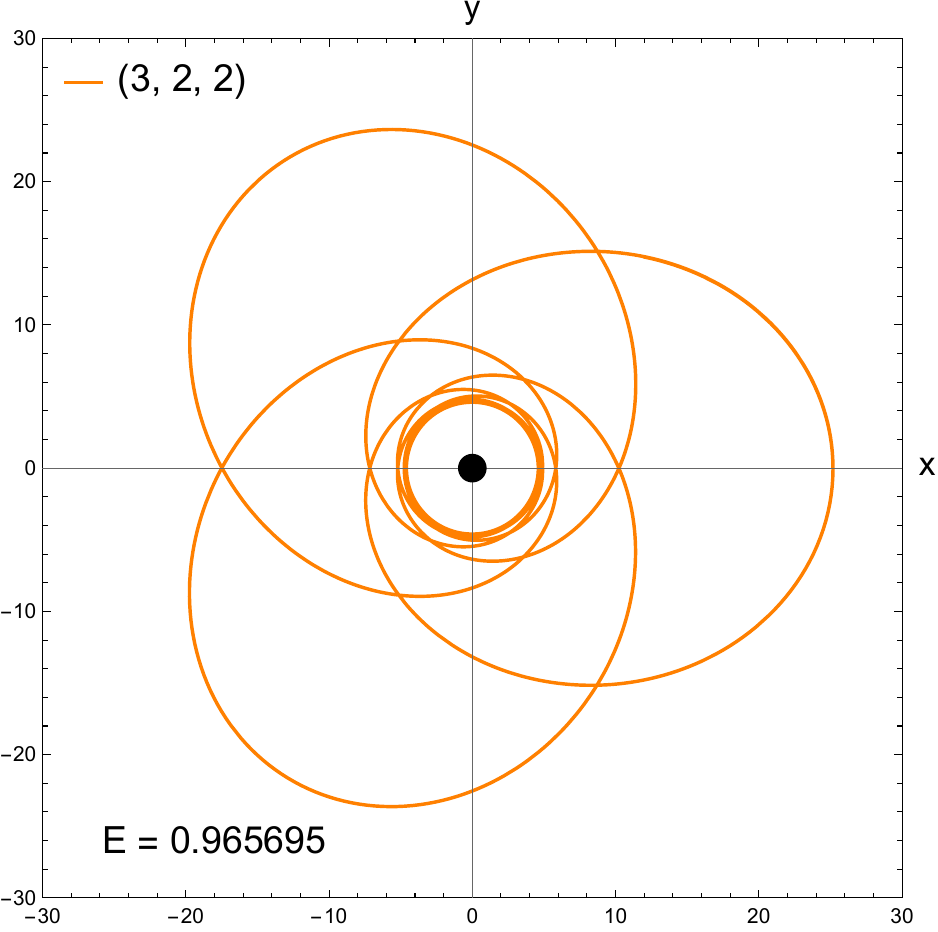} \hfill
    \includegraphics[width=0.32\textwidth]{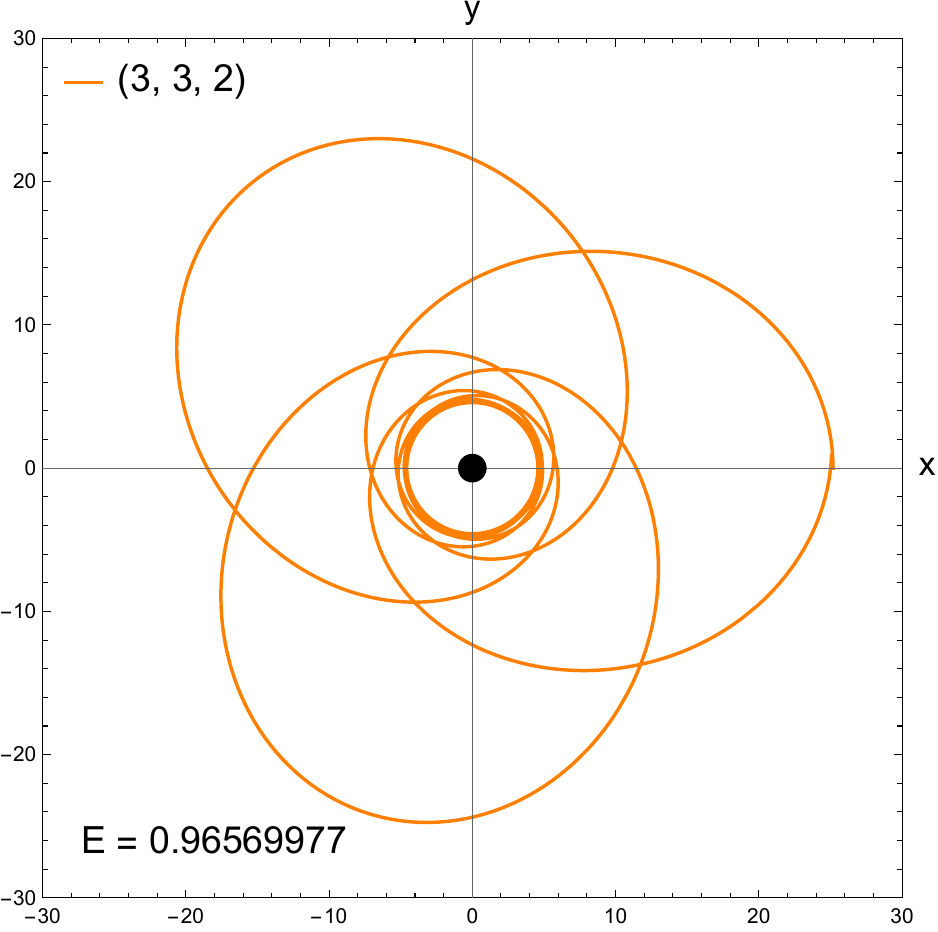} \\
    
    \vspace{0.2cm} 
    \includegraphics[width=0.32\textwidth]{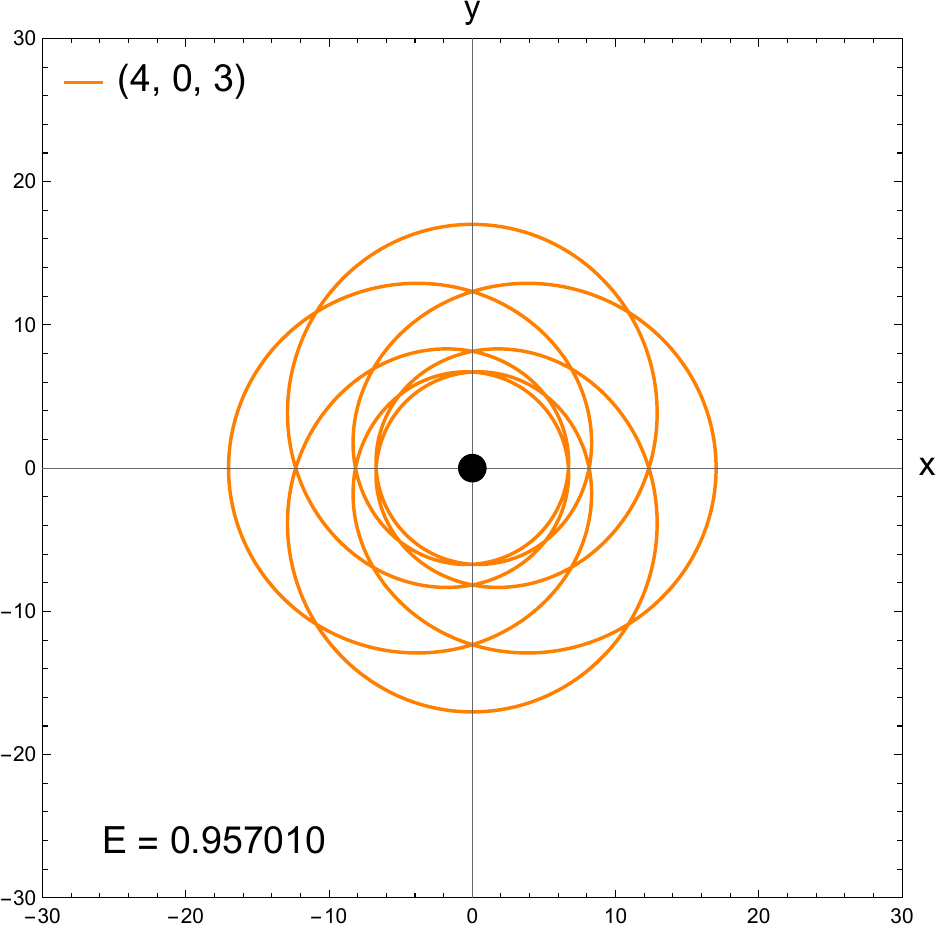} \hfill
    \includegraphics[width=0.32\textwidth]{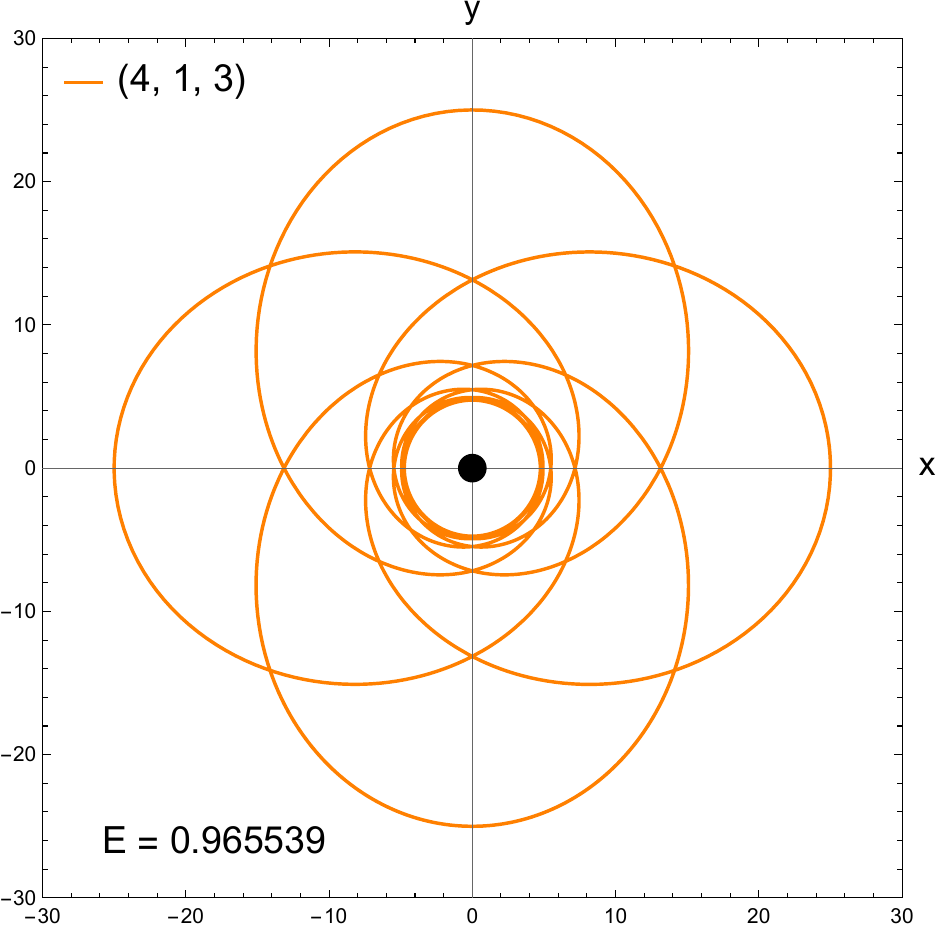} \hfill
    \includegraphics[width=0.32\textwidth]{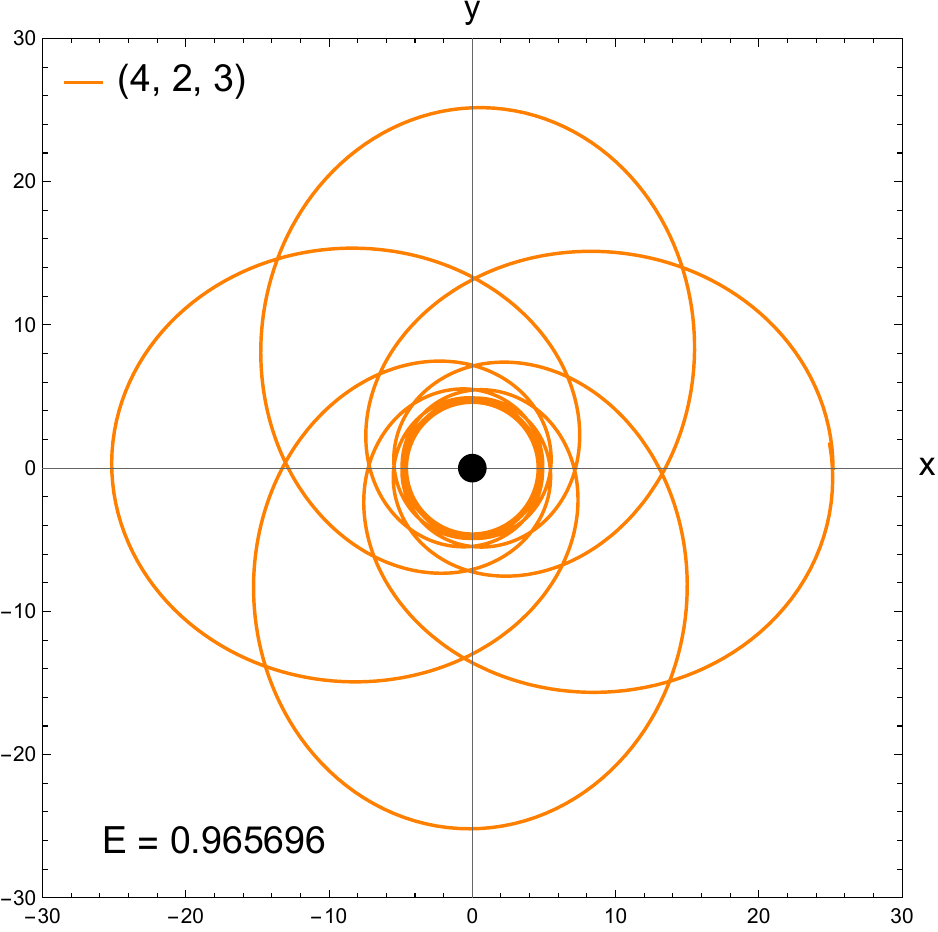}
    
    \caption{Periodic orbits corresponding to various $(z, w, v)$ values around a Schwarzschild BH in a King DM halo. Parameters: $\rho_s = 0.1$, $r_s = 0.3$, and $L = \tfrac{1}{2}(L_{\mathrm{MBO}} + L_{\mathrm{ISCO}})$.}
   \label{fig:periodic}
\end{figure*}
\begin{figure*}[htbp]
   \centering
    \includegraphics[width=0.32\textwidth]{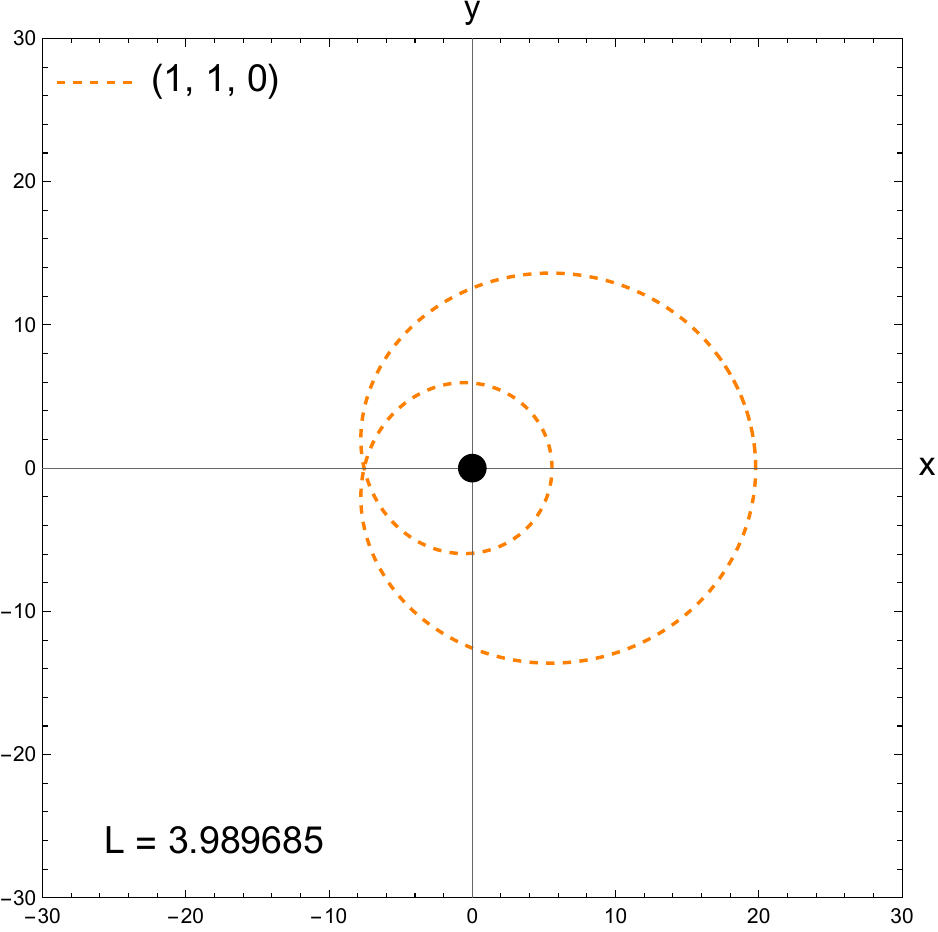} \hfill
    \includegraphics[width=0.32\textwidth]{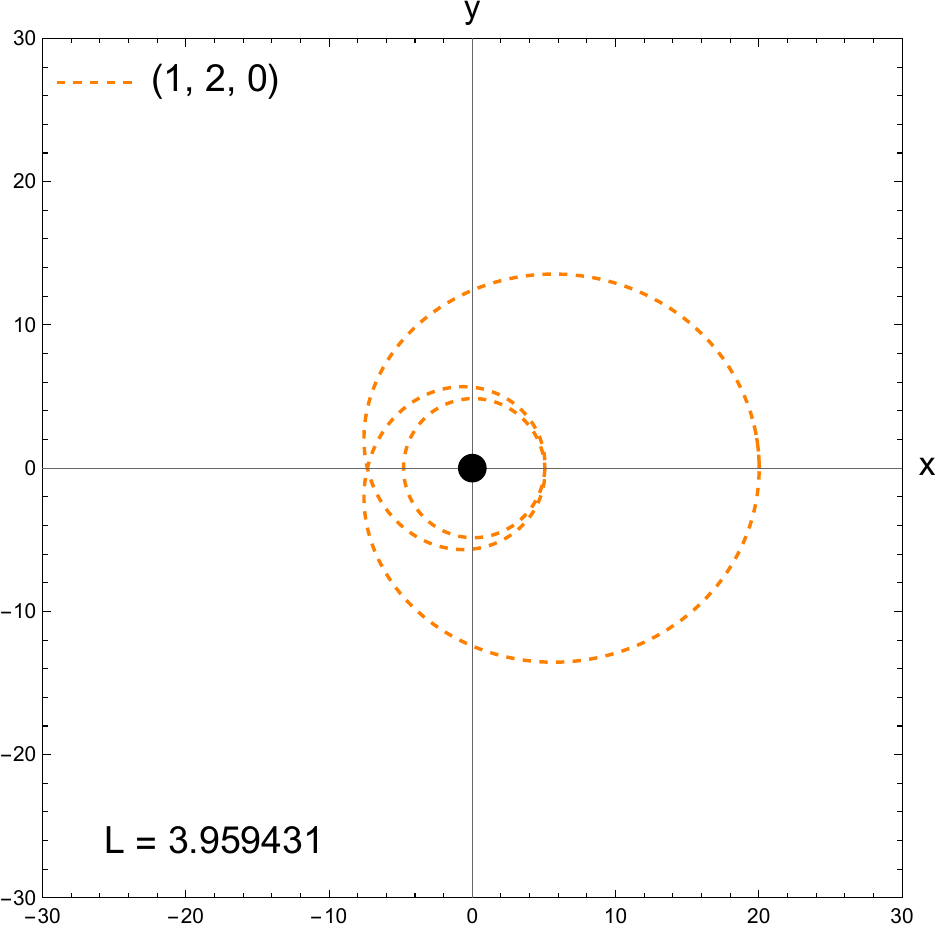} \hfill
    \includegraphics[width=0.32\textwidth]{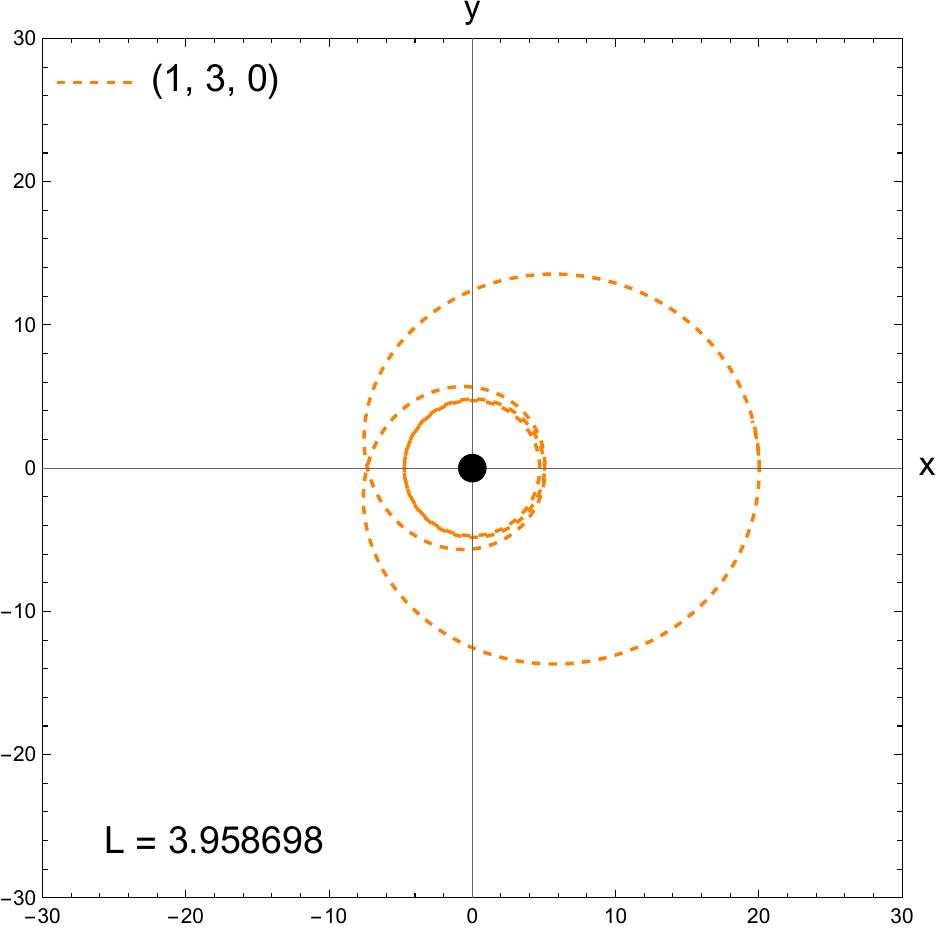} \\
    
    \vspace{0.2cm} 
    \includegraphics[width=0.32\textwidth]{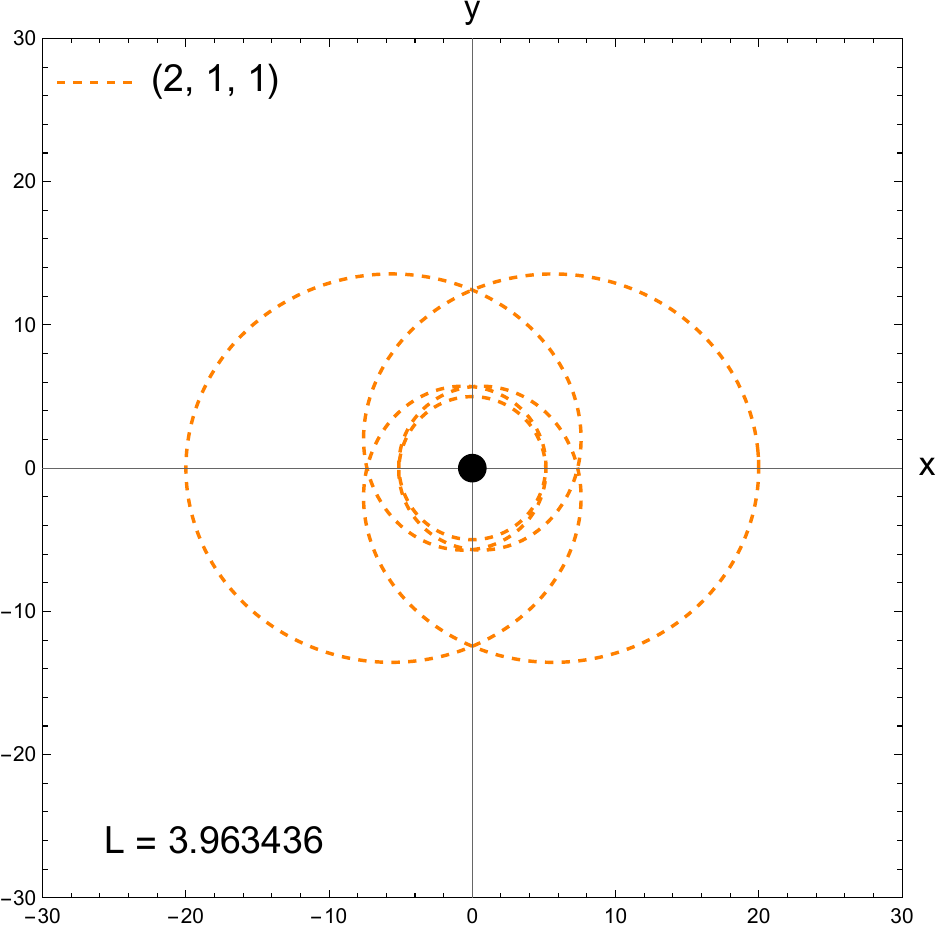} \hfill
    \includegraphics[width=0.32\textwidth]{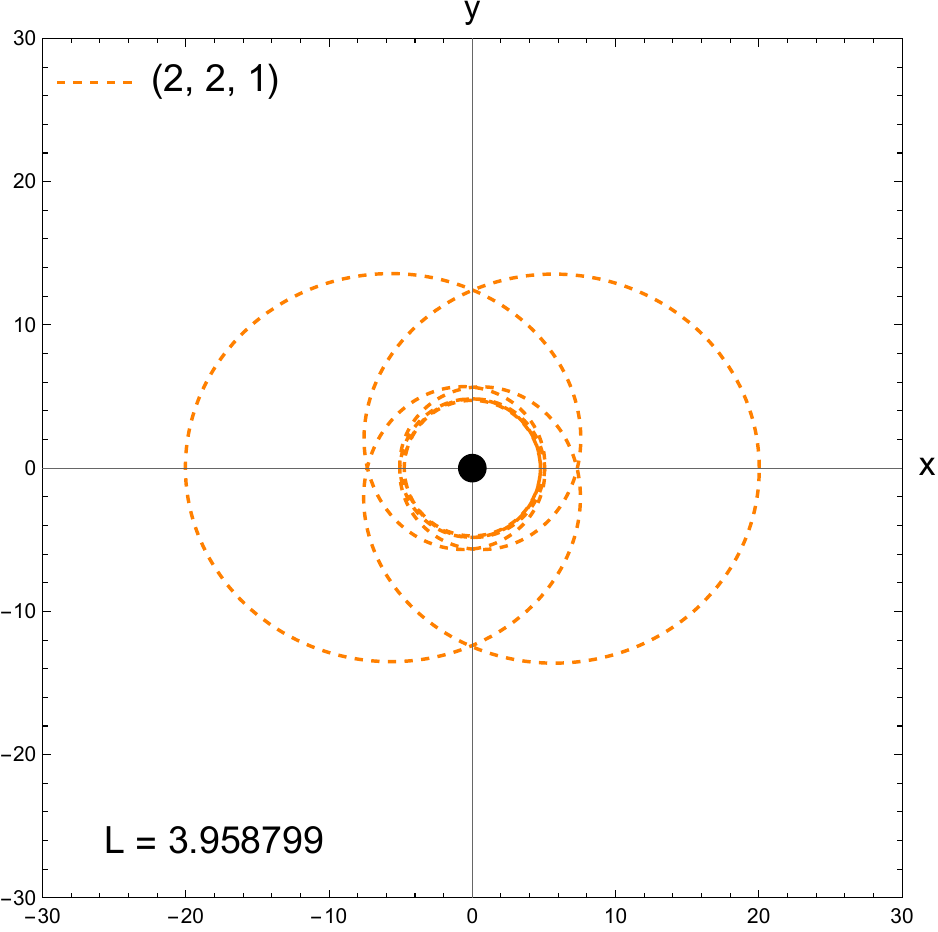} \hfill
    \includegraphics[width=0.32\textwidth]{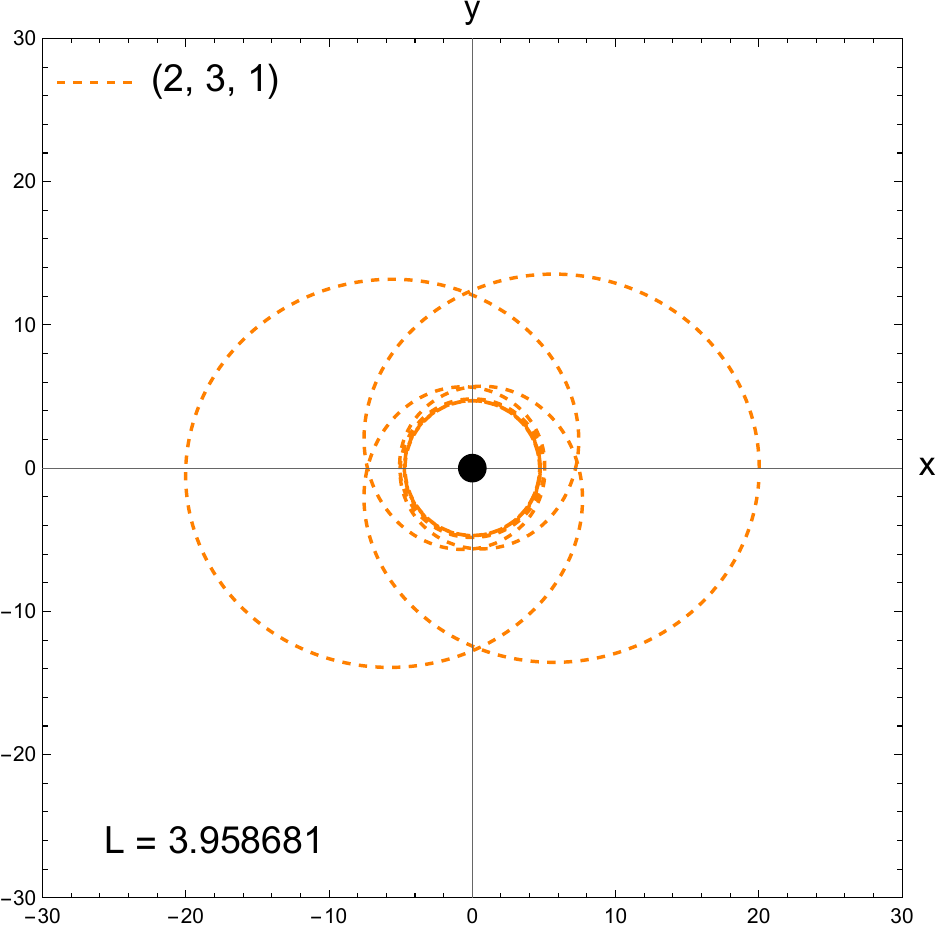} \\
    
    \vspace{0.2cm} 
    \includegraphics[width=0.32\textwidth]{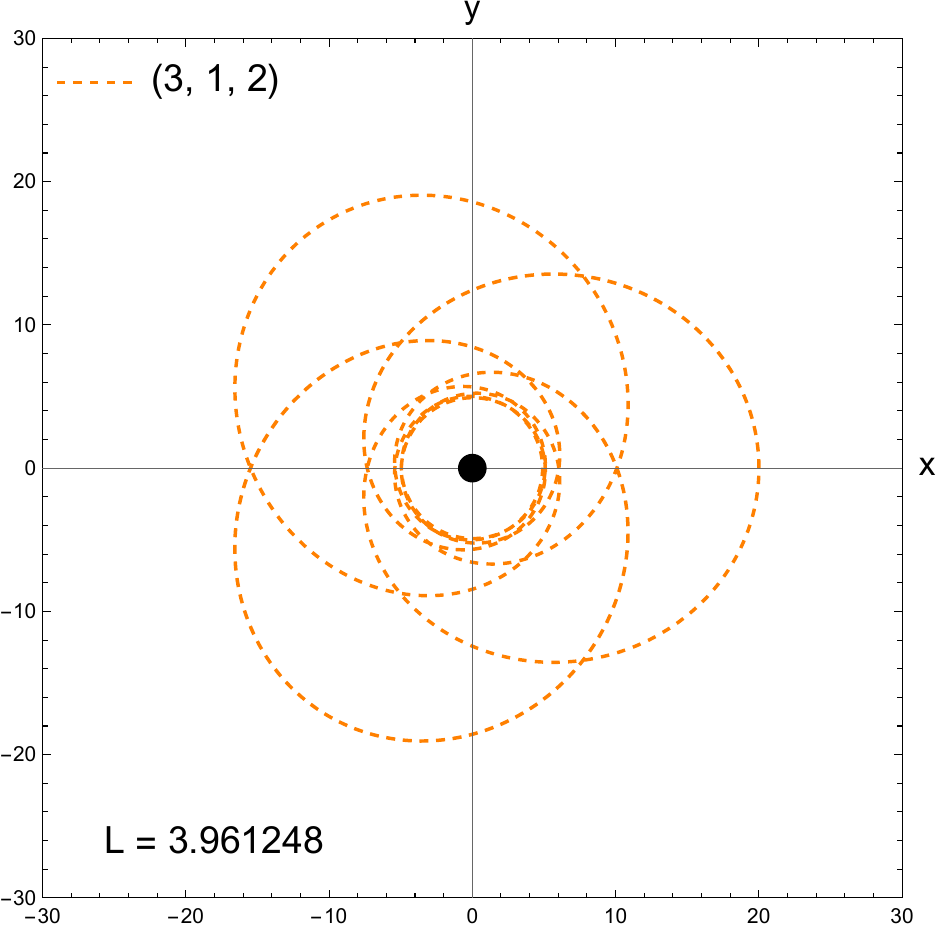} \hfill
    \includegraphics[width=0.32\textwidth]{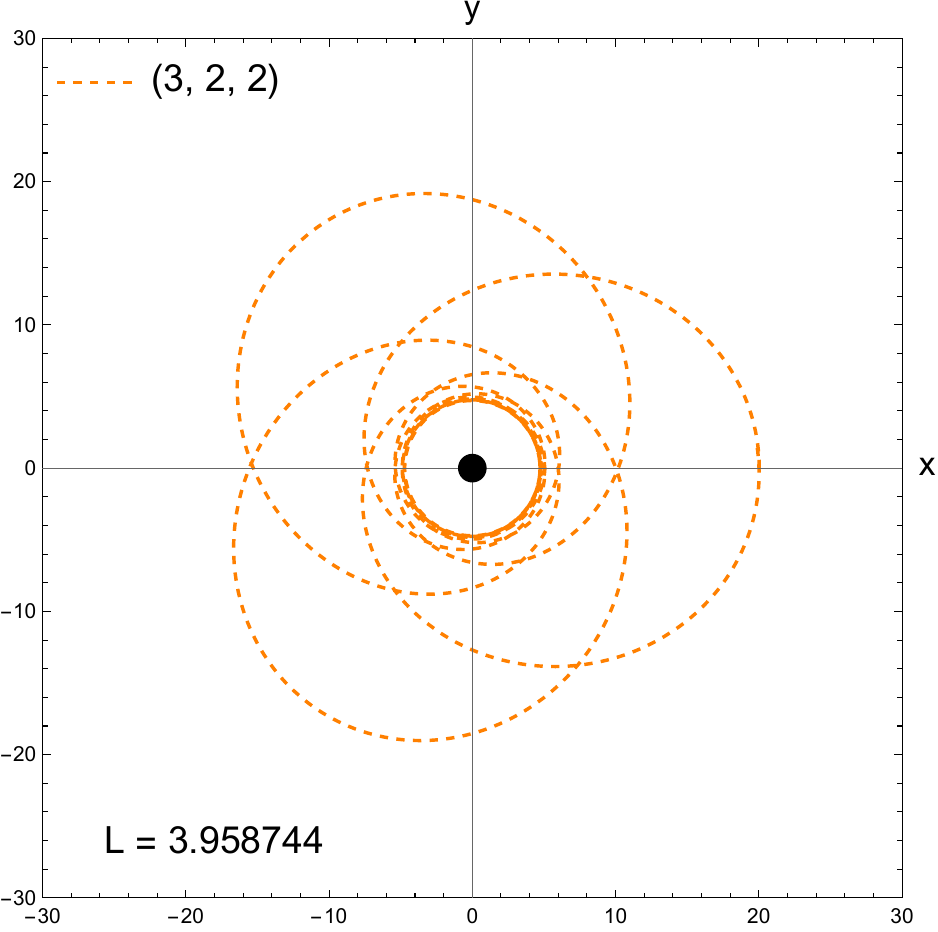} \hfill
    \includegraphics[width=0.32\textwidth]{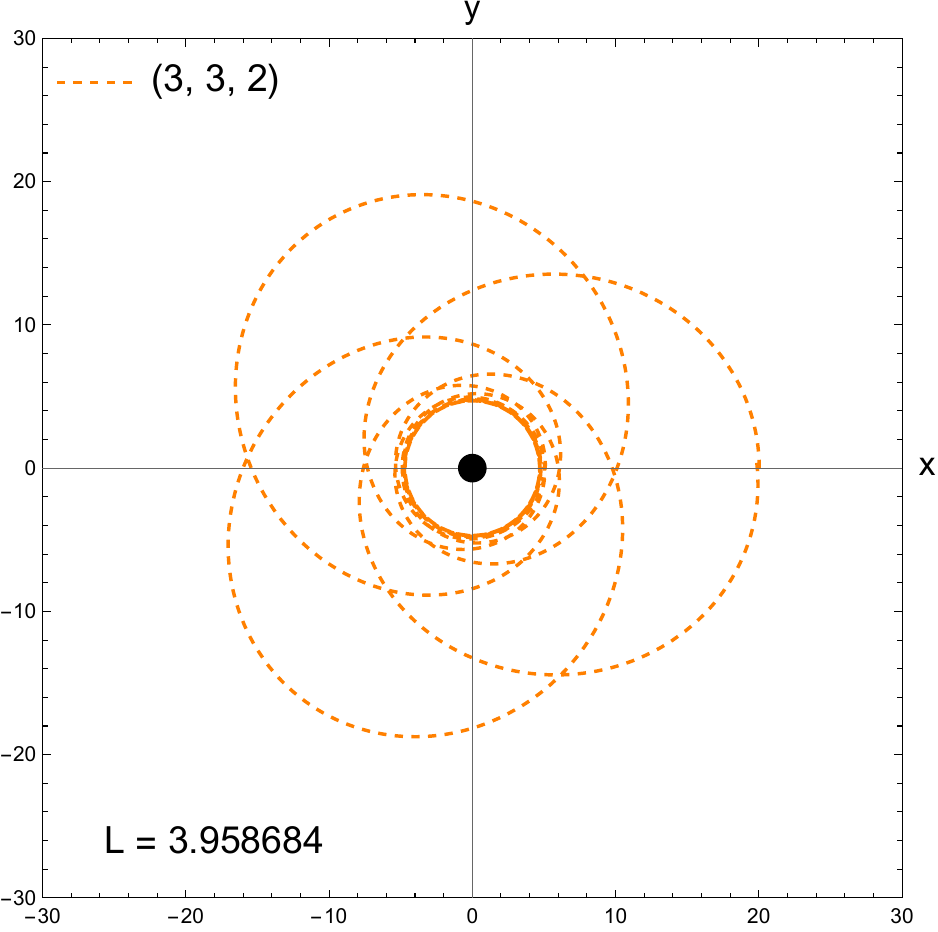} \\
    
    \vspace{0.2cm} 
    \includegraphics[width=0.32\textwidth]{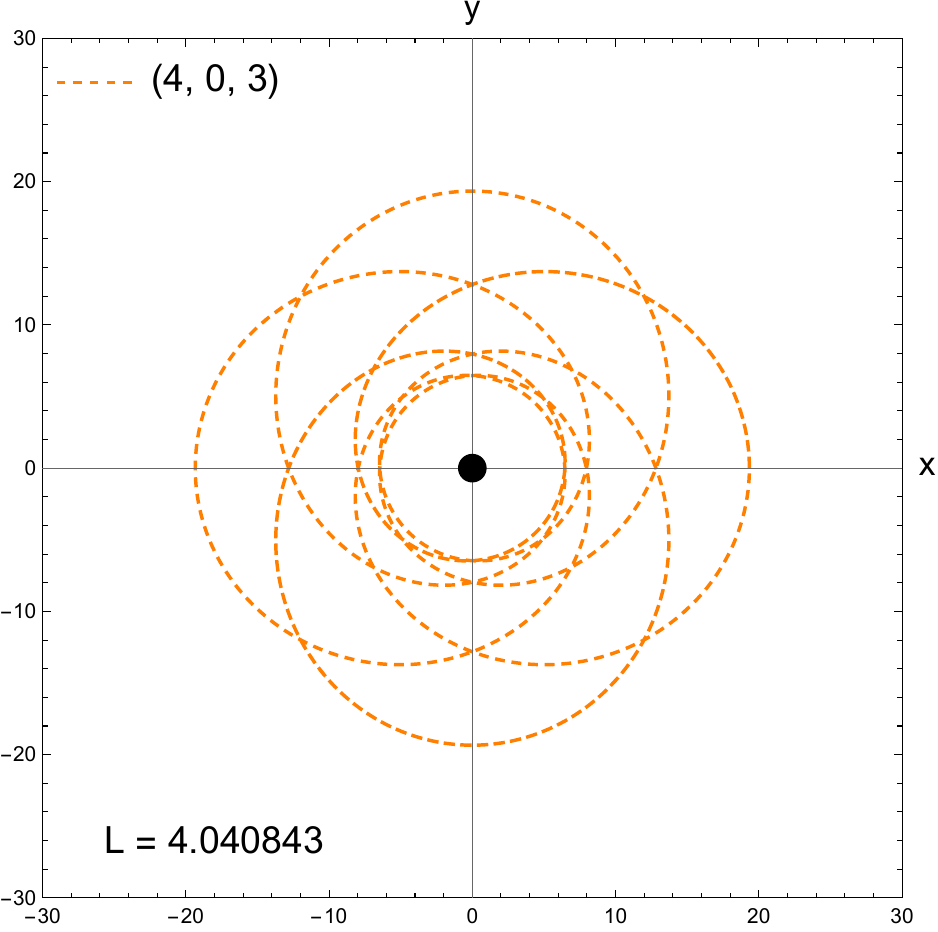} \hfill
    \includegraphics[width=0.32\textwidth]{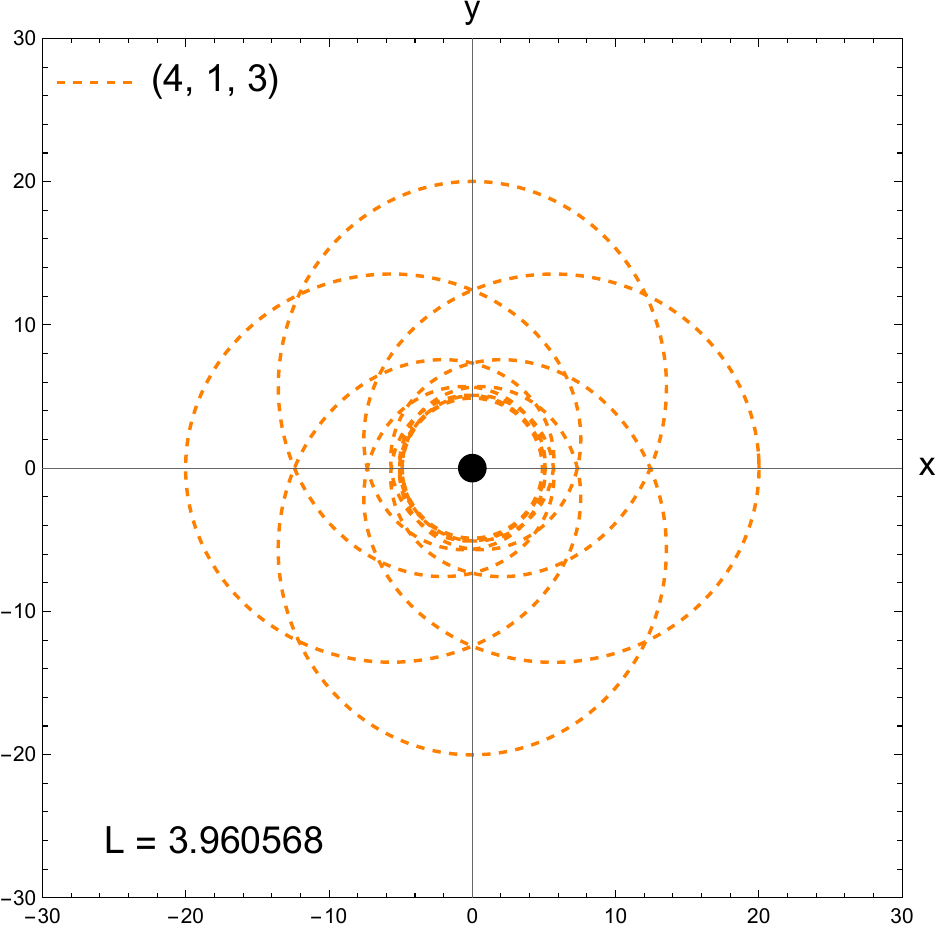} \hfill
    \includegraphics[width=0.32\textwidth]{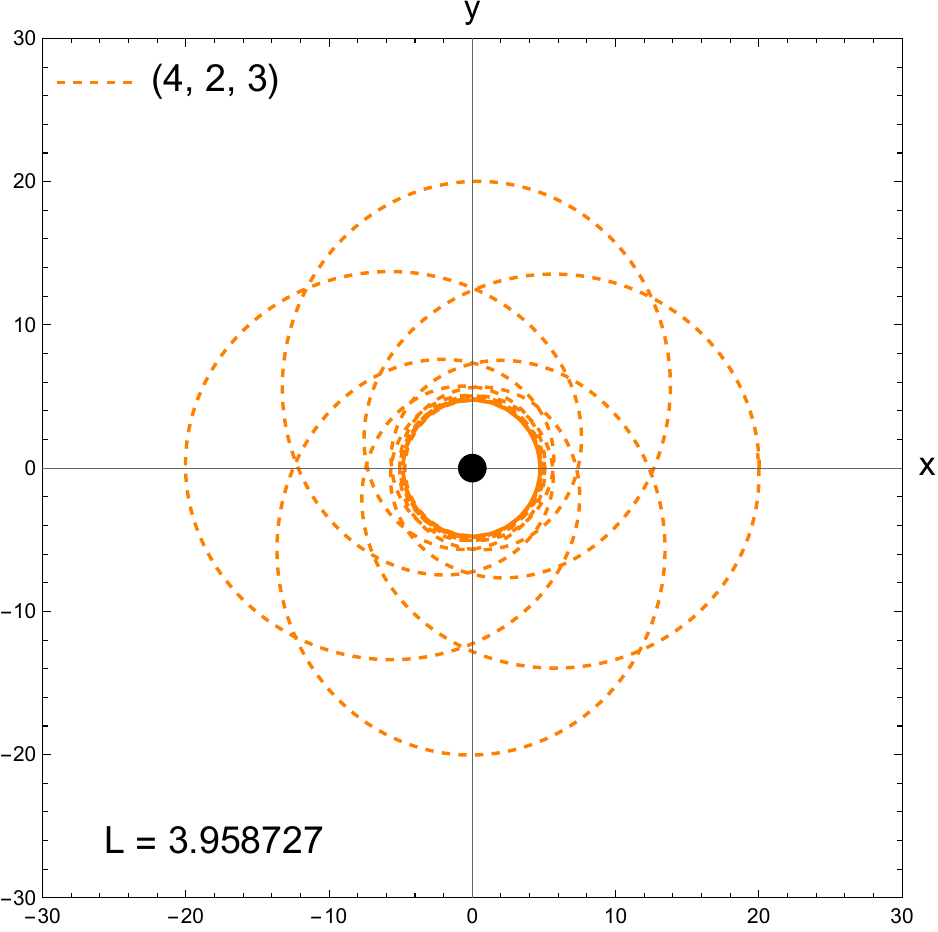}
    
    \caption{Periodic orbits corresponding to various $(z, w, v)$ values around a Schwarzschild BH in a King DM halo. Parameters: $\rho_s = 0.1$, $r_s = 0.3$, and $E = 0.96$.}
    \label{fig:periodicL}
\end{figure*}

\section{Periodic orbits around the Schwarzschild BH surrounded by dark matter halo}\label{Sec:III}

In order to explore the impact of the King DM halo on particle motion, we also analyze the corresponding periodic orbits. Among the bound orbits, periodic orbits play a special role. For periodic orbits, the fundamental orbital frequencies maintain a rational relationship. Each orbit can be uniquely described by a triplet of integers $(z, w, v)$, referred to as the zoom, whirl, and vertex numbers, respectively.
The rational form of this relation is expressed as~\cite{Levin_2008}
\begin{equation}
    q=\frac{\omega_{\phi}}{\omega_r}-1=w+\frac{v}{z}\, ,
\end{equation}
where $\omega_{\phi}$ and $\omega_{r}$ denote the angular and radial frequencies, respectively. From the equations of motion, the rational number can be written as~\cite{2025JCAP...01..091Y,SHABBIR2025101816,JIANG2024101627}:
\begin{equation}
    q=\frac{1}{\pi}\int^{r_2}_{r_1} \frac{\dot{\phi}}{\dot{r}}-1=\frac{1}{\pi} \int^{r_2}_{r_1}\frac{1}{r^2\sqrt{E^2-V_{eff}}}\, ,
\end{equation}
where $r_1$ and $r_2$ are the periapsis and apoapsis radii, respectively.

Figure~\ref{fig:q} illustrates the dependence of the rational number $q$ on the energy and orbital angular momentum of the particle for different values of the King DM halo parameters. From the top panel of Fig.~\ref{fig:q}, it is evident that the rational number $q$ increases with increasing energy $E$, exhibiting a sharp rise as $E$ approaches its maximum value. Additionally, as the King DM halo parameters increase, the corresponding $q$–$E$ curves shift toward lower energy values. The bottom panel shows the dependence of $q$ on the orbital angular momentum $L$. In the bottom panel, $q$ is plotted against the orbital angular momentum $L$, showing a sharp rise as $L$ approaches its minimum and a gradual decrease at higher values of $L$.

The energies $E$ of periodic orbits with different $(z, w, v)$ values are obtained numerically for a fixed angular momentum $L = \tfrac{1}{2}(L_{\mathrm{MBO}} + L_{\mathrm{ISCO}})$, as shown in Table~\ref{table1}. Using these data, Fig.~\ref{fig:comparison} illustrates a comparison between the periodic orbit $(1, 1, 1)$ in the Schwarzschild spacetime and the same orbit when the BH is embedded in the King DM halo. Furthermore, Fig.~\ref{fig:periodic} depicts periodic orbits with different $(z, w, v)$ configurations for $L = \tfrac{1}{2}(L_{\mathrm{MBO}} + L_{\mathrm{ISCO}})$, $r_s = 0.3$, and $\rho_s = 0.1$. As can be seen from Table~\ref{table1} and Fig.~\ref{fig:comparison}, the presence of DM leads to an outward expansion of the periodic orbits and a corresponding decrease in their energies compared to the pure Schwarzschild case.

Table~\ref{table2} lists the orbital angular momentum $L$ for periodic orbits characterized by different $(z, w, v)$ parameters, assuming $E = 0.96$ and $r_s = 0.2$. To further illustrate, we present in Fig.~\ref{fig:periodicL} the periodic orbits for different $(z, w, v)$ configurations with $\rho_s = 0.1$ and $r_s = 0.3$. As observed, orbits with higher zoom numbers $z$ exhibit increasingly complex geometrical structures, while orbits with larger whirl numbers $w$ make more revolutions around the central BH between successive apoapses.
\begin{figure*}
\centering
  \includegraphics[scale=0.72]{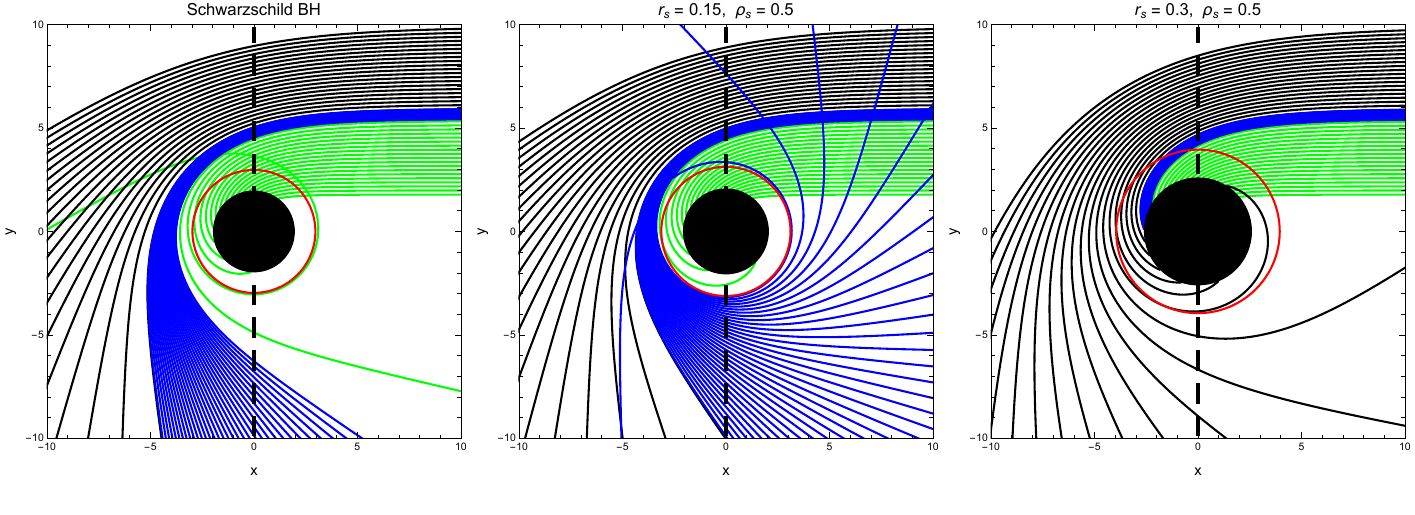}
  \caption{The behavior of photon trajectories near the BH within the King DM halo is plotted for different values of $r_s$
and fixed $\rho_s$. In all plots, the colored lines represent different ranges of the impact parameter $b$, e.g., green corresponds to  $1.8<b\leq5.5$, blue to $5.5<b\leq6$, while black to $b>6$. We note that the red ring indicates the location of the photon sphere.}
\label{fig:ray1}
\end{figure*}
\begin{figure*}
   \centering
  \includegraphics[scale=0.6]{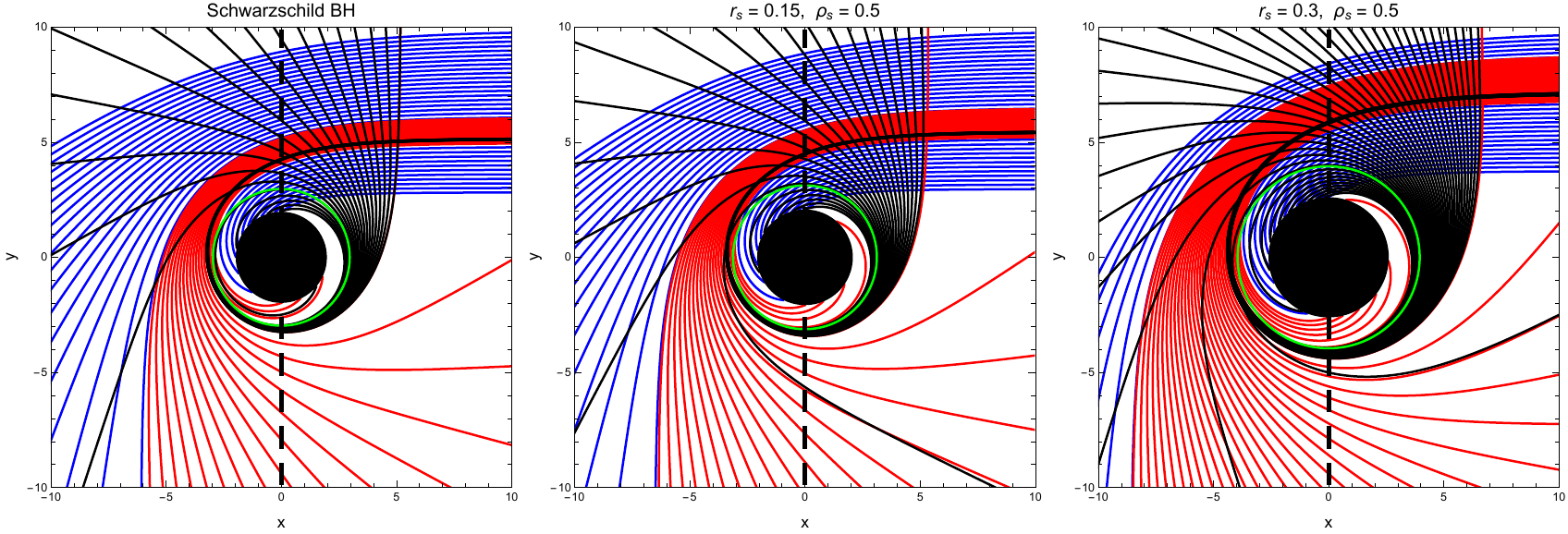}
 \caption{The behavior of photon trajectories near the BH within the King DM halo is plotted for different values of $r_s$
and fixed $\rho_s$ for the ranges of the total number of revolutions, $\eta$, defining the number of times the ray crosses the vertical axis. In all plots, the colored lines represent different ranges of $\eta$, e.g., blue corresponds to $1/4<\eta<3/4$, red to $3/4<\eta<5/4$, and black to $\eta>5/4$. We note that the green ring indicates the location of the photon sphere.}
\label{fig:ray2}
\end{figure*}
\begin{figure} 
    \centering
    \includegraphics[scale=0.41]{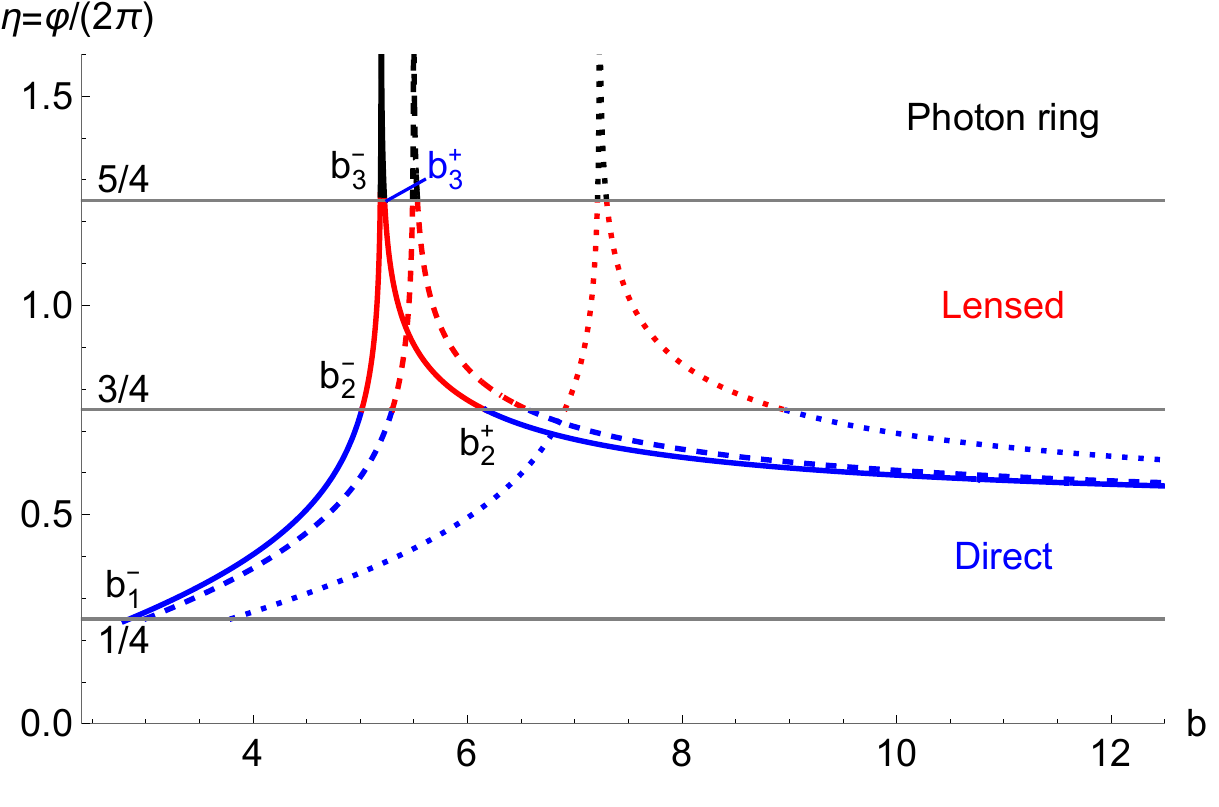}
    \caption{The plot illustrates the dependence of the total number of revolutions $\eta$ on the impact parameter $b$. The solid, dashed, and dotted curves correspond to the Schwarzschild BH case, and to the cases with $r_s=0.15$ and $r_s=0.3$, respectively, for the fixed $\rho_s=0.5$.}
 \label{F_b1}
\end{figure}

\section{Null geodesics around the BH surrounded by the King DM halo }\label{Sec:IV}

We now turn our attention to the null geodesics around a Schwarzschild BH embedded in the King DM halo. The four-momentum of a massless particle (photon) can be expressed as follows:    
\begin{equation} \label{momentum-photon}
    p^\alpha = dx^\alpha/d\lambda \, .
\end{equation}  
For simplicity, we restrict the photon motion to the equatorial plane ($\theta = \pi/2$). Using Eqs.~\eqref{eq:eqmotion}, \eqref{momentum-photon}, and the normalization condition, we then obtain the following equations governing light propagation:
\begin{equation}
    \frac{dt}{d\lambda}=\frac{E}{f(r)}, \quad
    \frac{d\phi}{d\lambda}=\frac{L}{r^2}\ ,
\end{equation}    
\begin{equation}
    \left(\frac{dr}{d\lambda}\right)^2=E^2-f(r)\frac{L^2}{r^2}\ ,
\end{equation}
where $\lambda$ is the affine parameter, which is related to the proper time $\tau$ by $\lambda = \tau / m$.
\begin{table*}
\centering
\caption{The values of event horizon $r_h$ and the photon sphere radius $r_{ph}$, and the impact parameter $b_n^\pm$ for different values of DM halo parameter $r_s$. Note that we set the DM halo density profile $\rho_s=0.5$.}
\label{tab:nb}
\renewcommand{\arraystretch}{1.3}
\resizebox{0.8\textwidth}{!}{
\begin{tabular}{|c|c|c|c|c|c|c|c|c|}
\hline
\textbf{BHs} & \textbf{$r_h$} & \textbf{$r_{ph}$} & \textbf{$b^-_1$} & \textbf{$b^-_2$} & \textbf{$b^-_3$} & \textbf{$b_c$} & \textbf{$b^+_3$} & \textbf{$b^+_2$} \\
\hline
SchwBH & 2& 3& 2.848 & 5.015 & 5.188\ & $5.196$ & 5.228& 6.168 \\
\hline
\textbf{$r_s=0.15$} & 2.099& 3.153& 2.993& 5.299& 5.489&  5.498& 5.534& 6.571\\
\hline
%\textbf{$r_s=0.3, \rho_s=0.25$} & 2.295& 3.459& 3.285 & 5.9 & 6.134& 6.147 & 6.194 & 7.41\\
%\hline
\textbf{$r_s=0.3$} & 2.637& 3.991& 3.788 & 6.907& 7.213 & 7.232& 7.298& 8.958 \\
\hline
\end{tabular}
}
\end{table*}

By applying the transformation $\lambda' = L\lambda$ and substituting the impact parameter $b = L/E$, we obtain the following the equations:
\begin{equation}
    \frac{dt}{d\lambda'}=\frac{1}{bf(r)}, \quad
    \frac{d\phi}{d\lambda'}=\frac{1}{r^2}\ ,
\end{equation} 
\begin{equation}
    \left(\frac{dr}{d\lambda'}\right)^2=\frac{1}{b^2}-\frac{f(r)}{r^2}\ .
    \label{radialM}
\end{equation} 
From the above equations, one can Eq.~\eqref{radialM} in terms of the azimuthal angle $\phi$ as follows: 
\begin{equation}
    \frac{1}{r^4}\left(\frac{dr}{d\phi}\right)^2=\frac{1}{b^2}-\frac{f(r)}{r^2}=V_{eff}(r)\, ,
    \label{radialPhi}
\end{equation}
where $V_{\mathrm{eff}}$ is the effective potential for a massless particle. The critical impact parameter $b_c$ can be determined by substituting the photon-sphere radius $r = r_{\mathrm{ph}}$ into the condition $V_{\mathrm{eff}}(r) = 0$, yielding
\begin{equation}
b_c = \frac{r_{ph}}{\sqrt{f(r_{ph})}}\ .
\end{equation}
With the substitution $r=1/u$, Eq.~\eqref{radialPhi} becomes:
\begin{equation}
    \left( \frac{du}{d\phi} \right)^2 = \frac{1}{b^2} - u^2 f\left(u \right)\equiv G(u)\ .
\end{equation}
Using the metric function given in Eq.~\eqref{metric}, we obtain the complete expression for $G(u)$
\begin{align}
    G(u)=&\frac{1}{b^2}-u^2+2 M u^3 -\frac{8 \pi  \rho_s r_s^3}{\sqrt{r_s^2+\frac{1}{u^2}}}u^2\nonumber\\&-8 \pi  \rho_s r_s^3 u^3 \log \left(\frac{\sqrt{r_s^2+\frac{1}{u^2}}-\frac{1}{u}}{r_s}\right)\ .
\end{align}
The total azimuthal deflection $\varphi$ is represented by the following expression as a function of the impact parameter $b$:
\begin{equation}
\varphi = 
\begin{cases} 
\displaystyle \ \ \
\int\limits_0^{u_h} \frac{du}{\sqrt{G(u)}}\ ,  \quad & b < b_c\  \\[12pt]
\displaystyle
2 \int\limits_0^{u_{\min}} \frac{du}{\sqrt{G(u)}}\ , \quad & b > b_c\ ,
\end{cases}
\end{equation}
where $u_h \equiv 1/r_h$ denotes the inverse of the event horizon radius, and the turning point $u_{\mathrm{min}}$ is determined by selecting the smallest positive root of the equation $G(u) = 0$.

Parallel rays coming from infinity experience different deflection angles near the black hole depending on their impact parameters $b$. This phenomenon is illustrated in Fig.~\ref{fig:ray1}, which depicts the trajectories of photons near the BH. In the Figure, the colors represent different ranges of $b$: green for $1.8<b\leq5.5$, blue for $5.5<b\leq6$, and black for $b>6$. By setting $(r_s, \rho_s)$ to $(0.15, 0.5)$, $(0.3, 0.25)$, and $(0.3, 0.5)$, the corresponding critical impact parameters are obtained as $b_c = 5.498$, $6.147$, and $7.232$, respectively. In the figure, the central black region corresponds to the event horizon, and the red ring represents the photon sphere. The radii of both increase with larger DM halo parameters. As the DM halo parameters increase, the blue rays bend more strongly and fall into the Schwarzschild black hole surrounded by a King DM halo.

It should be emphasized that the photon trajectories around a BH can be classified into three distinct trajectories, such as direct, lensed, and photon rings (see, e.g., \cite{Gralla_2019}). To define the boundaries of these distinct trajectories, we employ the following equation for the total number of revolutions ($\eta = \frac{\varphi}{2\pi}$)~\cite{Peng_2021}
\begin{eqnarray}
    \eta = \frac{2N-1}{4}, \quad N = 1, 2, 3, \cdots\ ,
\end{eqnarray}
where $N$ denotes the number of intersection points. To gain a better understanding of the total number of revolutions, we show the photon trajectories near the BH in the presence of the King DM halo in Fig.~\ref{fig:ray2}. As can be seen in Fig.~\ref{fig:ray2}, $\eta$ represents the number of times the ray crosses the vertical axis. This classification can also be described using the impact parameter $b$, whose values for the Schwarzschild BH case and different scale radii $r_s$ are listed in Table~\ref{tab:nb}. The classification boundaries, expressed in terms of the total number of revolutions $\eta$, are given as follows:
\begin{itemize}
\item Direct trajectory: \( N = 1 \Rightarrow \frac{1}{4} < \eta < \frac{3}{4} \), corresponding to \( (b_1^{-},\, b_2^{-}) \cup (b_2^{+},\, \infty) \), shown by blue curves in Fig.~\ref{fig:ray2}.
\item Lensed trajectory: \( N = 2 \Rightarrow \frac{3}{4} < \eta < \frac{5}{4} \), \( (b_2^{-},\, b_3^{-}) \cup (b_3^{+},\, b_2^{+}) \), 
red curves.
\item Photon ring trajectory: \( N \ge 3 \Rightarrow \eta > \frac{5}{4} \), \( (b_3^{-},\, b_3^{+}) \),  
black curves.
\end{itemize}

Here we note that the data presented in Table~\ref{tab:nb} is reflected in Figs.~\ref{fig:ray2} and~\ref{F_b1}. In Fig.~\ref{F_b1}, we show the dependence of the total number of revolutions $\eta$ on the impact parameter $b$. We note that, in Fig.~\ref{F_b1}, the peaks represent the critical value of the impact parameters $b_c$, while three horizontal lines exhibit three fixed values, such as $\eta=1/4$, $3/4$, and $5/4$. For being more informative, the points where the curve intersects these horizontal lines are labeled ($b_1^-$, $b_2^-$, $b_2^+$) and ($b_3^-$, $b_3^+$), respectively. It is obvious that increasing the DM halo parameters shifts the critical impact parameter to its larger values, indicating an enhancement of the gravitational field, as shown in Fig.~\ref{F_b1}. In Table~\ref{tab:nb}, we tabulate the values of physical quantities, $r_h$, $r_{ph}$, and $b_n^\pm$, for different values of the King DM halo parameter. As can be seen in Table~\ref{tab:nb}, these quantities increase with increasing the scale radius $r_s$ of the DM halo. Notably, once $b$ approaches the critical value, $b\to b_c$), the total number of revolutions $\eta$ tends to infinity, $\eta\to\infty$, as seen in Fig.~\ref{F_b1}. One can deduce that, as the King DM halo parameters $r_s$ and $\rho_s$ allow the geometry of the BH surrounded by the King DM halo to be more noticeable from the Schwarzschild BH. In particular, the photon sphere and the photon sphere radius become larger. To be more informative, the critical impact parameter refers to $b_c\approx5.196$ for the Schwarzschild BH case, while the photon ring lies within $5.188 <b<5.228$. In the presence of the Kind DM halo, specifically for $r_s = 0.3$ and $\rho_s=0.5$, the critical impact parameter and the photon ring increase to approximately $b_c\approx7.232$ and $7.213 <b <7.298$, respectively.
\begin{figure*}[!htb]
    \centering
    \includegraphics[width=0.48\linewidth]{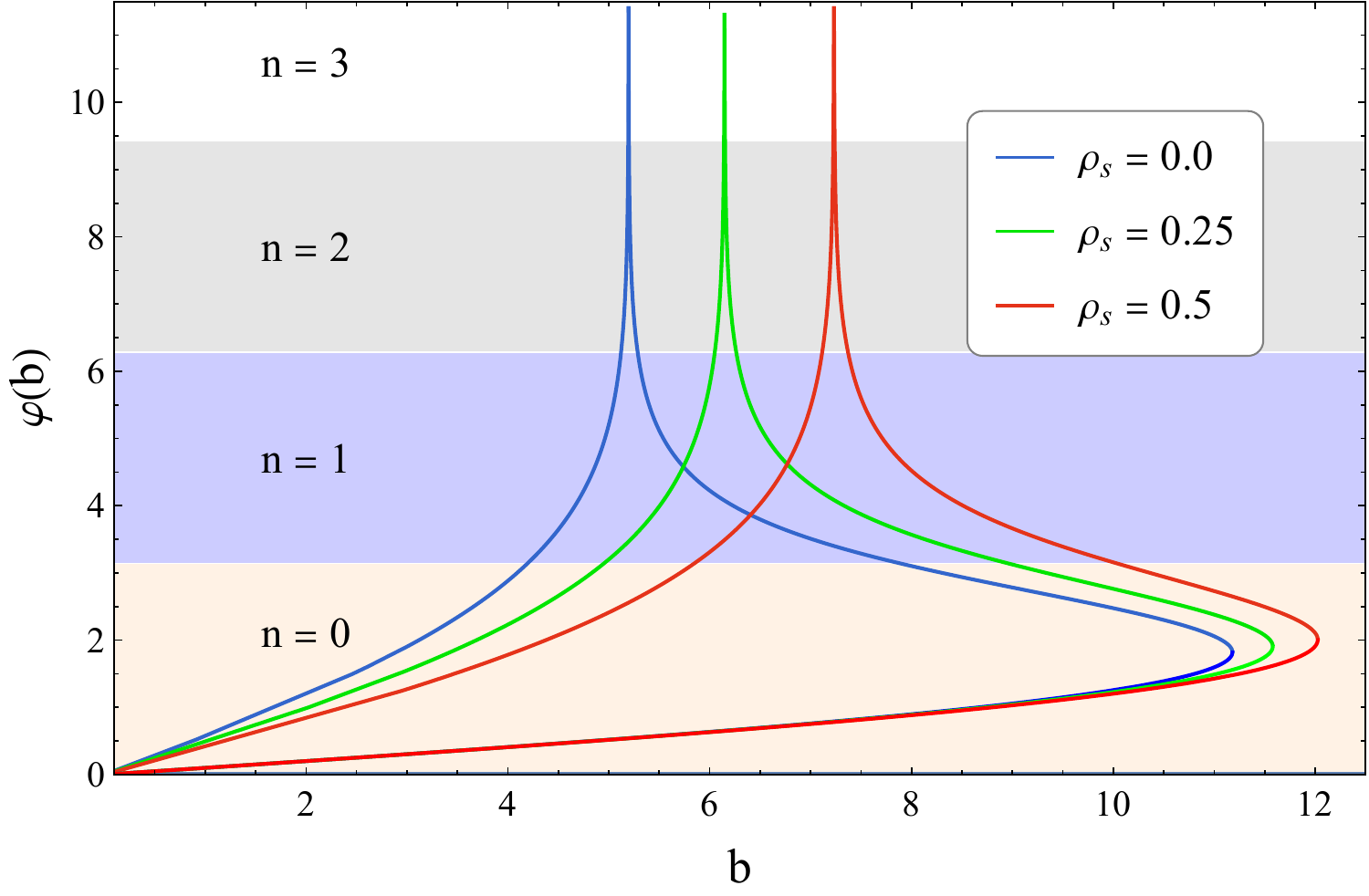}\hspace{0.5cm}
    \includegraphics[width=0.48\linewidth]{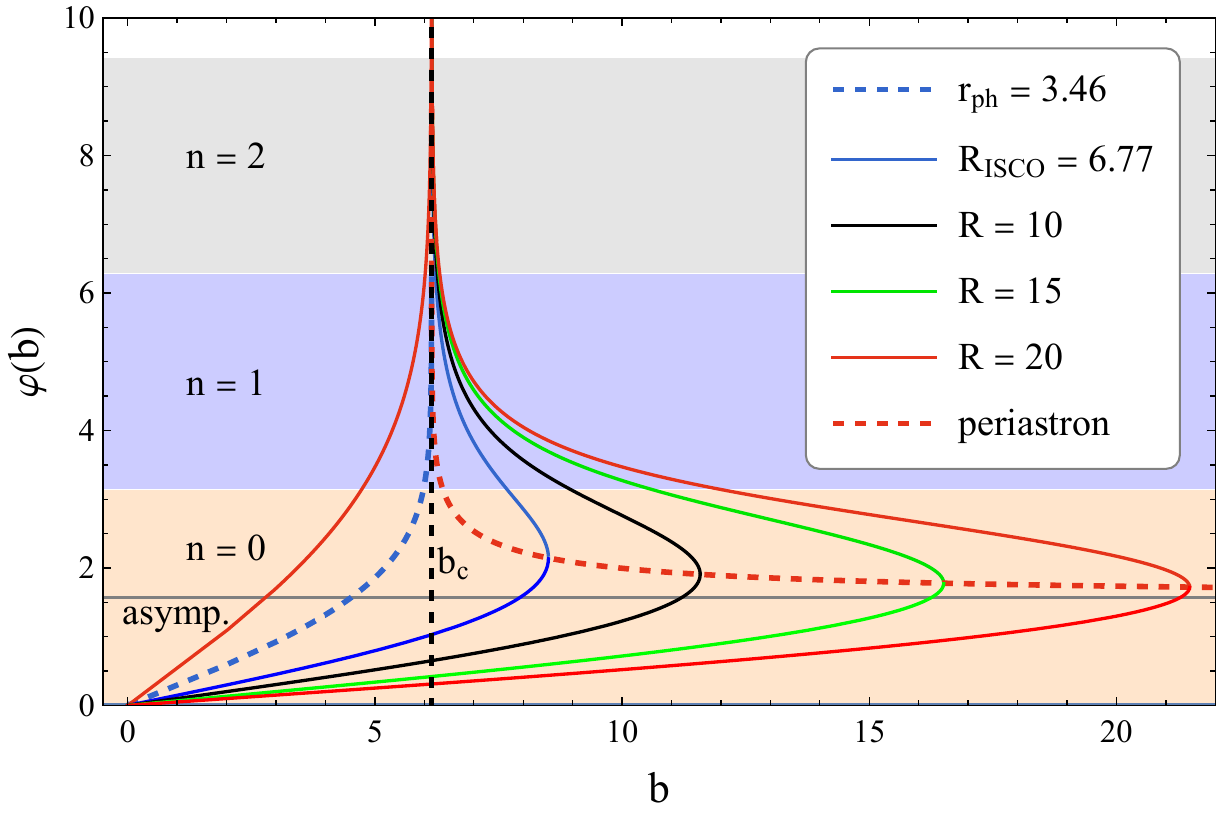}
    \caption{Left panel: Deflection angle of photons, $\varphi(b)$, plotted as a function of the impact parameter $b$ for different values of $\rho_s$, with $r_s = 0.3$ and fixed radius $R = 10$. Right panel: Deflection angle of photons $\varphi(b)$ plotted as a function of the impact parameter $b$ for different radii $R$, with $\rho_s = 0.25$ and $r_s = 0.3$.}
    \label{fig:FBline1}
\end{figure*}

\section{IMAGES AND PHYSICAL PROPERTIES OF THIN ACCRETION DISK around BH surrounded by King DM halo}\label{Sec:V}

In this section, we investigate the radiation properties of the accretion disk around a Schwarzschild BH embedded in a King DM halo. This accretion disk is composed of high-temperature gas, dust, and plasma orbiting the BH. The Novikov–Thorne model~\cite{Novikov:1973kta, Cui24} is a well-established theoretical approach that describes the radiation properties of geometrically thin, optically thick accretion disks. Because the disk is geometrically thin ($h \ll r$), its vertical thickness is negligible compared with its radial extent. Observationally, the main quantities characterizing accretion disks are their electromagnetic flux, temperature profile, and luminosity. In the following, we investigate the influence of the DM halo parameters on the disk’s image, energy flux, and gravitational redshift.
\begin{figure*} 
    \centering
    \includegraphics[scale=0.55]{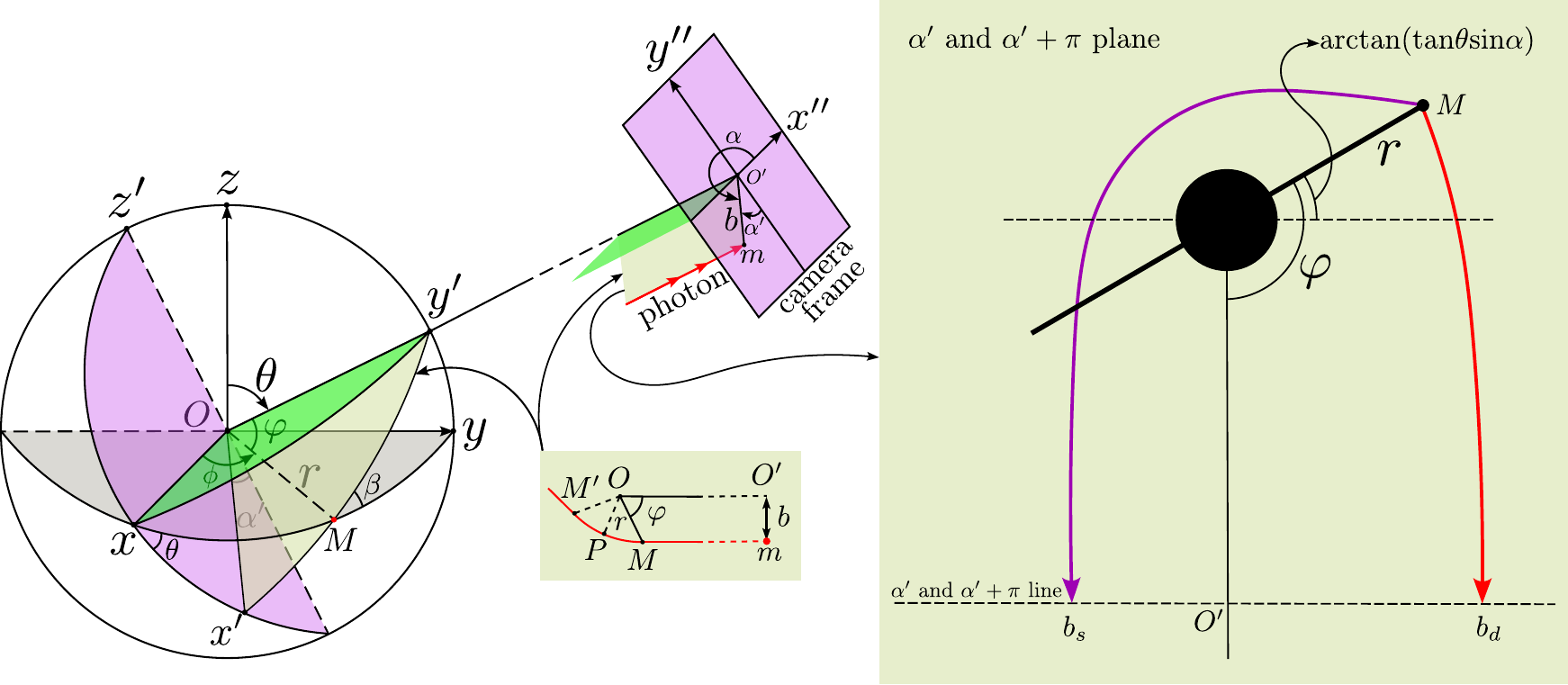}
    
    \caption{Schematic representation of the coordinate system used to construct the accretion disk image \cite{XamidovAccretion2025}.
    }
 \label{fig:coordinate}
\end{figure*}

In order to study the image formation of the thin accretion disk, we introduce an observational coordinate system, illustrated in Fig.~\ref{fig:coordinate}. In this system, the observer is assumed to be located at infinity in spherical coordinates $(r, \theta, \phi)$ centered on the BH, where $(r = 0)$ denotes the origin. In the camera frame $O'x''y''$, a photon emitted from $m(b, \alpha)$ travels perpendicular to the image plane toward the BH and intersects the disk at $M(r, \pi/2, \phi)$. Based on the principle of optical path reversibility, the trajectory of a photon emitted from $M(r, \pi/2, \phi)$ on the accretion disk is identical to that of a photon emitted from the image point $m(b, \alpha)$ in the camera frame.

Our analysis involves two image types. A direct image $(b_d, \alpha)$ is produced when a photon emitted from a point on the accretion disk reaches the camera plane directly. In contrast, a secondary image $(b_s, \alpha + \pi)$ is formed when a photon emitted in the opposite direction from the same point on the disk is deflected by gravitational lensing and subsequently reaches the camera (see the right panel of Fig.~\ref{fig:coordinate}).

Using the sine theorem for spherical triangles, we can derive the following relations from $\triangle Myy'$ and $\triangle Mxx'$~\cite{XamidovAccretion2025}:
\begin{eqnarray}
    \frac{\sin(\varphi)}{\sin(\frac{\pi}{2})} = \frac{\sin(\frac{\pi}{2}-\theta)}{\sin(\beta)}\, ,
\end{eqnarray}
\begin{eqnarray}
    \frac{\sin(\frac{\pi}{2}-\varphi)}{\sin(\theta)} = \frac{\sin(\frac{\pi}{2}-\alpha)}{\sin(\beta)}\, .
\end{eqnarray}
From the preceding relations and the condition $\alpha + \alpha' = 3\pi/2$, the deflection angle $\varphi$ is obtained as:
\begin{equation}
\varphi = \frac{\pi}{2} + \arctan(\tan\theta \sin\alpha).
\end{equation}
\begin{figure*}[!htb]
    \centering
    \includegraphics[width=0.48\linewidth]{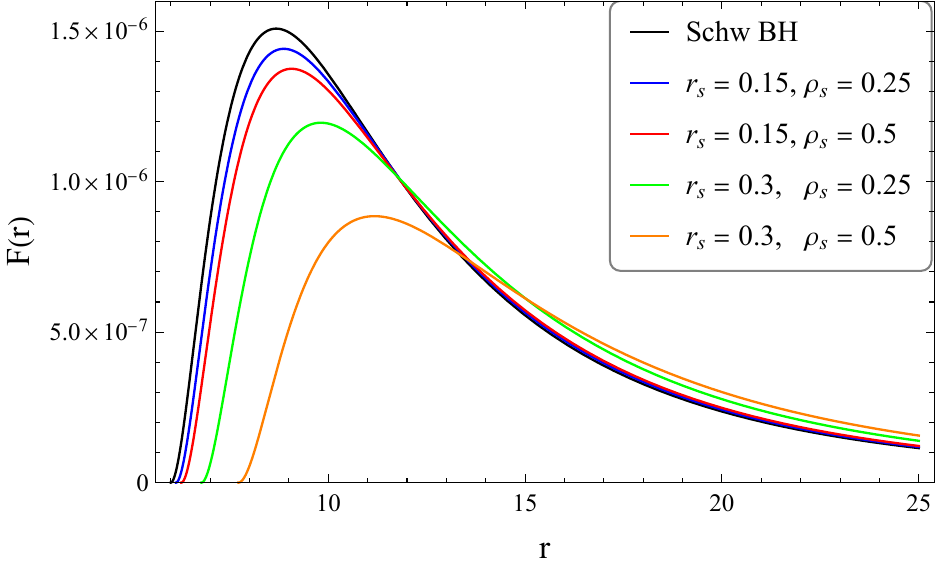}\hspace{0.5cm}
    \includegraphics[width=0.45\linewidth]{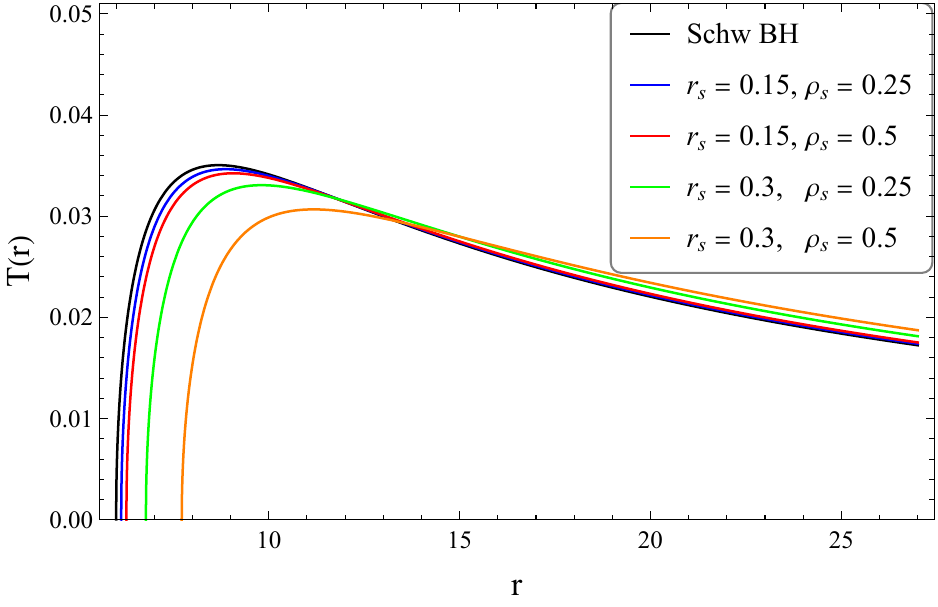}
    \caption{The distributions of energy flux $F(r)$ (left) and temperature $T(r)$ (right) on the accretion disk are plotted as a function of $r$.}
    \label{fig:Flux}
\end{figure*}

The angular deflection for the $n^{\text{th}}$-order image, denoted by $\varphi_n$, is given by~\cite{you2024}:
\begin{widetext}
\begin{equation} \label{Eq:nthimage}
\varphi_n = 
\begin{cases}
\frac{n}{2} 2\pi + (-1)^n \left[ \frac{\pi}{2} + \arctan(\tan\theta \sin\alpha) \right], & \text{for $n$ even}, \\[6pt]
\frac{n+1}{2} 2\pi + (-1)^n \left[ \frac{\pi}{2} + \arctan(\tan\theta \sin\alpha) \right], & \text{for $n$ odd} \, ,
\end{cases} 
\end{equation}
\end{widetext}
where $n=0$ and $n=1$ correspond to the direct and secondary images, respectively. A photon emitted from the camera plane at infinity with coordinates $(b, \alpha)$ and arriving at a circular orbit of radius $r$ on the accretion disk experiences a total deflection angle given by:
\begin{equation} \label{Eq:phi1disk}
\varphi_1(b) = \int^{u_r}_{0} \frac{1}{\sqrt{G(u)}} du, 
\end{equation}
\begin{equation}\label{Eq:phi2disk}
\varphi_2(b) = 2 \int^{u_{\min}}_{0} \frac{1}{\sqrt{G(u)}} du - \int^{u_r}_{0} \frac{1}{\sqrt{G(u)}} du.
\end{equation}

\begin{figure*}
\begin{tabular}{ccc}
  \includegraphics[scale=0.3]{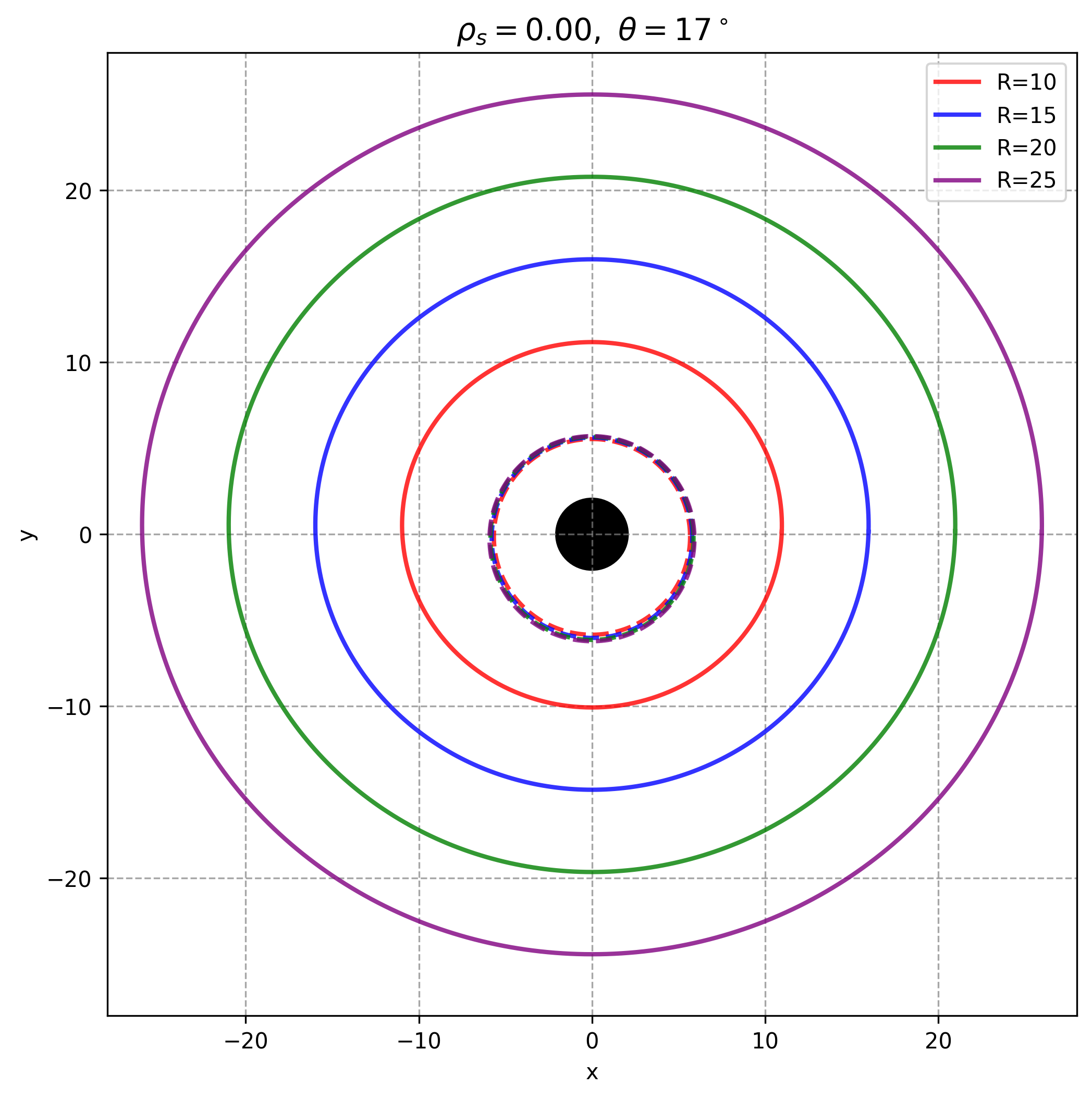}\hspace{-0.1cm}
  \includegraphics[scale=0.3]{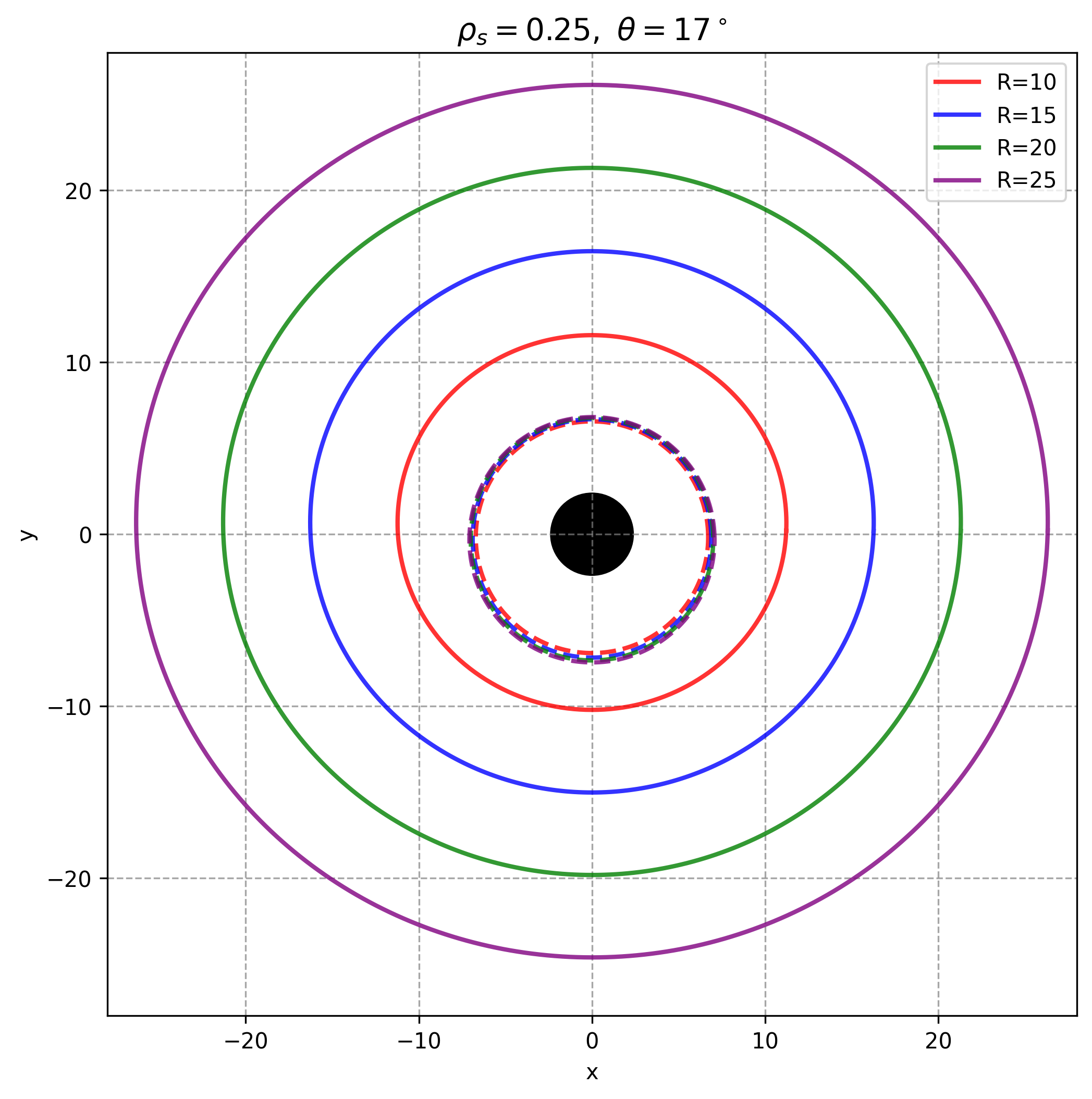}\hspace{-0.1cm}
  \includegraphics[scale=0.3]{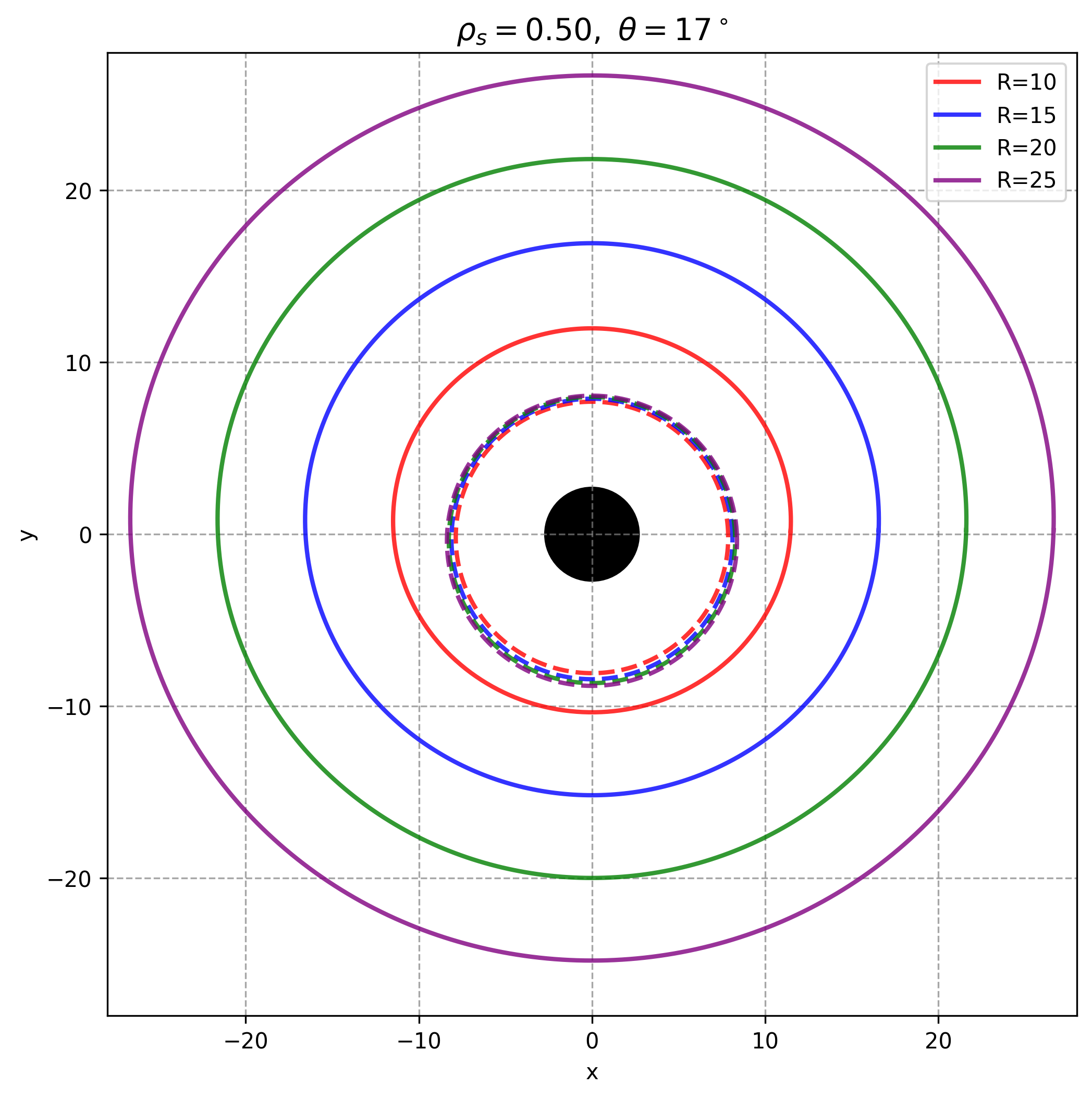}\\
  \includegraphics[scale=0.3]{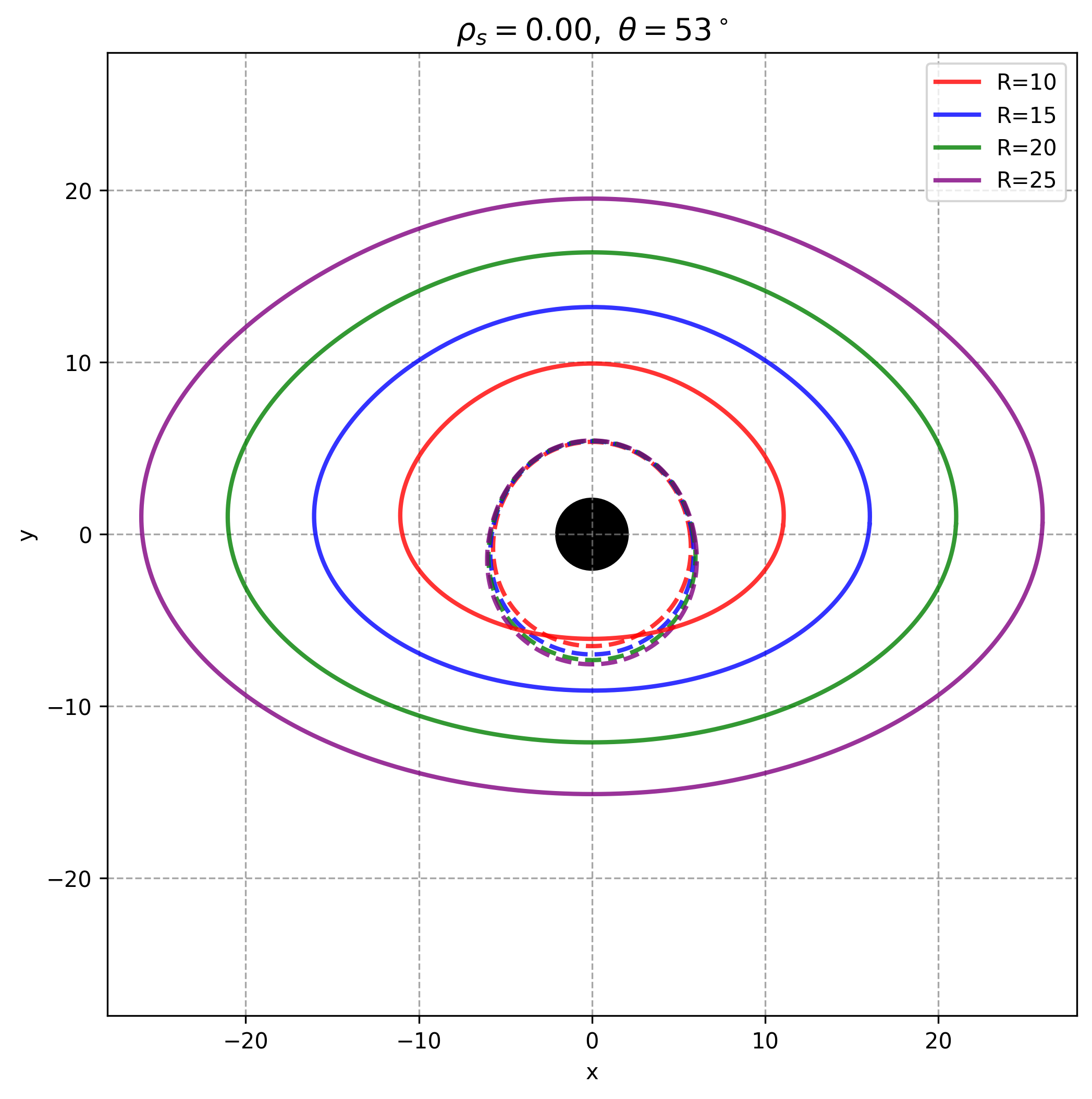}\hspace{-0.1cm}
  \includegraphics[scale=0.3]{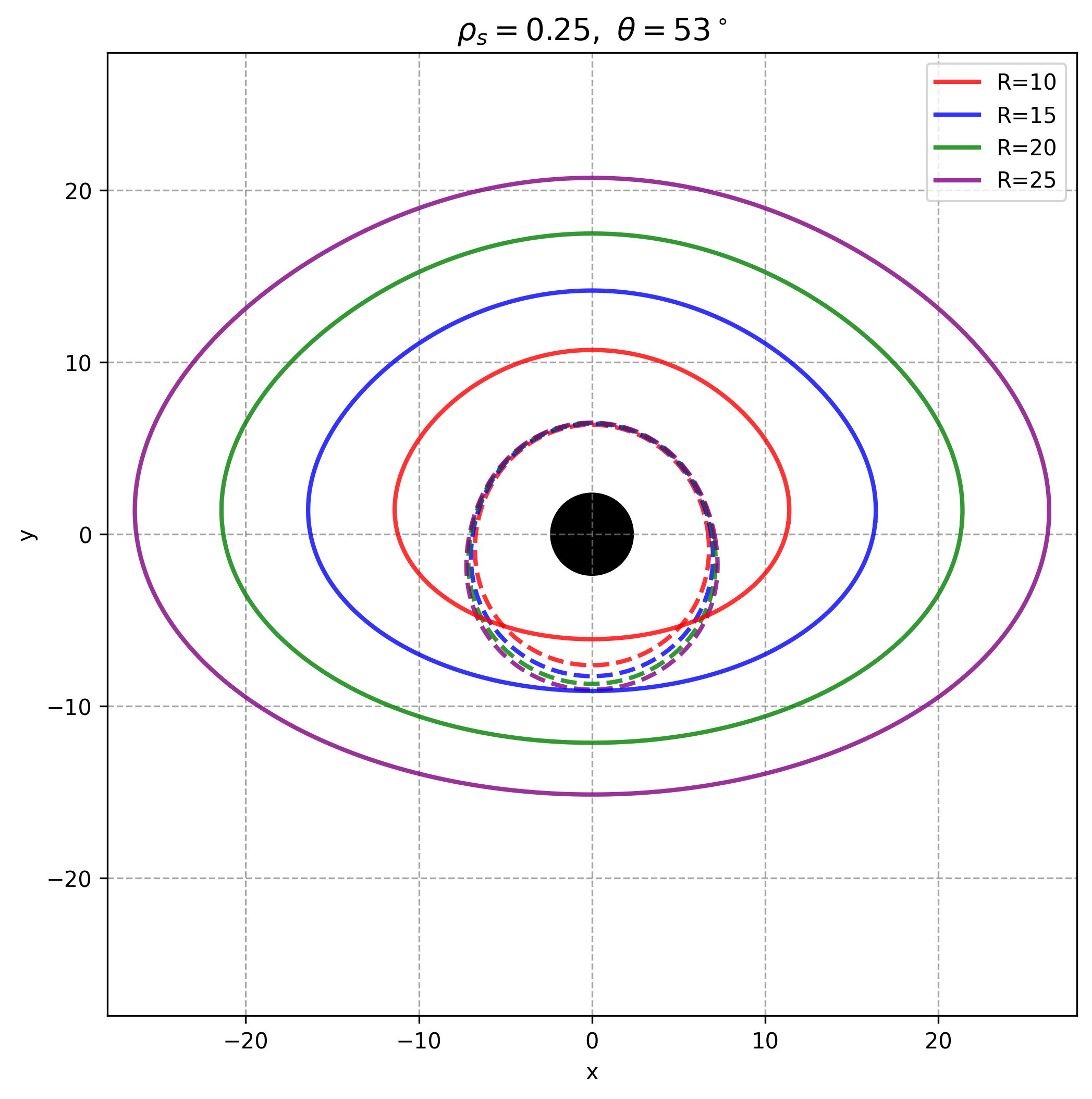}\hspace{-0.1cm}
  \includegraphics[scale=0.3]{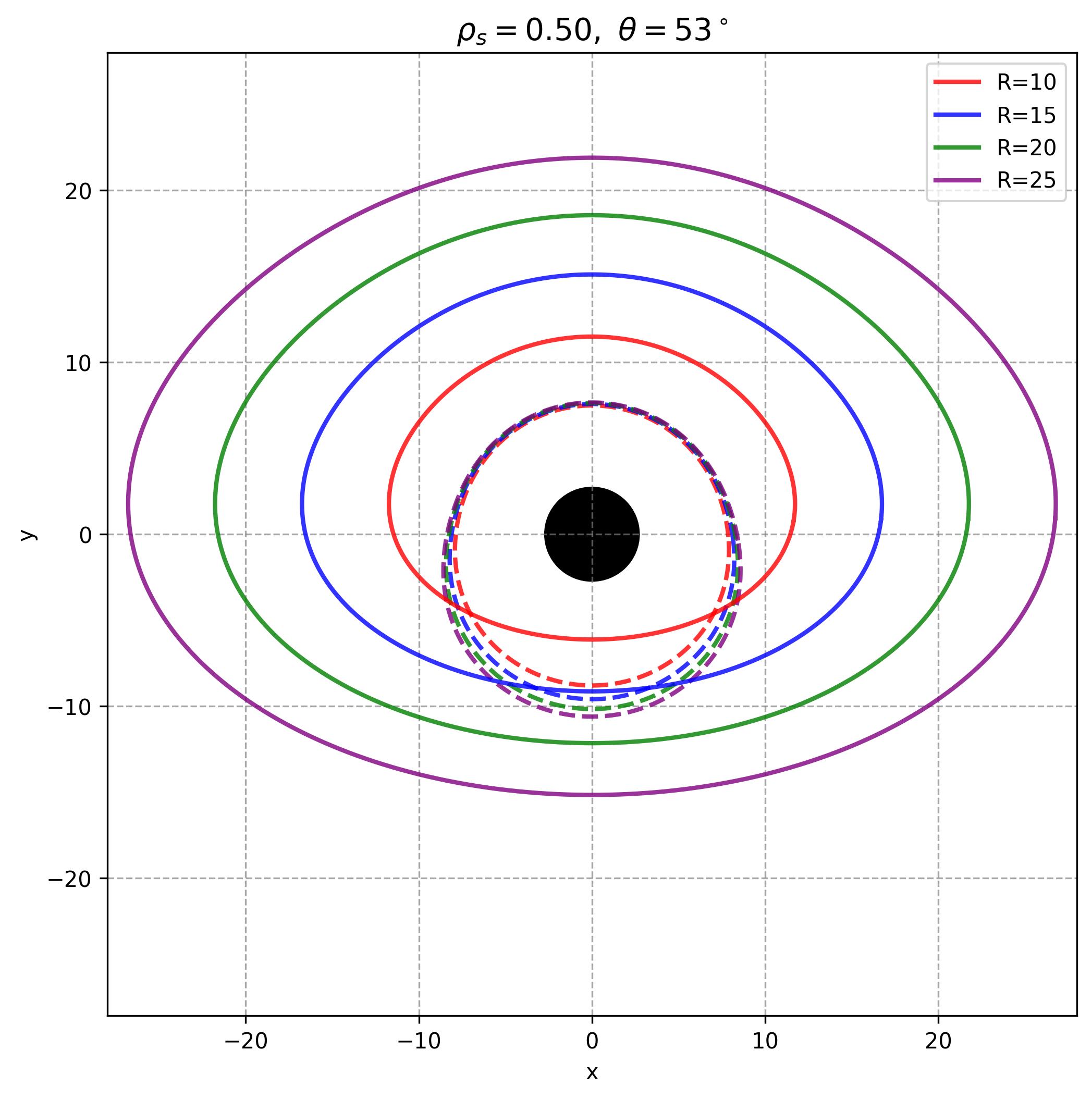}\\
  \includegraphics[scale=0.3]{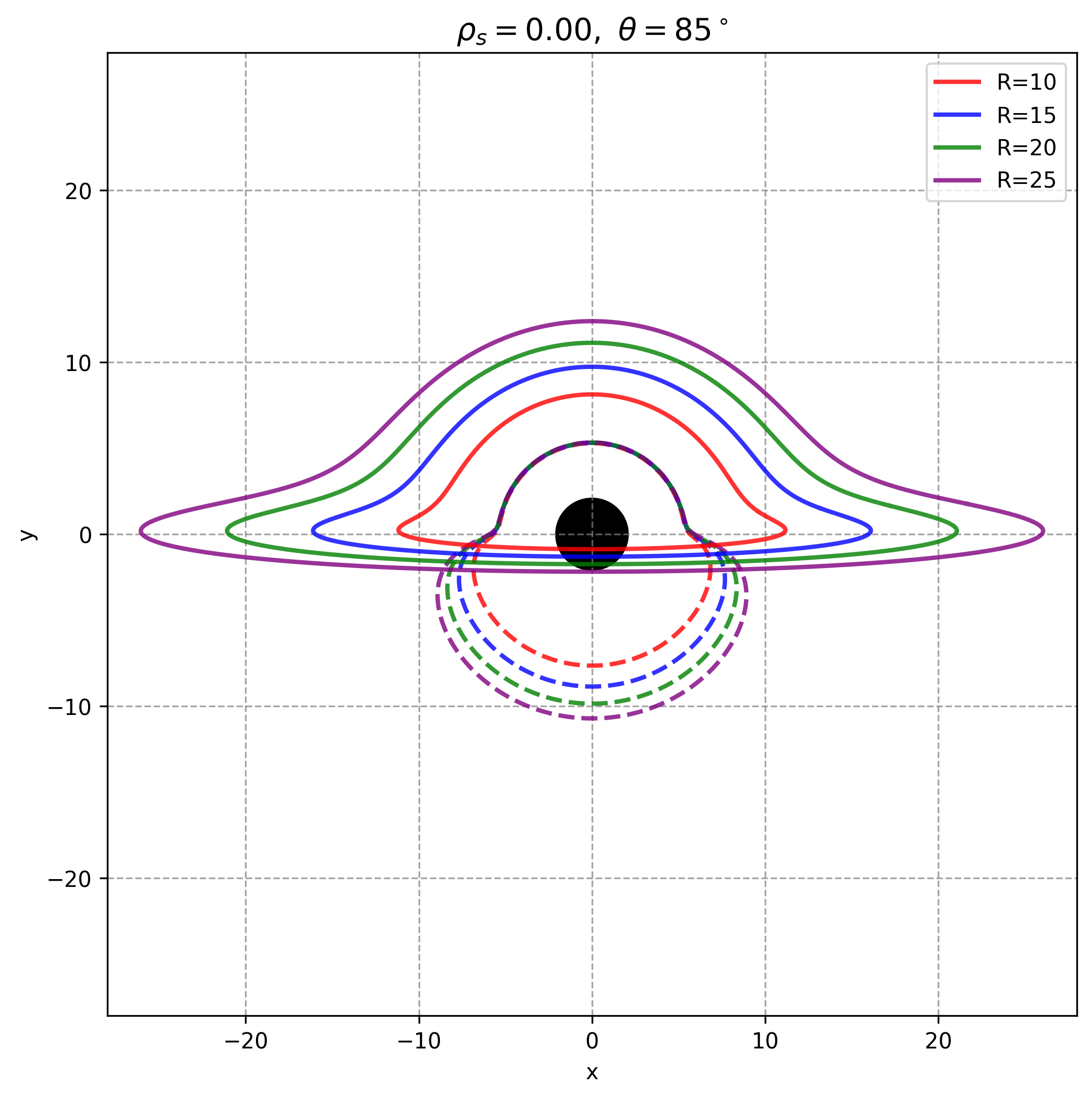}\hspace{-0.1cm}
  \includegraphics[scale=0.3]{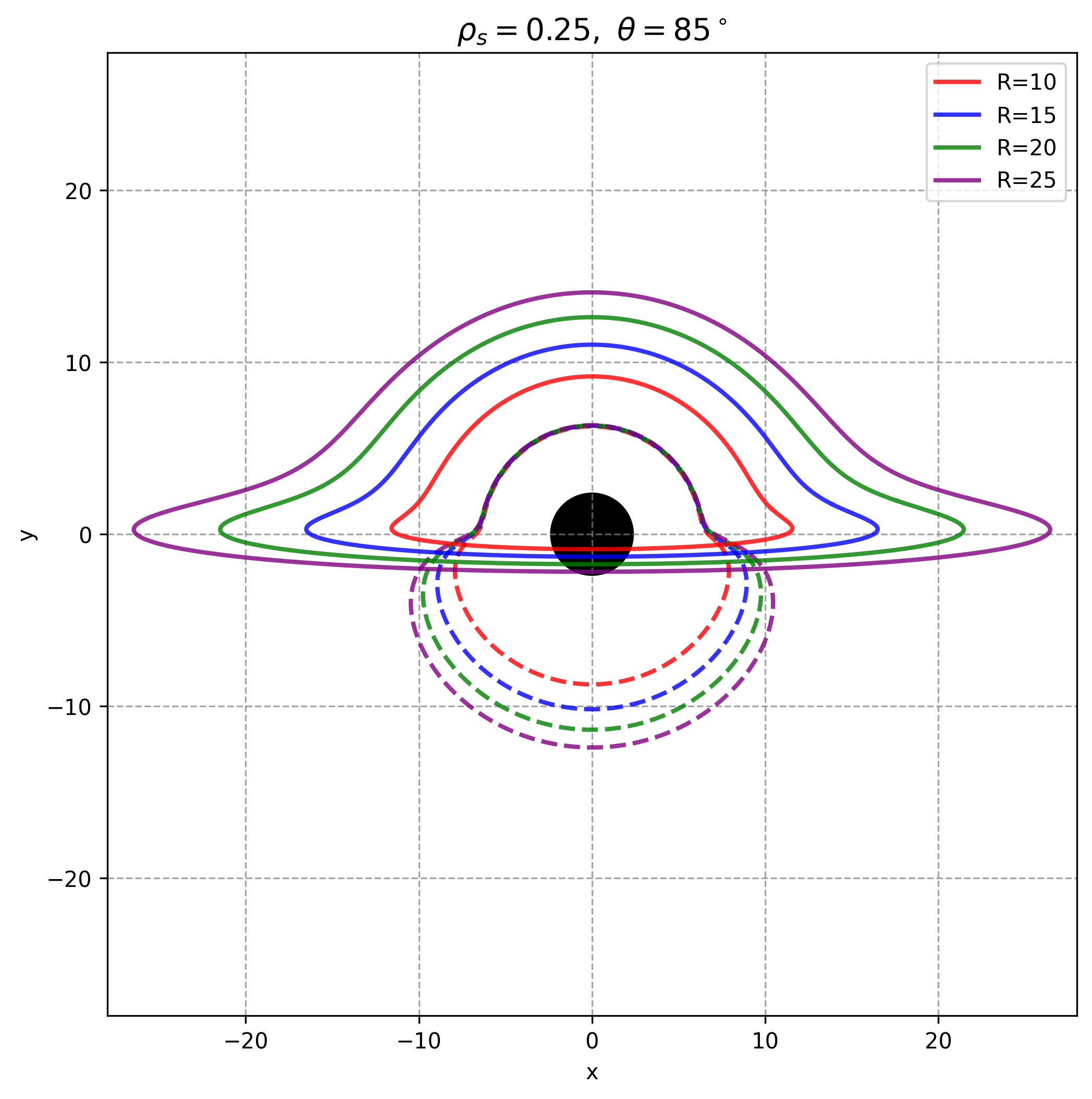}\hspace{-0.1cm}
  \includegraphics[scale=0.3]{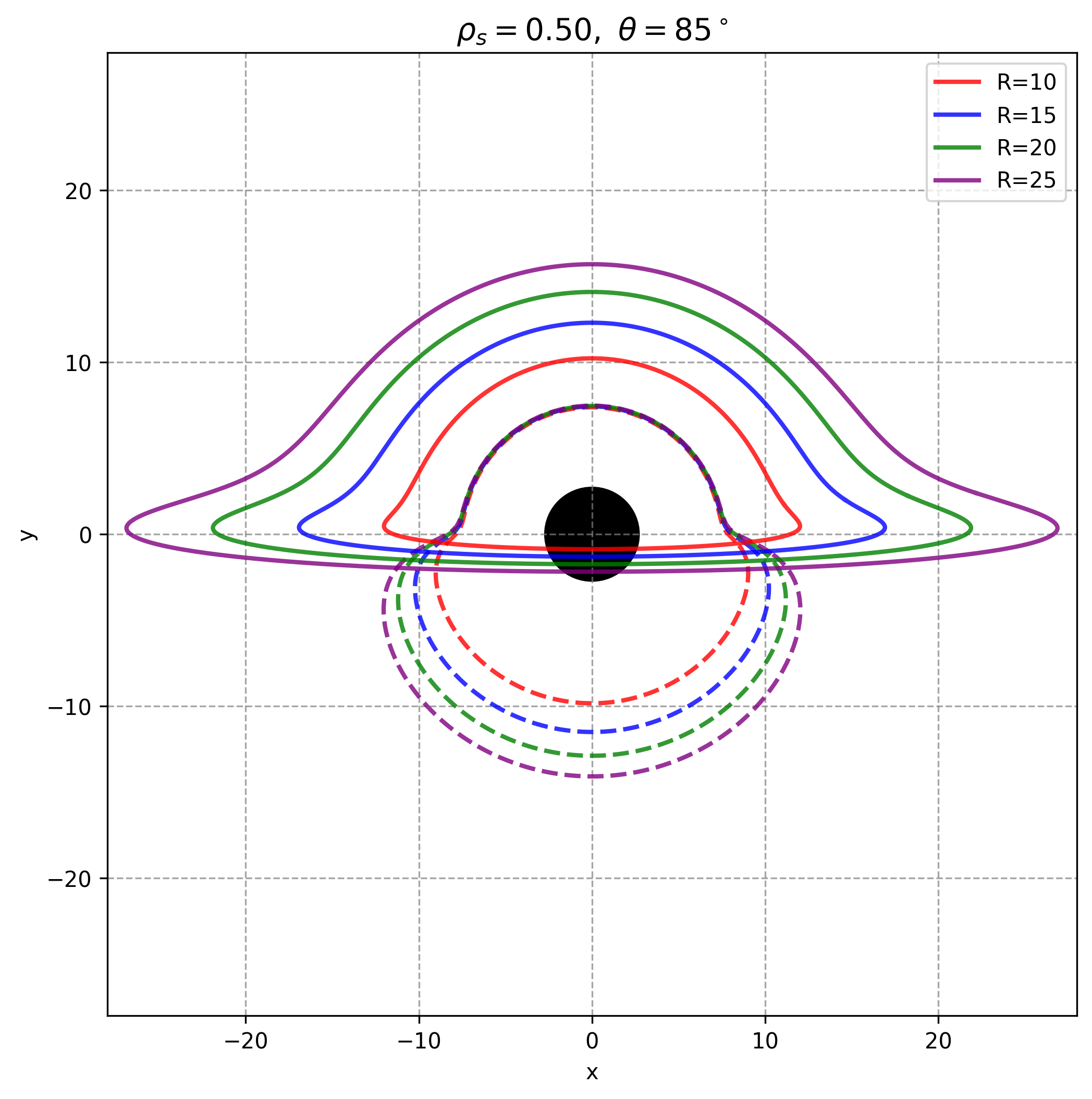}
  \end{tabular}
	\caption{\label{fig:accrthin} Direct and secondary images of a thin accretion disk surrounding a Schwarzschild BH embedded in the King DM halo. The columns (top to bottom) represent inclination angles of $17^\circ$, $53^\circ$, and $85^\circ$, whereas the rows (left to right) correspond to $\rho_s$ values of $0.0$, $0.25$, and $0.50$, assuming $r_s = 0.3$.}
\end{figure*}

In Fig.~\ref{fig:FBline1}, we show the photon deflection angles $\varphi_1(b)$ and $\varphi_2(b)$ as functions of the impact parameter $b$, plotted for different values of $\rho_s$ and various radii. The colored regions represent the order of the images $n$. It is obvious that as the impact parameter approaches its critical value $b_c$, the deflection angle tends toward infinity. With increasing $\rho_s$, the corresponding $b_c$ shifts right to larger values of $b$. The curves intersect the region $n=1$ differently for various values of $\rho_s$, indicating that the secondary images also depend on $\rho_s$, as seen in the left panel of Fig.~\ref{fig:FBline1}. As seen in the figure, the maximum value of $b$ increases with larger values of $\rho_s$. The right panel of Fig.~\ref{fig:FBline1} shows the deflection angle for photons arriving from different radii of the accretion disk $(R=10,15,20)$, with the central density fixed at $\rho_s=0.25$. For these radii, we obtained the maximum values of $b$ as $b=11.58$, $16.51$, and $21.48$, respectively. From the right panel, it is evident that as the disk radius increases, the direct image $(n=0)$ changes noticeably, whereas the secondary images $(n\geq1)$ remain nearly unchanged. In addition, in the right panel of Fig.~\ref{fig:FBline1}, the red dashed curve indicates the periastron curve that crosses the intersection point of $\varphi_1$ and $\varphi_2$, determined by the integral equation $\varphi_3$, which asymptotically approaches the straight line $\pi/2$ at infinity:
\begin{equation}
\varphi_3(b) = \int^{u_{\min}}_{0} \frac{1}{\sqrt{G(u)}} du\, . 
\end{equation}
The coordinates $(b, \alpha)$ of the accretion disk in the observer’s plane are obtained by numerically solving Eqs.~\eqref{Eq:nthimage}, \eqref{Eq:phi1disk}, and~\eqref{Eq:phi2disk} for the primary and secondary images~\cite{Gyulchev20,Ziqiang25}. The direct and secondary images of stable circular orbits around a Schwarzschild BH surrounded by the King DM halo, viewed at different inclination angles, are shown in Fig.~\ref{fig:accrthin}. As seen in Fig.~\ref{fig:accrthin}, larger values of the King DM halo parameters result in larger primary and secondary images. This suggests that a larger impact parameter $b$ is required for photons to reach the same stationary circular orbit, indicating a strengthening of the gravitational field, consistent with the behavior observed for periodic orbits.

According to the standard Novikov–Thorne formulation~\cite{Novikov:1973kta,Shakura:1972te,Thorne:1974ve}, the radiant energy flux emitted from the accretion disk surface is given by
\begin{equation}
F(r) = -\frac{\dot{M}_0 \Omega_{,r}}{4\pi \sqrt{-g}(E - \Omega L)^2}
\int_{r_{\text{isco}}}^{r} (E - \Omega L)L_{,r}dr\, , 
\label{eq:flux}
\end{equation}
where $\dot{M}_0$ stands for the mass accretion rate, $g$ represents the determinant of the metric, and $E$, $L$, and $\Omega$ correspond to the specific energy, angular momentum, and angular velocity of matter on circular orbits.
For circular motion around a static, spherically symmetric BH, these quantities take the following form~\cite{Shapiro83,Shaymatov22a}:
\begin{equation}
E = -\frac{g_{tt}}{\sqrt{-g_{tt} - g_{\phi\phi}\Omega^2}}, 
\end{equation}
\begin{equation}
L = \frac{g_{\phi\phi}\Omega}{\sqrt{-g_{tt} - g_{\phi\phi}\Omega^2}}, 
\end{equation}
\begin{equation}
\Omega = \frac{d\phi}{dt} = \sqrt{-\frac{g_{tt,r}}{g_{\phi\phi,r}}}.
\end{equation}
Using the Stefan–Boltzmann law, one can express the relation between the energy flux $F(r)$ of the disk and its radiation temperature $T$ as follows~\cite{Heydari-Fard:2022xhr} 
\begin{equation}
    F(r)=\sigma_{SB}T^4(r)\ .
    \label{eq:T}
\end{equation}

The left panel of Fig.~\ref{fig:Flux} shows the radial distribution of the energy flux $F(r)$ on the accretion disk around a Schwarzschild BH surrounded by the King DM halo. The distribution begins at $r = R_{\mathrm{ISCO}}$. Increasing $r_s$ and $\rho_s$ reduces the flux magnitude and shifts its peak to larger radii, indicating a dimmer and cooler disk compared to the pure Schwarzschild case. The right panel of Fig.~\ref{fig:Flux} illustrates the temperature distribution $T(r)$ on the accretion disk, where the peak decreases and shifts outward, similar to the behavior of the energy flux.

\begin{figure*}
\begin{tabular}{ccc}
  \includegraphics[scale=0.35]{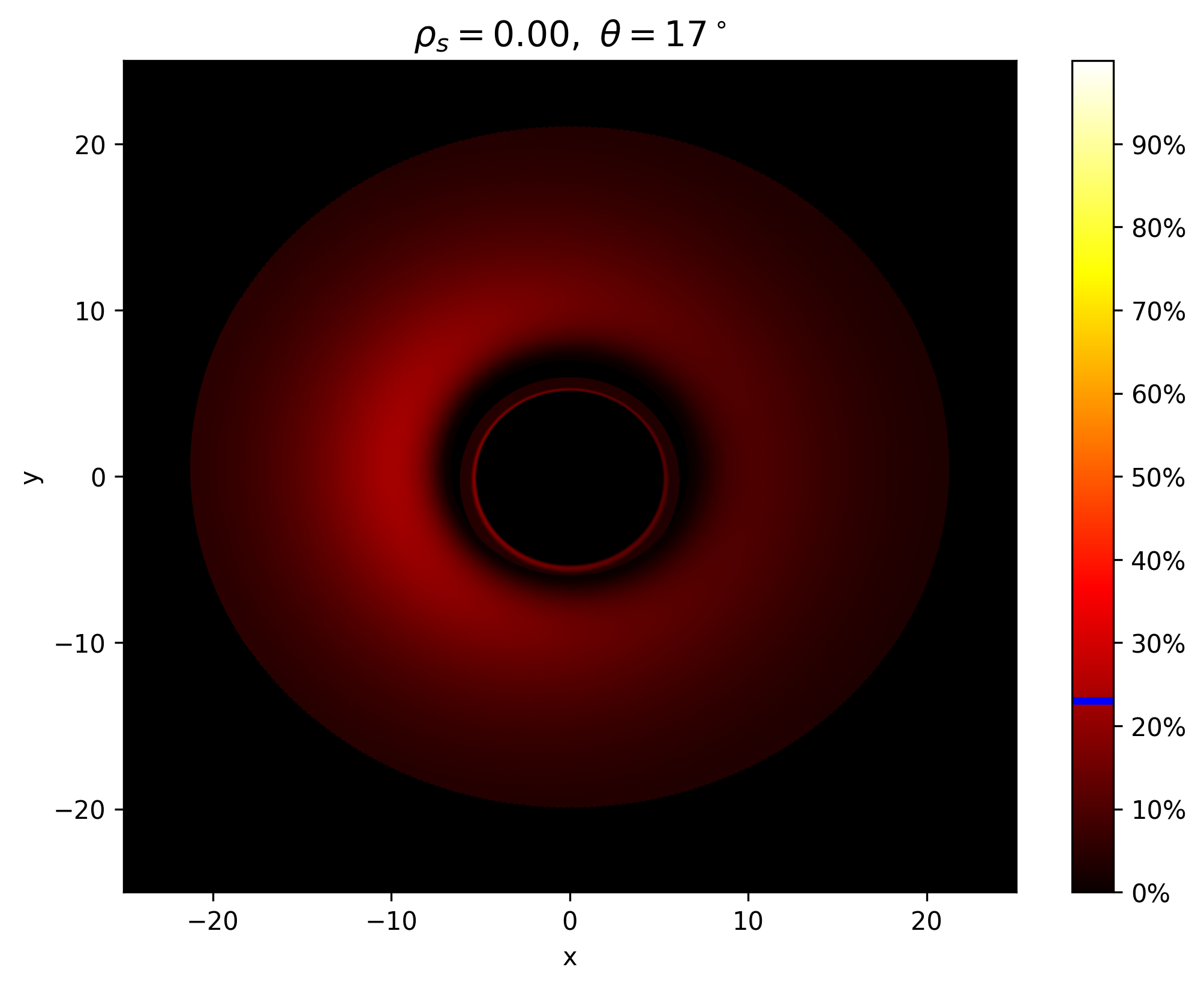}\hspace{-0.2cm}
   \includegraphics[scale=0.35]{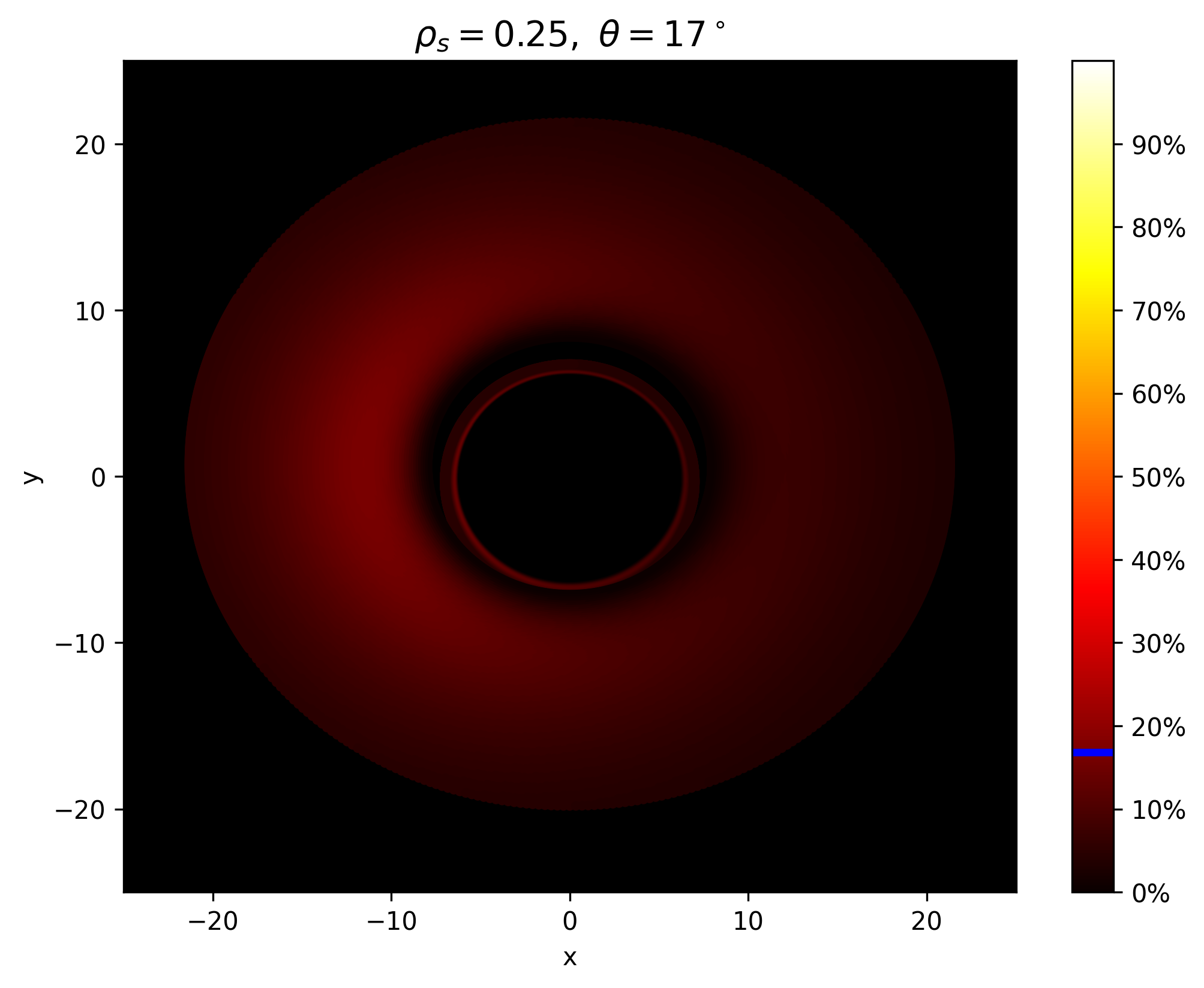}\hspace{-0.2cm}
  \includegraphics[scale=0.35]{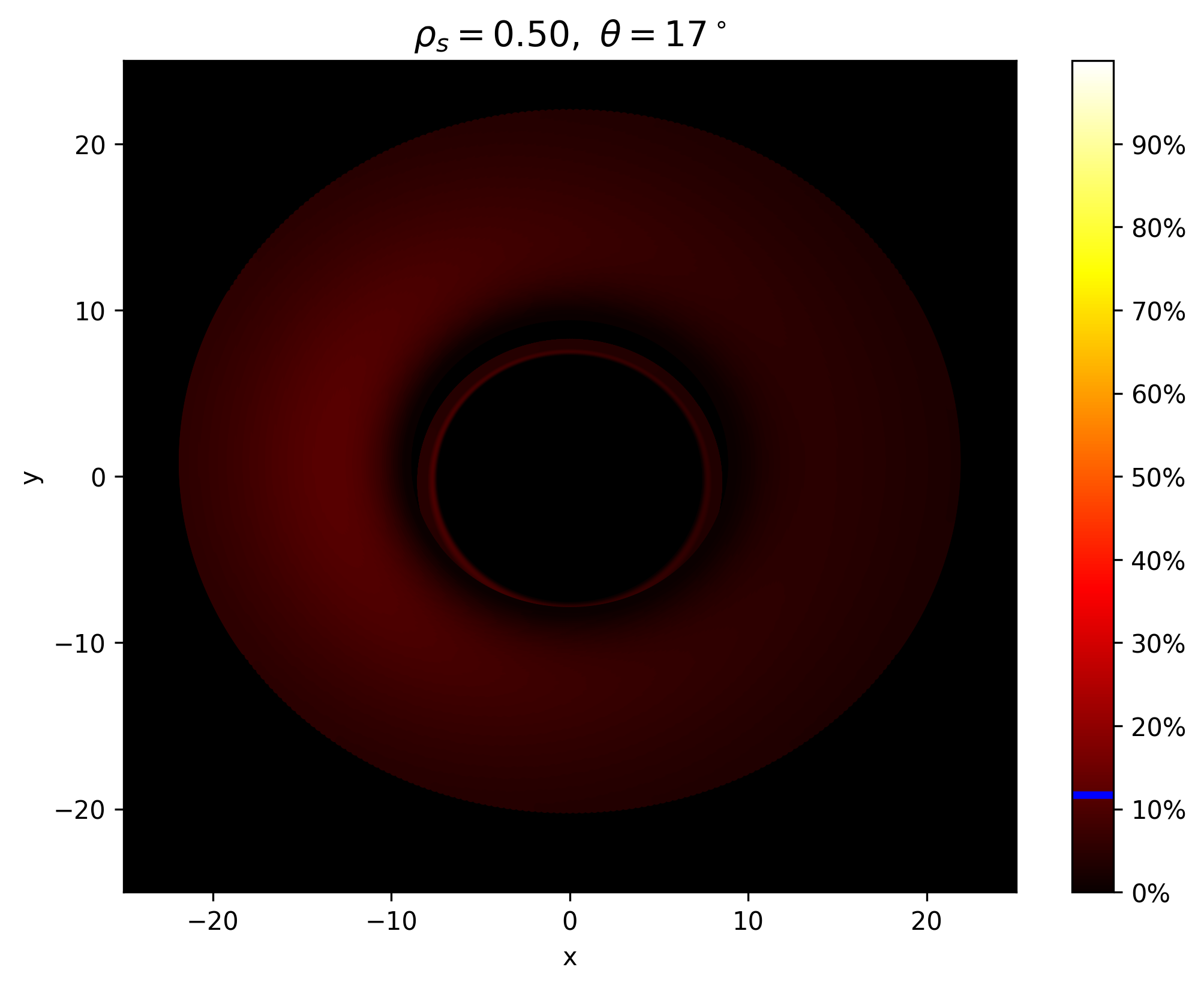}\\
  \includegraphics[scale=0.35]{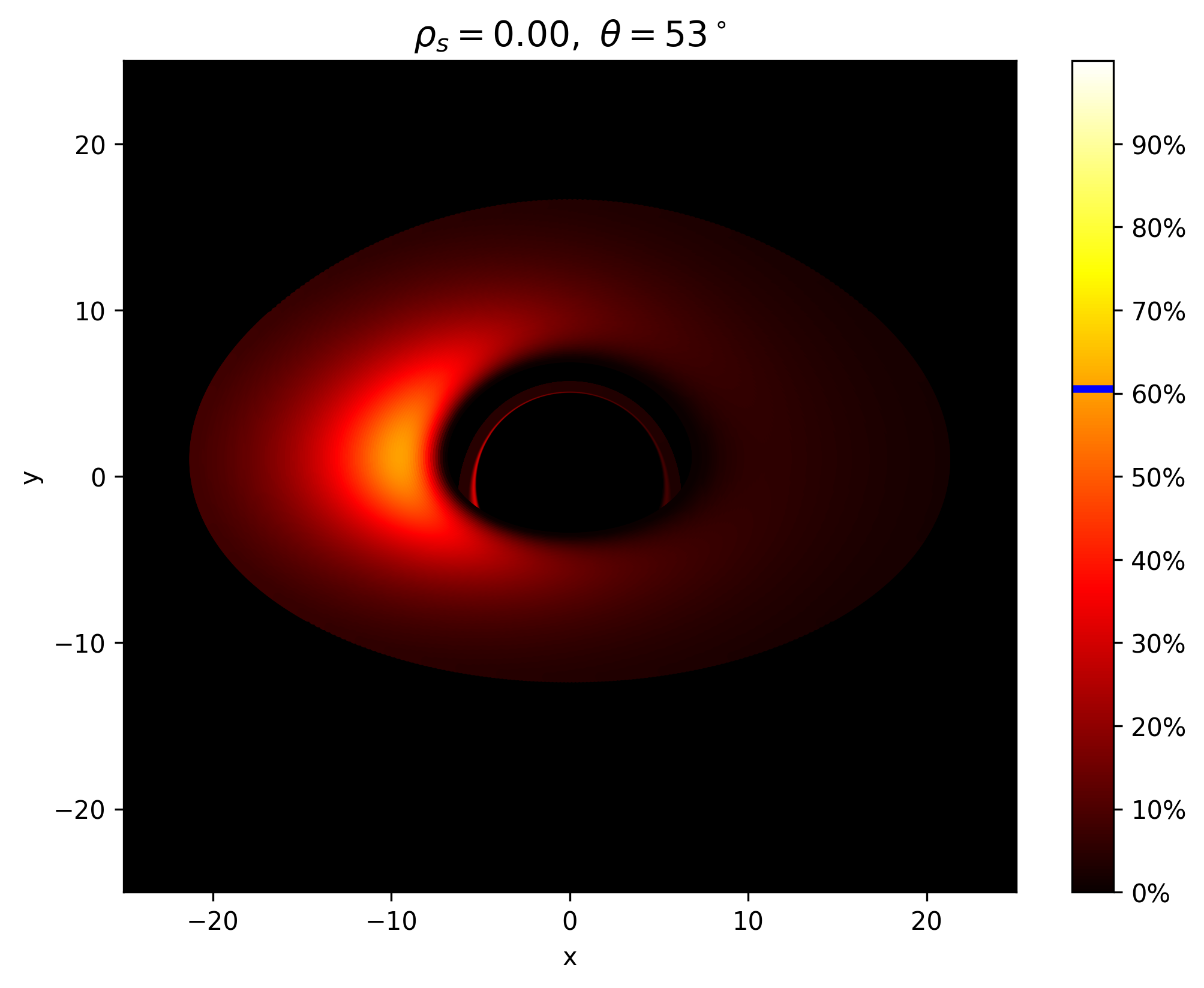}\hspace{-0.2cm}
  \includegraphics[scale=0.35]{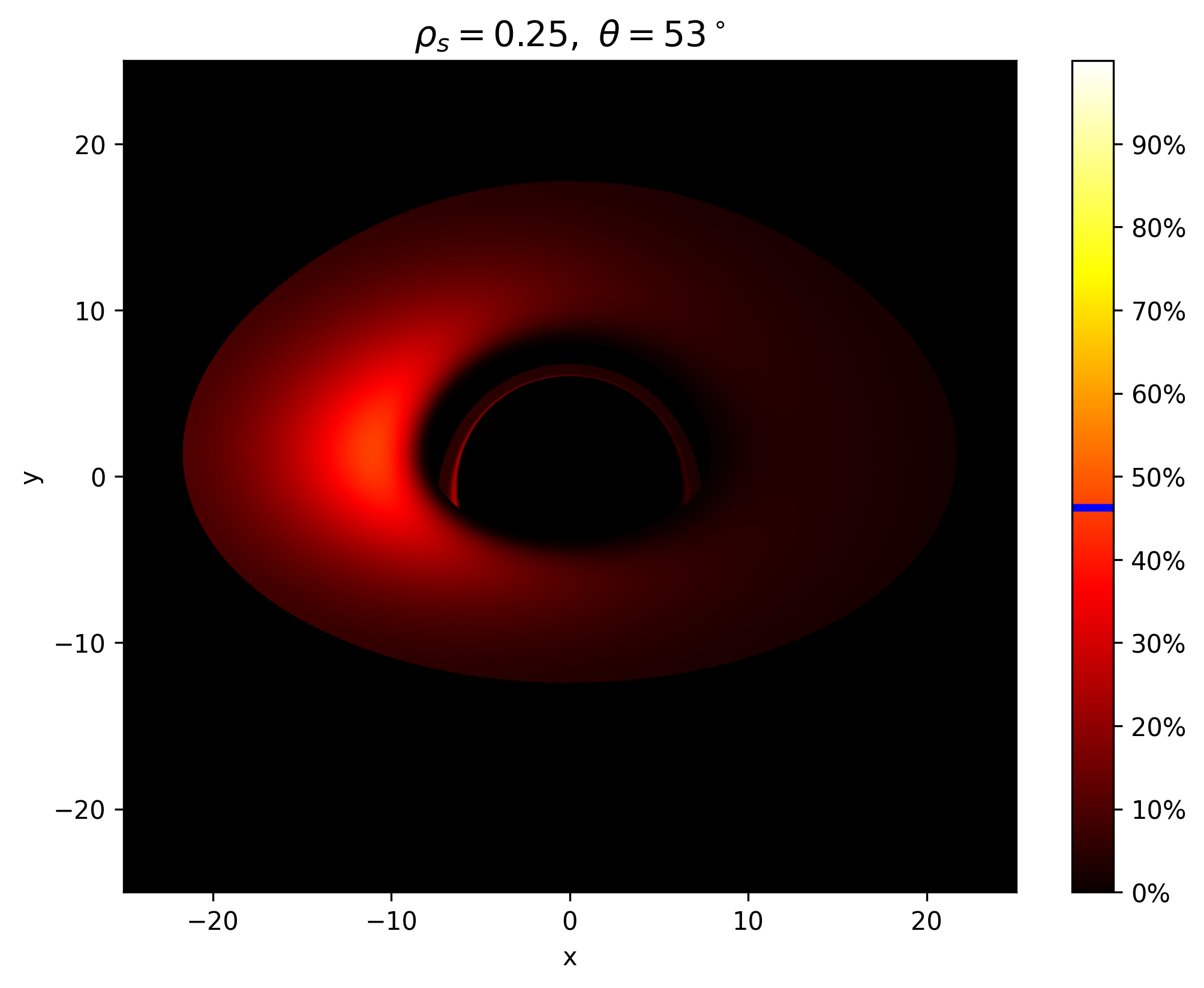}\hspace{-0.2cm}
  \includegraphics[scale=0.35]{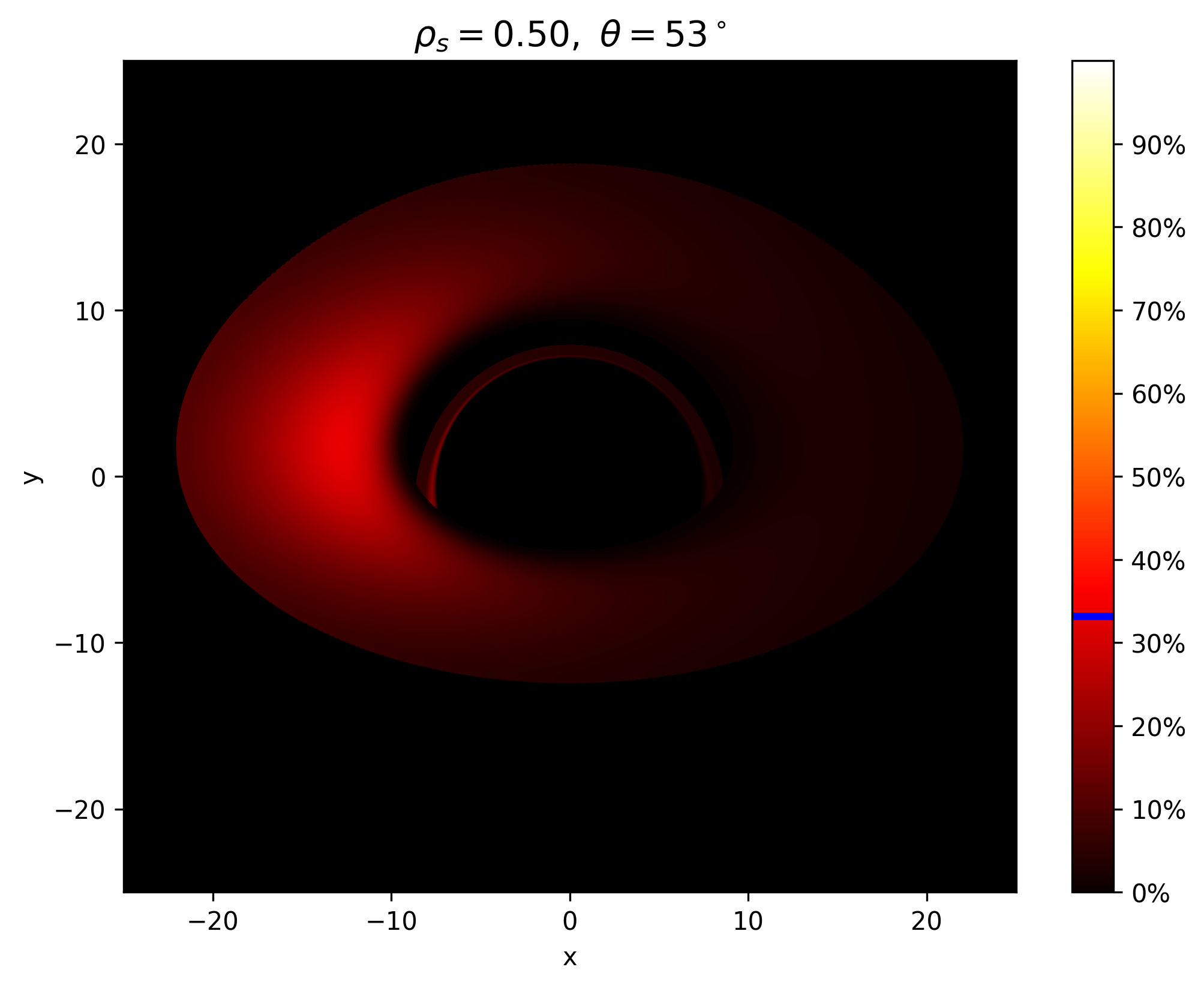}\\
  \includegraphics[scale=0.35]{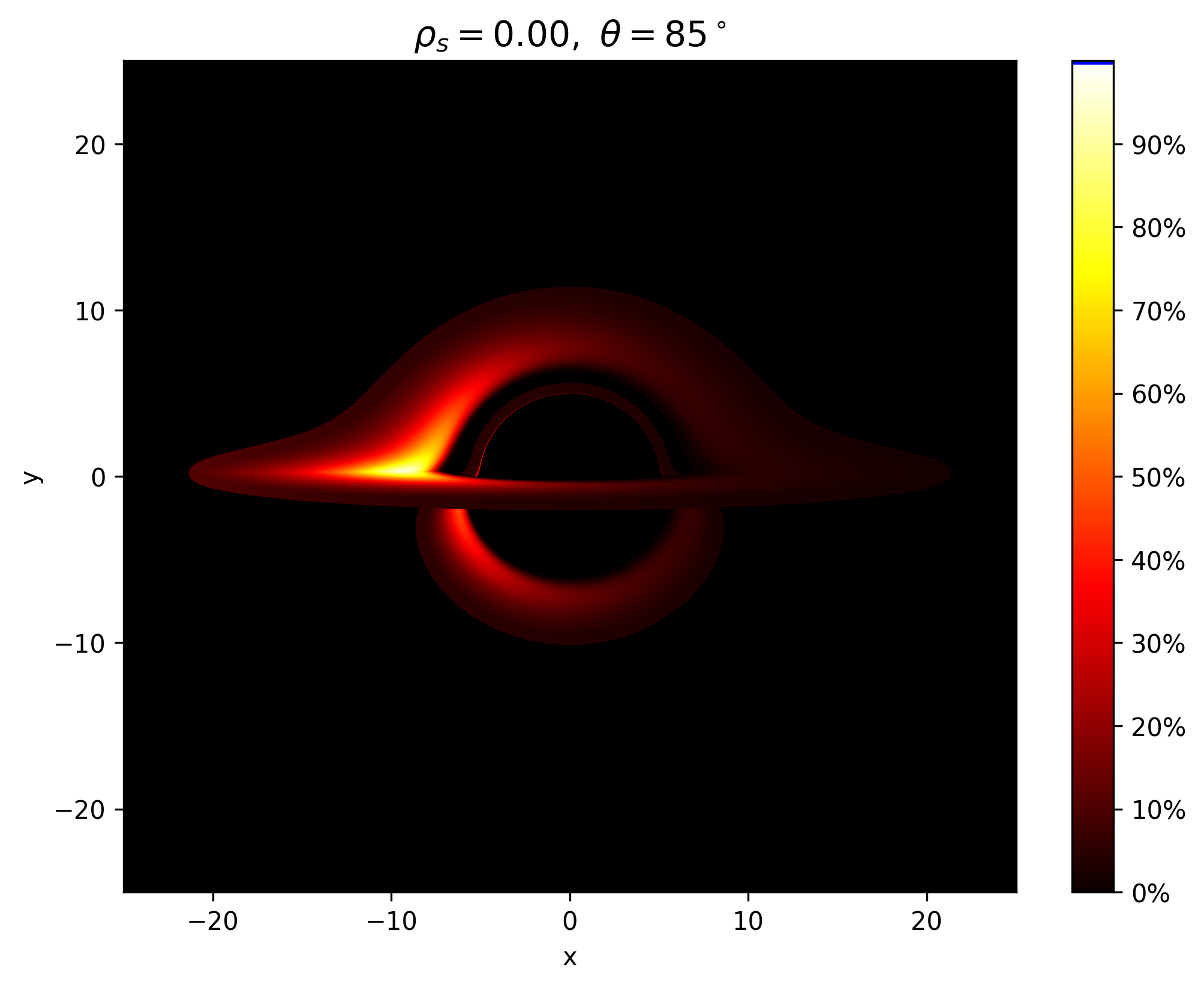}\hspace{-0.2cm}
  \includegraphics[scale=0.35]{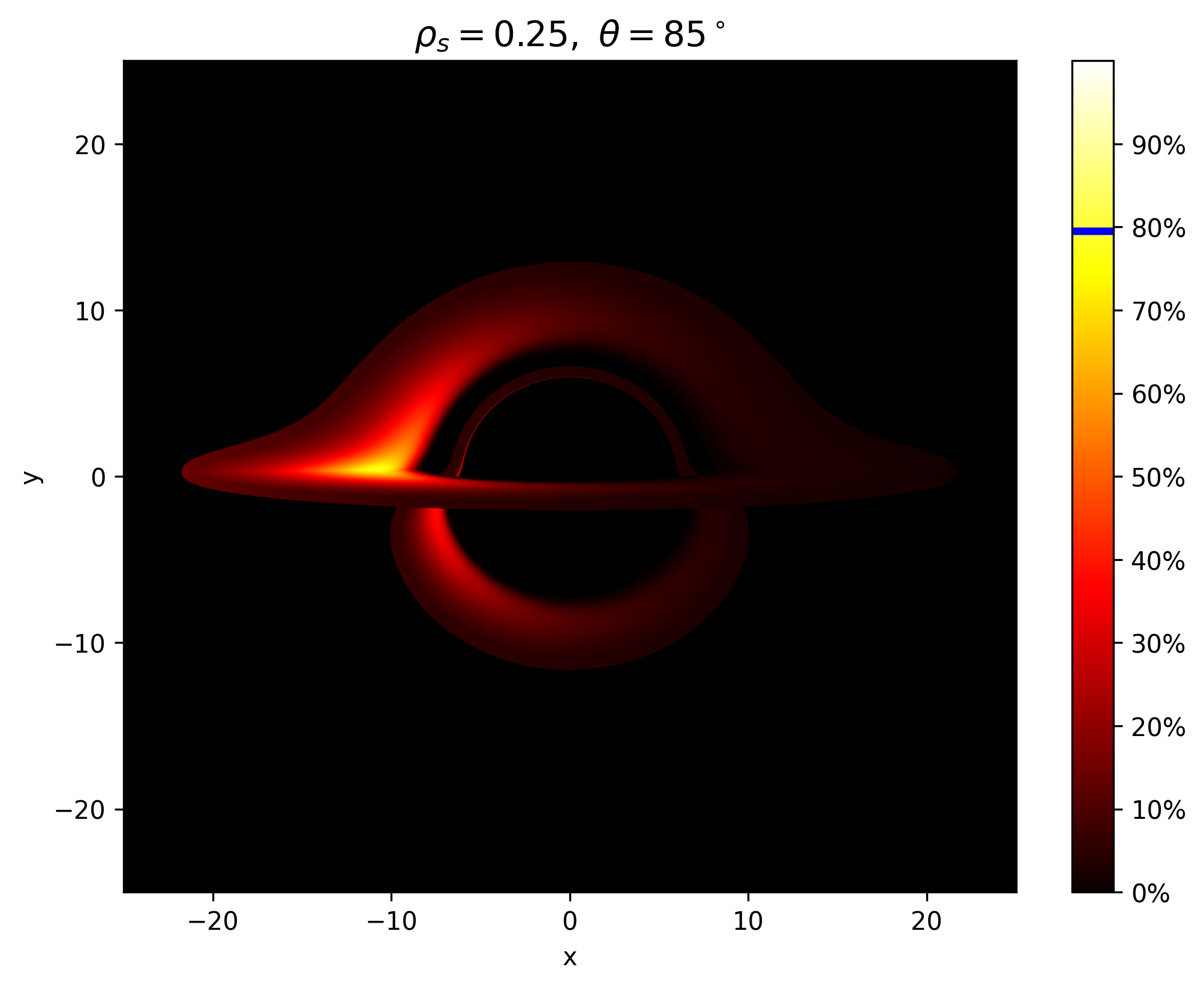}\hspace{-0.2cm}
  \includegraphics[scale=0.35]{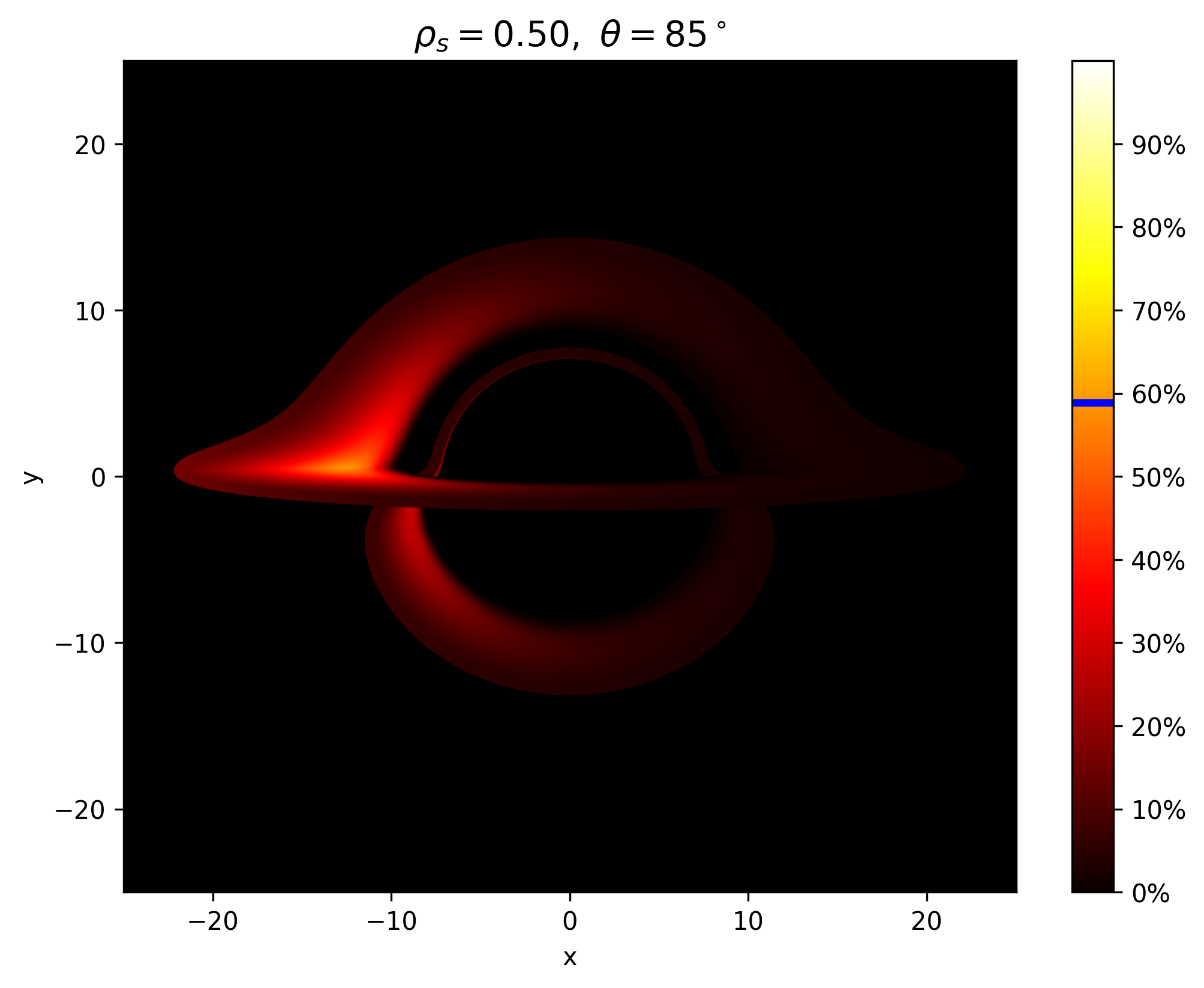}
  \end{tabular}
	\caption{\label{fig:fluxobs} Distribution of the observed flux $F_{obs}$ in the direct and secondary images of a Schwarzchild BH surrounded by a King DM halo is shown for different inclination angles $\theta$ and central densities $\rho_s$, with the scale radius fixed at $r_s = 0.3$.}
\end{figure*}

Due to the Doppler effect and gravitational redshift, the energy flux detected by a distant observer differs from that emitted by the accretion disk~\cite{Ellis2009}
\begin{equation}
F_{obs} = \frac{F(r)}{(1+z)^4} \, ,
\end{equation}
where $z$ represents the redshift factor, which can be expressed as~\cite{Luminet1979}:
\begin{equation}
1 + z =
\frac{1 + \Omega b \sin \theta \cos\alpha}{\sqrt{-g_{tt} - g_{\phi\phi}\Omega^2}}. 
\end{equation}

\begin{figure*}
\begin{tabular}{ccc}
  \includegraphics[scale=0.35]{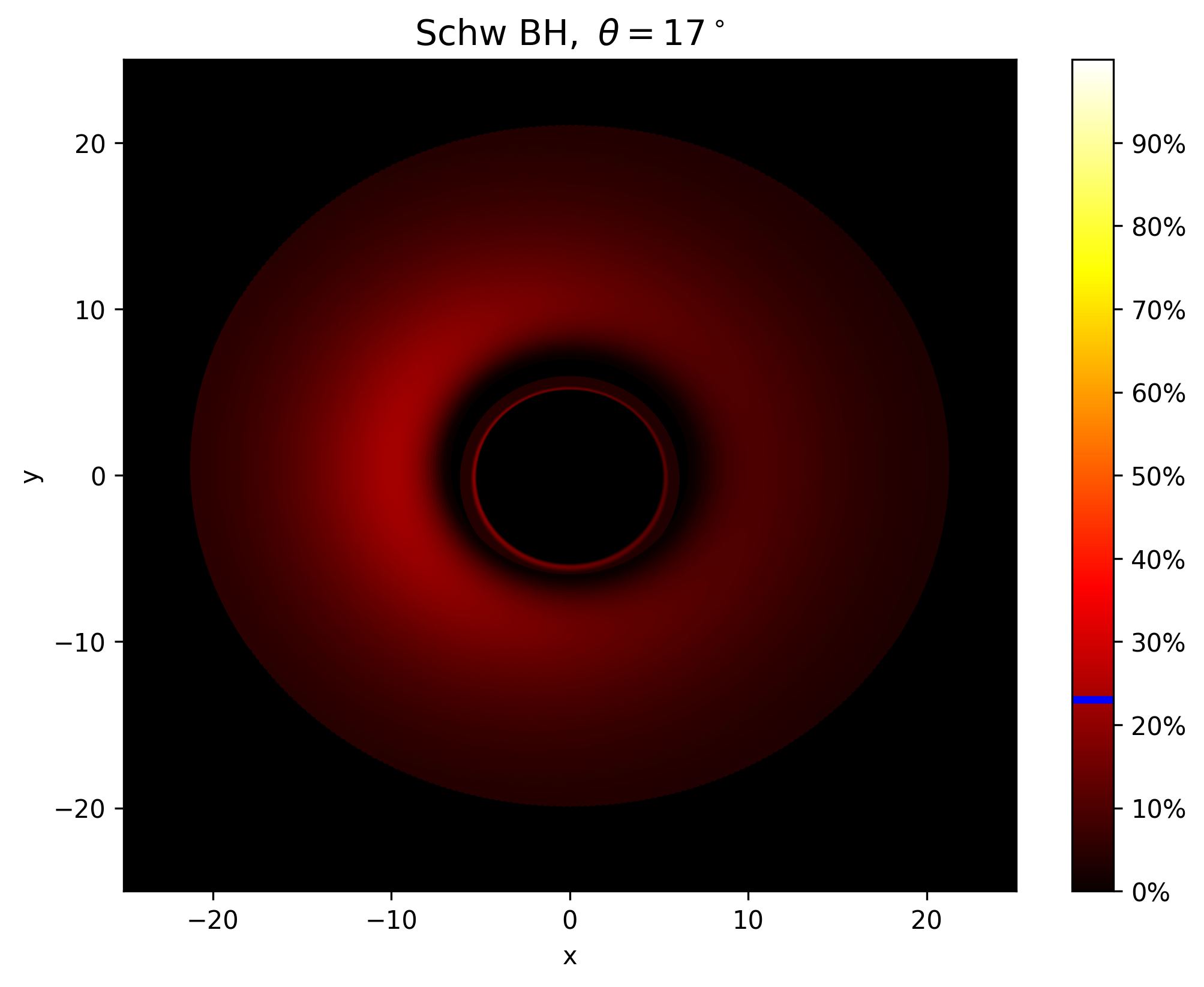}\hspace{-0.2cm}
   \includegraphics[scale=0.35]{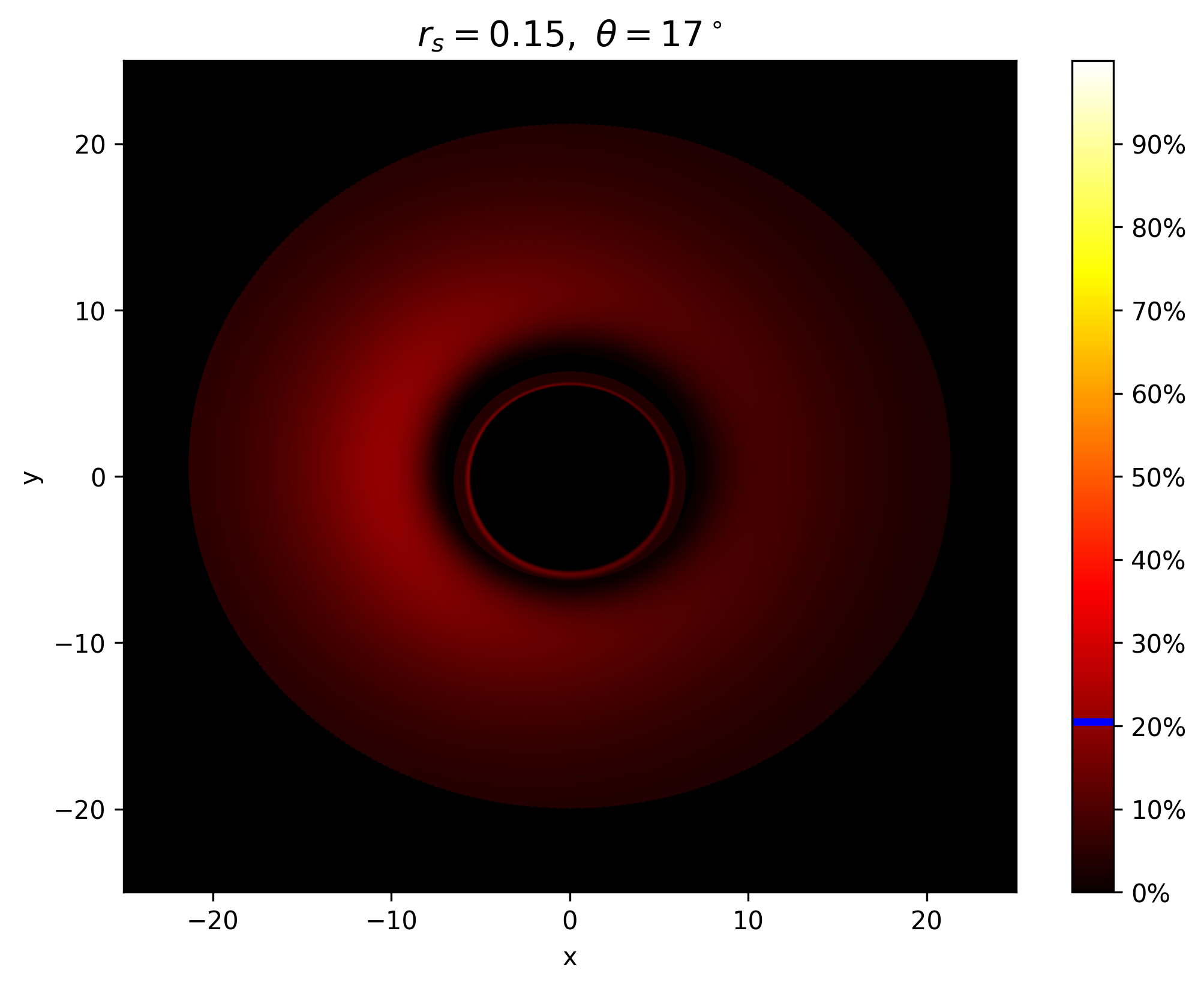}\hspace{-0.2cm}
  \includegraphics[scale=0.35]{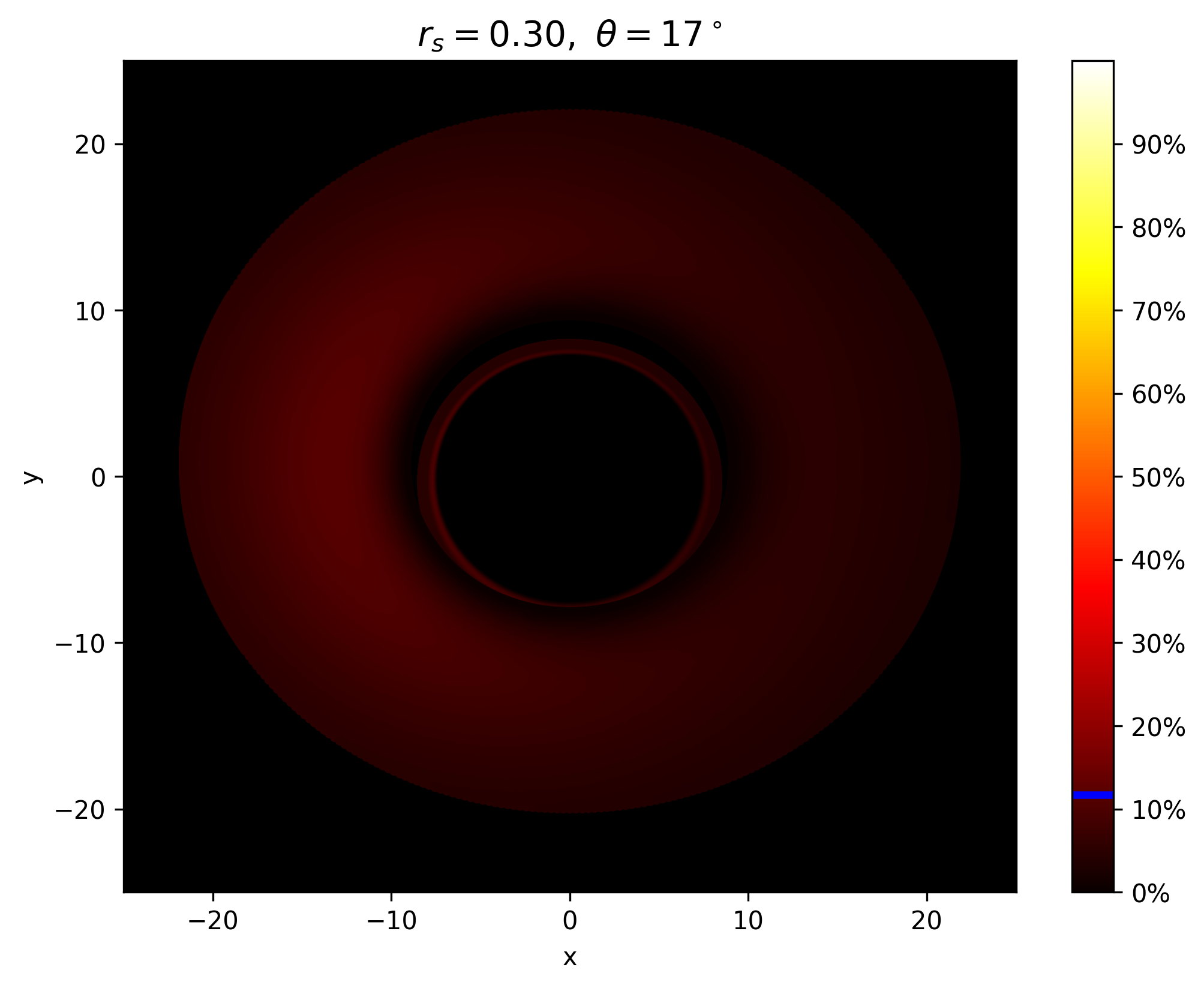}\\
  \includegraphics[scale=0.35]{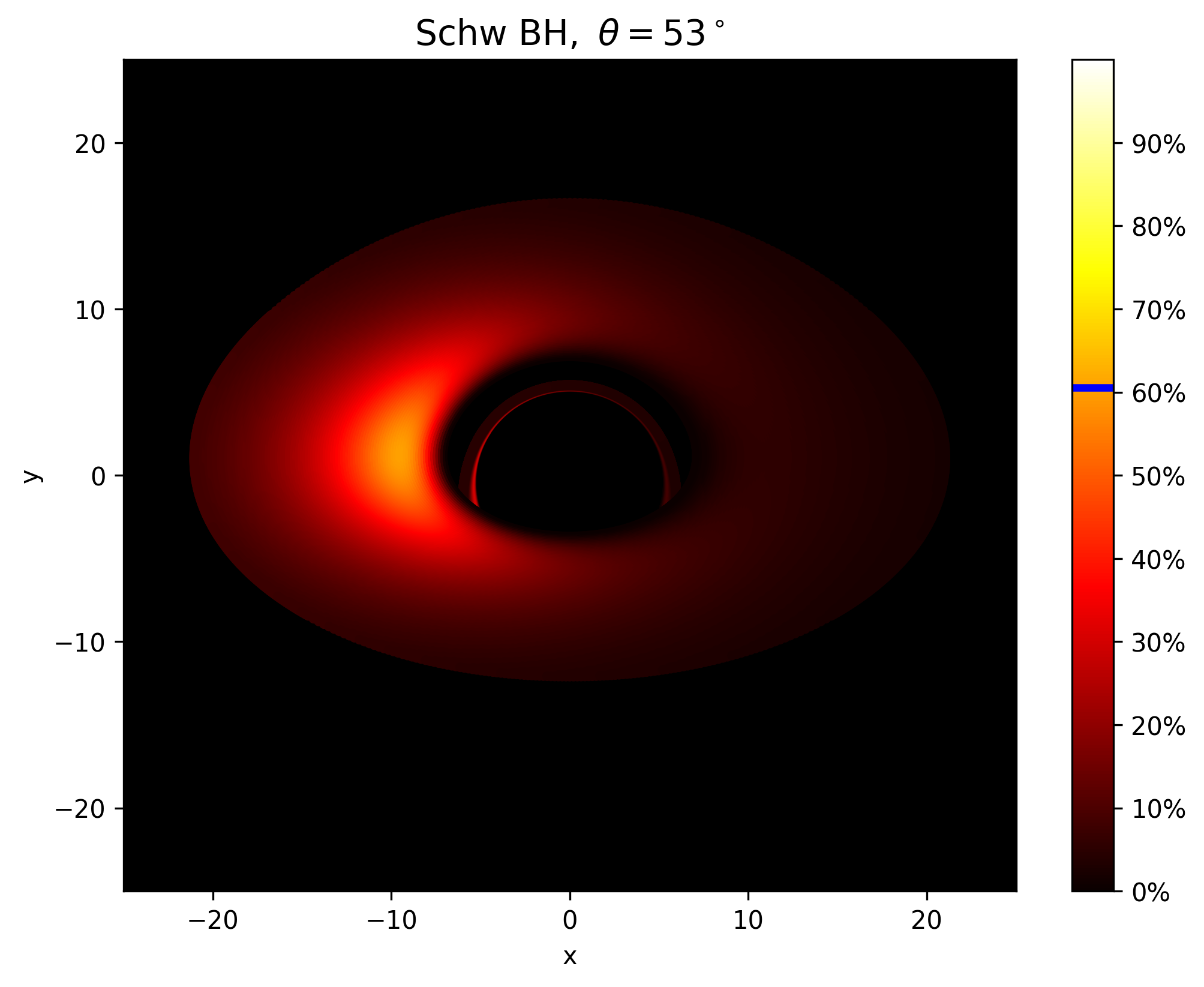}\hspace{-0.2cm}
  \includegraphics[scale=0.35]{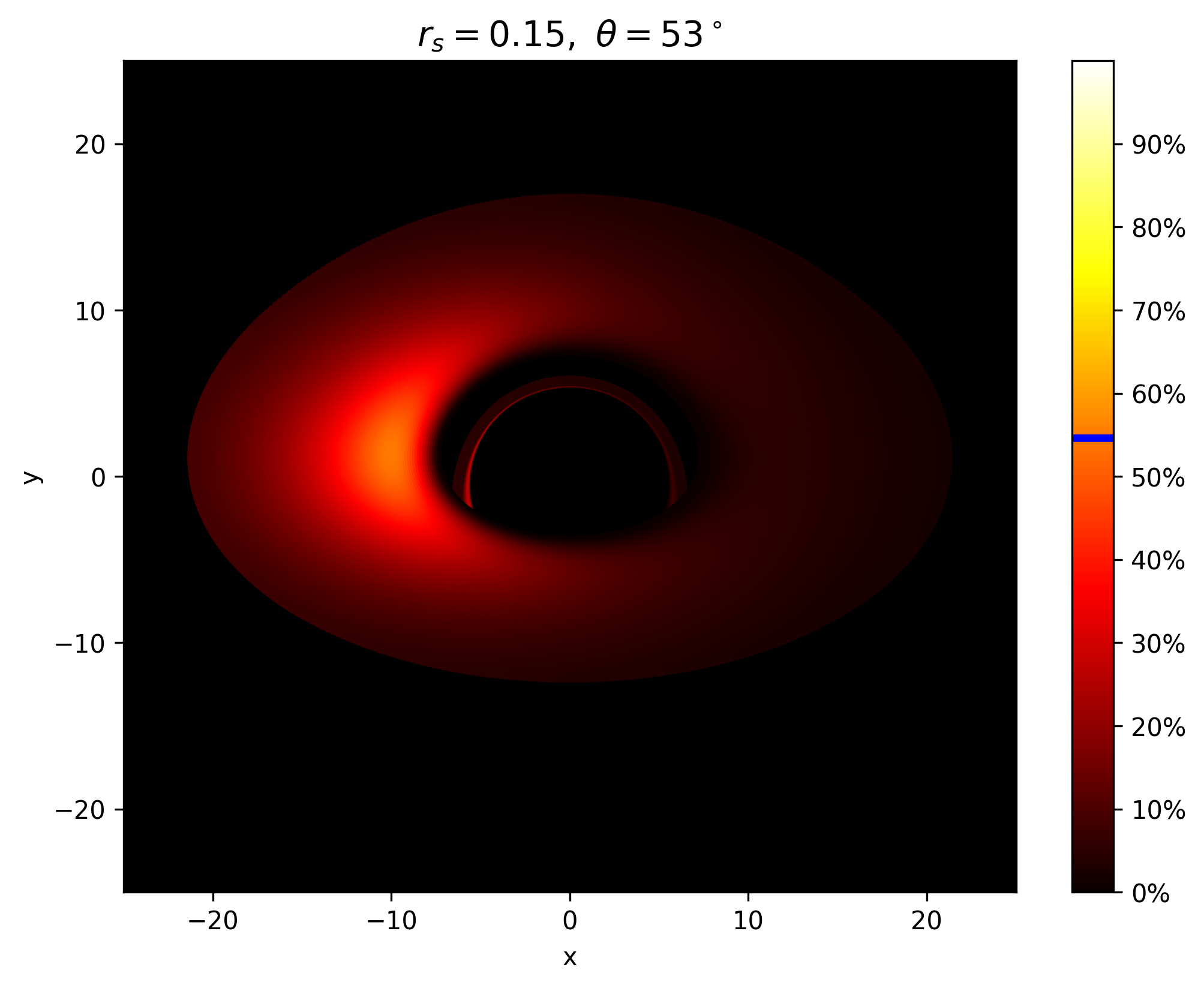}\hspace{-0.2cm}
  \includegraphics[scale=0.35]{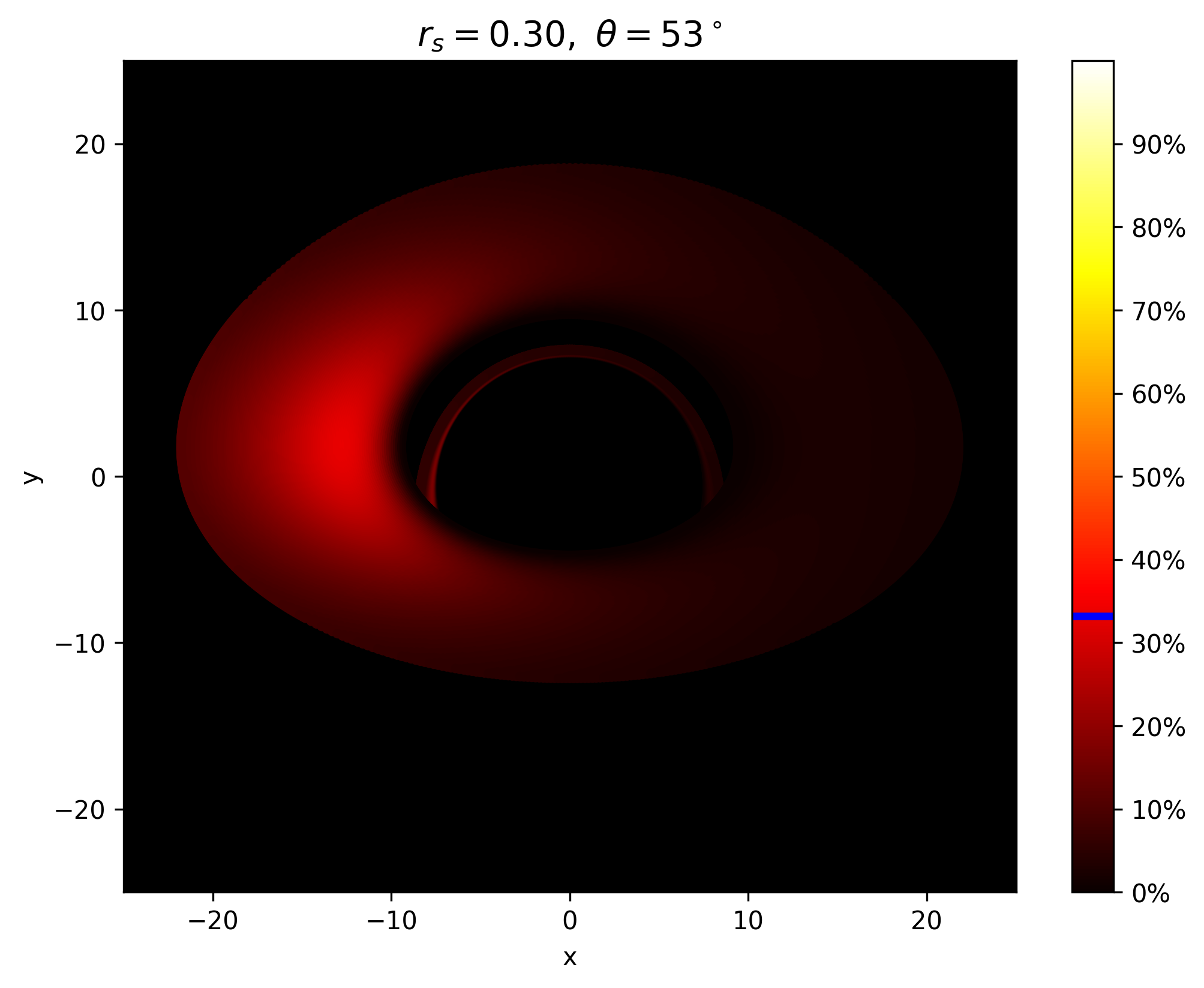}\\
  \includegraphics[scale=0.35]{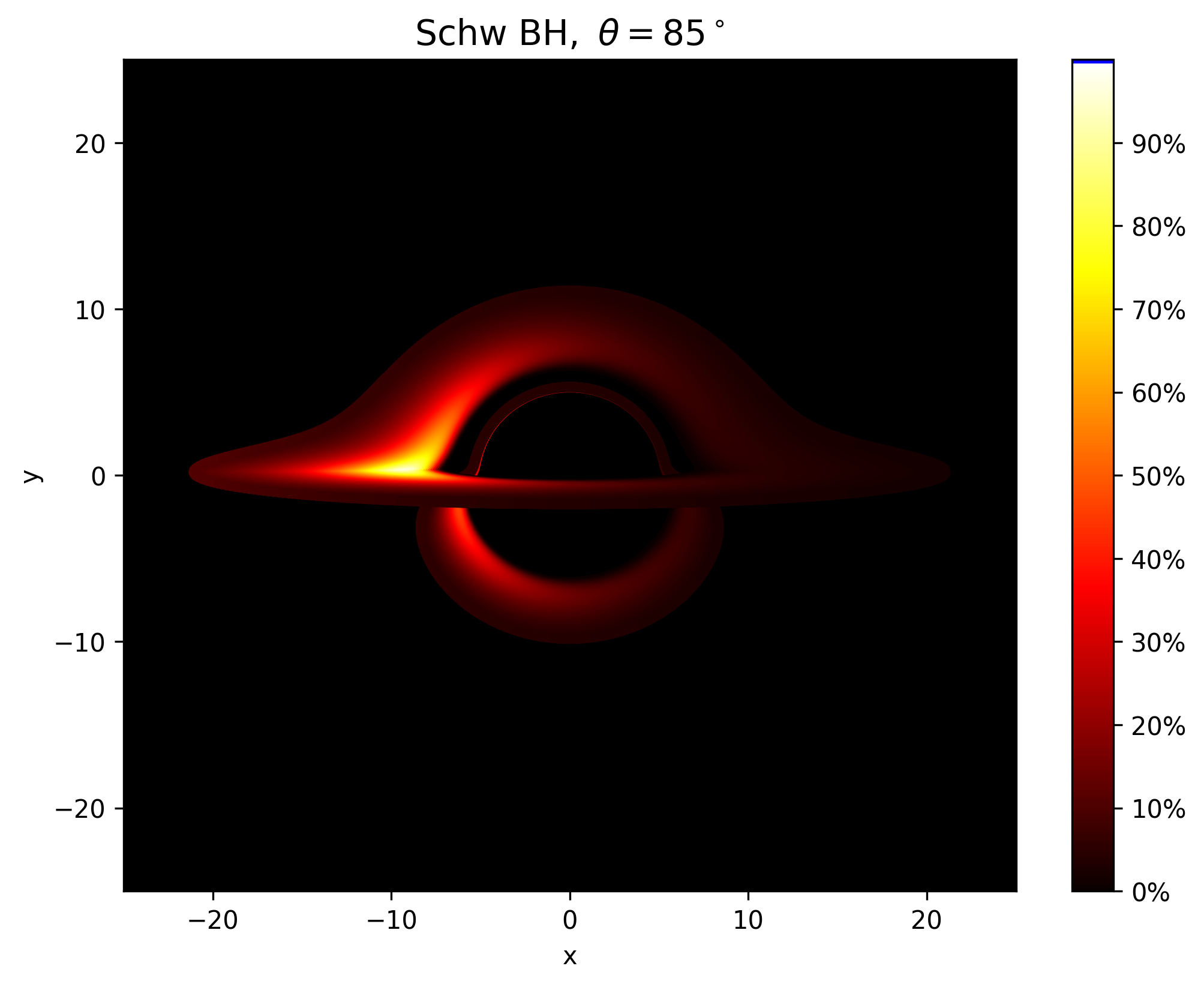}\hspace{-0.2cm}
  \includegraphics[scale=0.35]{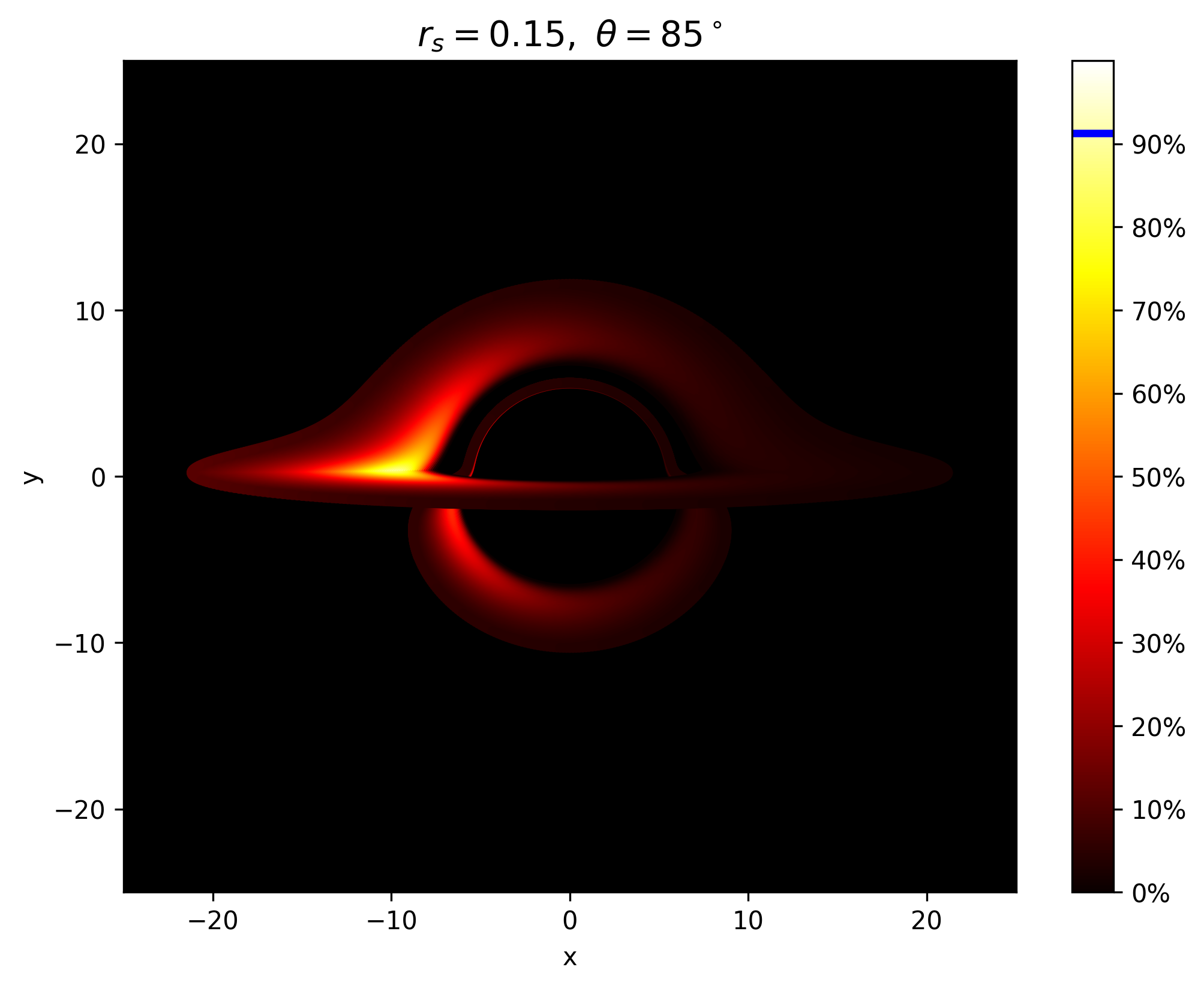}\hspace{-0.2cm}
  \includegraphics[scale=0.35]{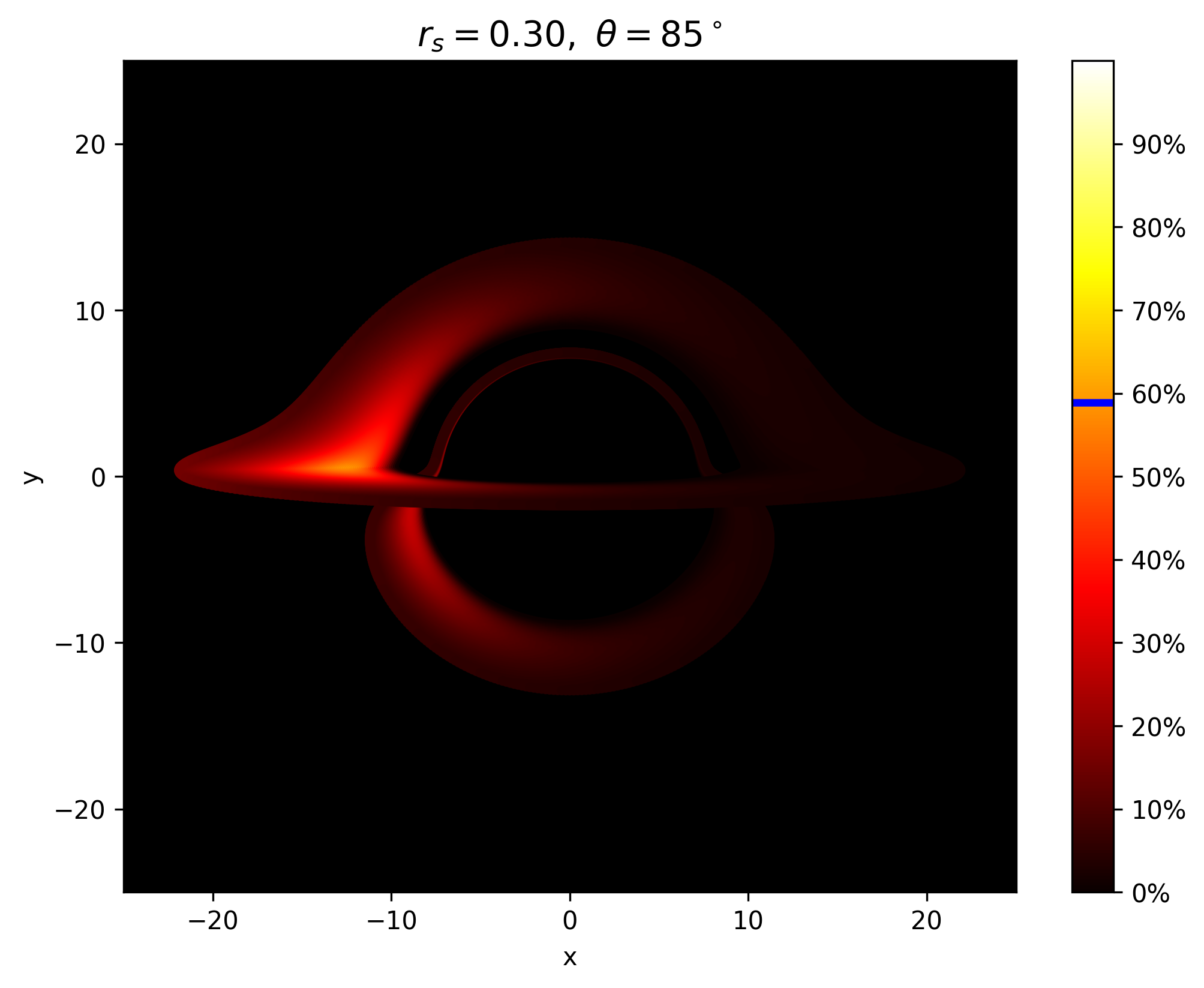}
  \end{tabular}
	\caption{\label{fig:fluxobs2} Distribution of the observed flux $F_{obs}$ in the direct and secondary images of a Scwarzchild BH surrounded by a King DM halo is shown for different inclination angles $\theta$ and scale radii $r_s$, with the central density fixed at $\rho_s = 0.5$.}
\end{figure*}

We calculated the observed energy flux $F_{obs}$ at each point in the accretion disk image on the camera plane and normalized it to the Schwarzschild black hole case at $\theta = 85^\circ$, setting it to 100\%. The normalized energy flux values for all accretion disk images were then visualized using the hot colormap. The image of the accretion disk is constructed for the region between $R_{\mathrm{ISCO}}$ and $R = 25$, as shown in Figs.~\ref{fig:fluxobs} and~\ref{fig:fluxobs2}. In Fig.~\ref{fig:fluxobs}, the scale radius is fixed at $r_s = 0.3$. For $\rho_s = 0.5$ and $\theta = 85^\circ$, the observed flux $F_{obs}$ reaches approximately 58\% of that of the Schwarzschild BH at the same inclination angle. In Fig.~\ref{fig:fluxobs2}, where the central density is fixed at $\rho_s = 0.5$, a similar trend is observed: as both DM halo parameters increase, the observed flux $F_{obs}$ decreases and appears dimmer compared to the Schwarzschild case.

\begin{figure*}
\begin{tabular}{ccc}
  \includegraphics[scale=0.34]{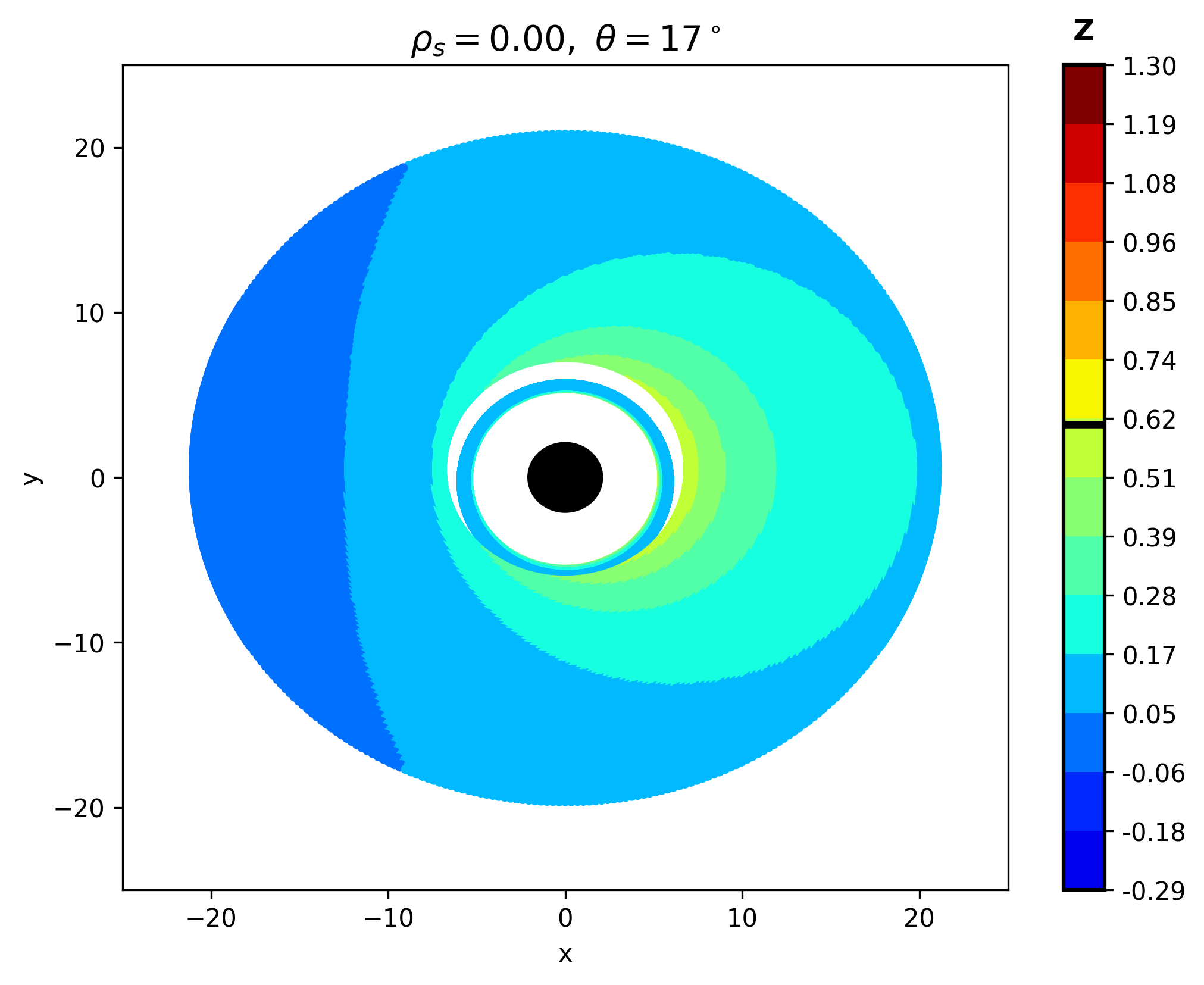}\hspace{-0.2cm}
  \includegraphics[scale=0.34]{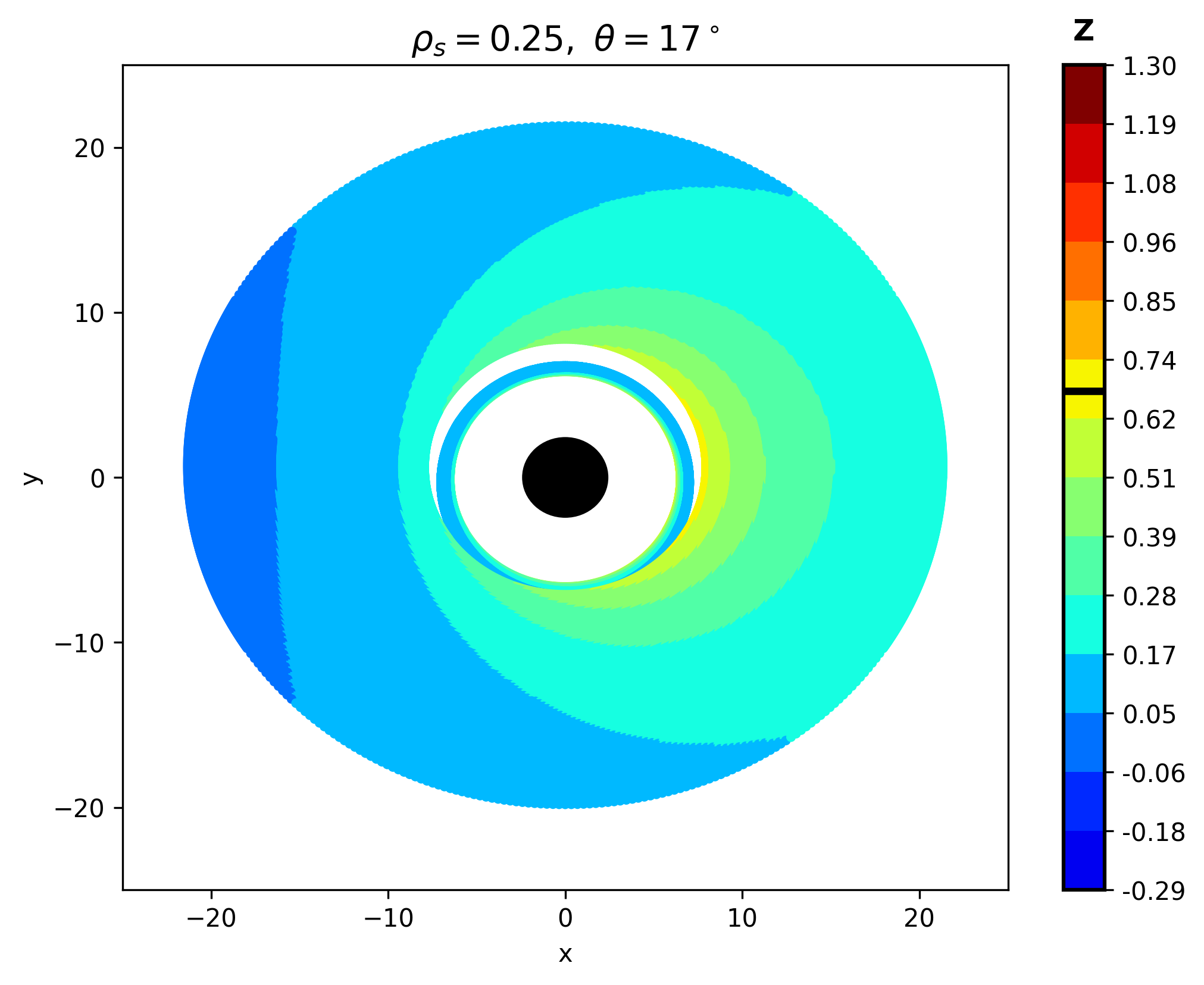}\hspace{-0.2cm}
  \includegraphics[scale=0.34]{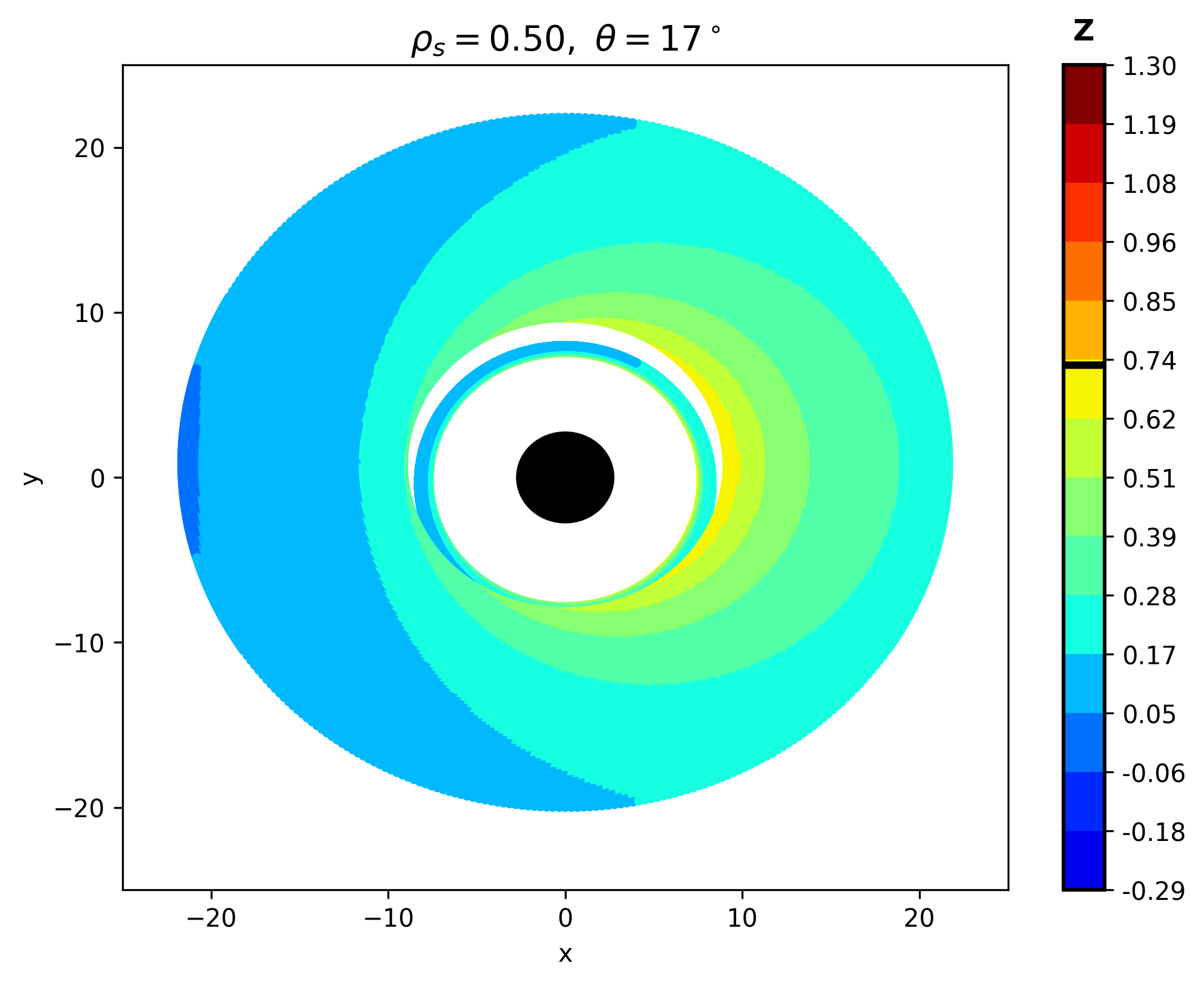}\\
  \includegraphics[scale=0.34]{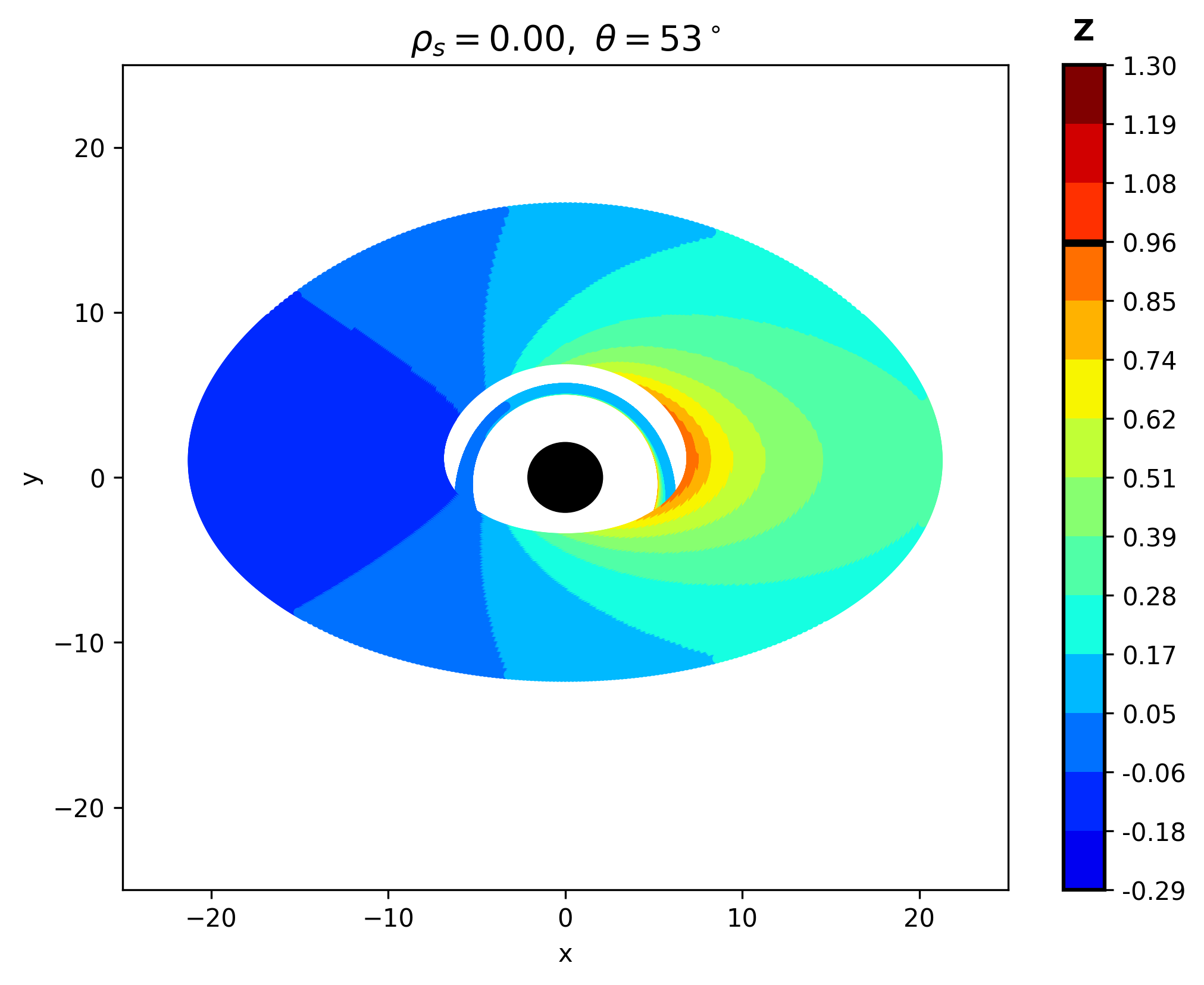}\hspace{-0.2cm}
  \includegraphics[scale=0.34]{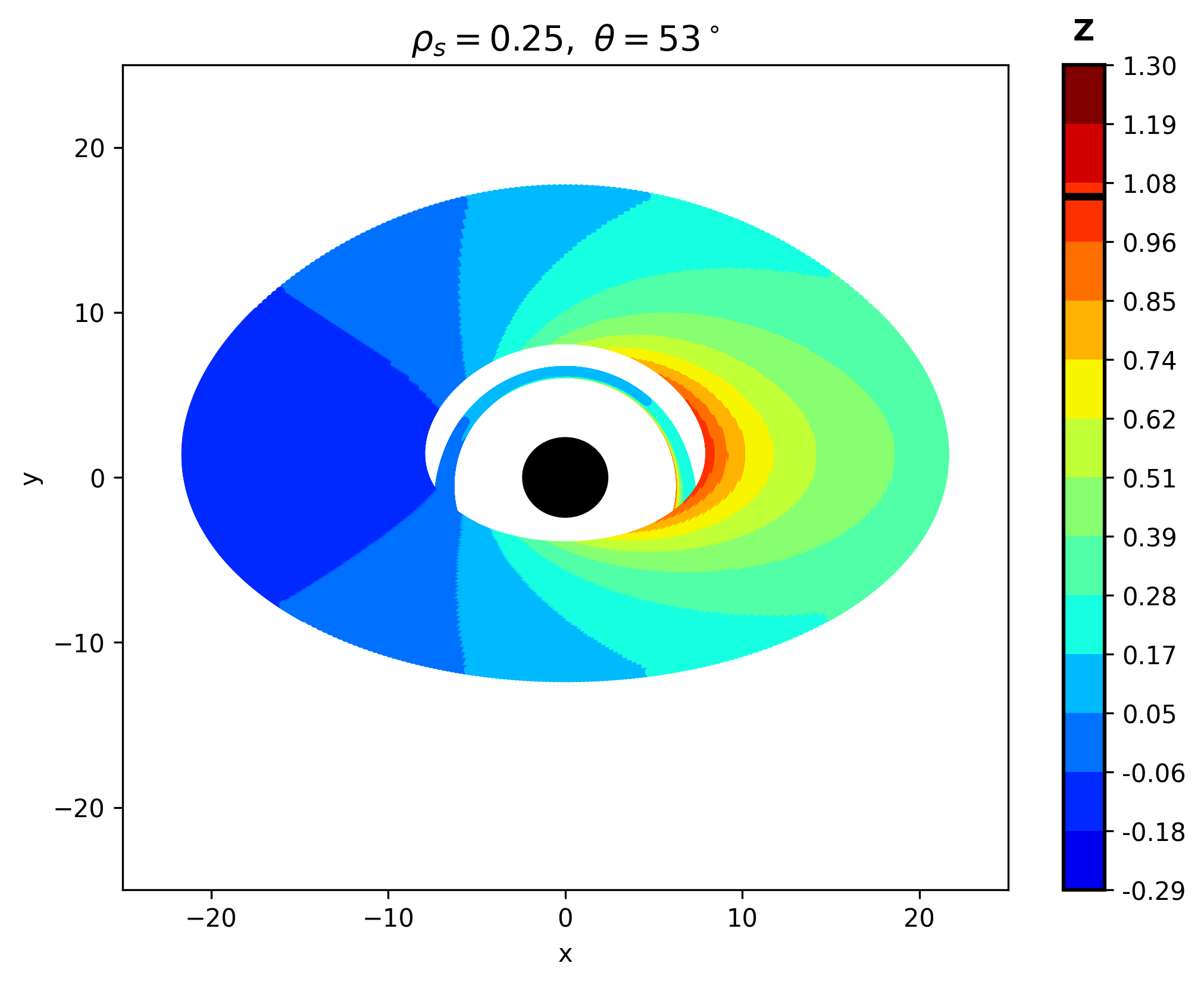}\hspace{-0.2cm}
  \includegraphics[scale=0.34]{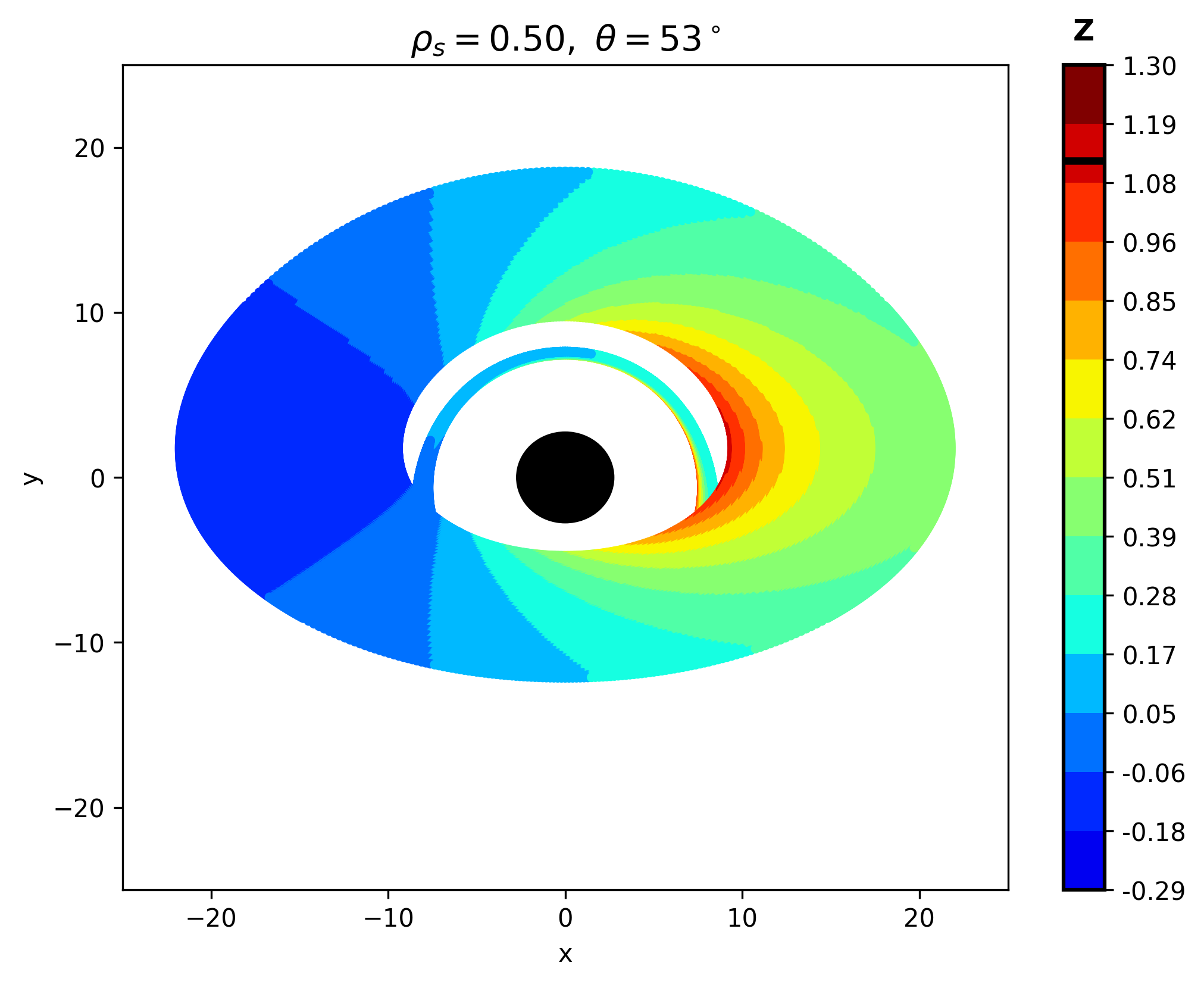}\\
  \includegraphics[scale=0.34]{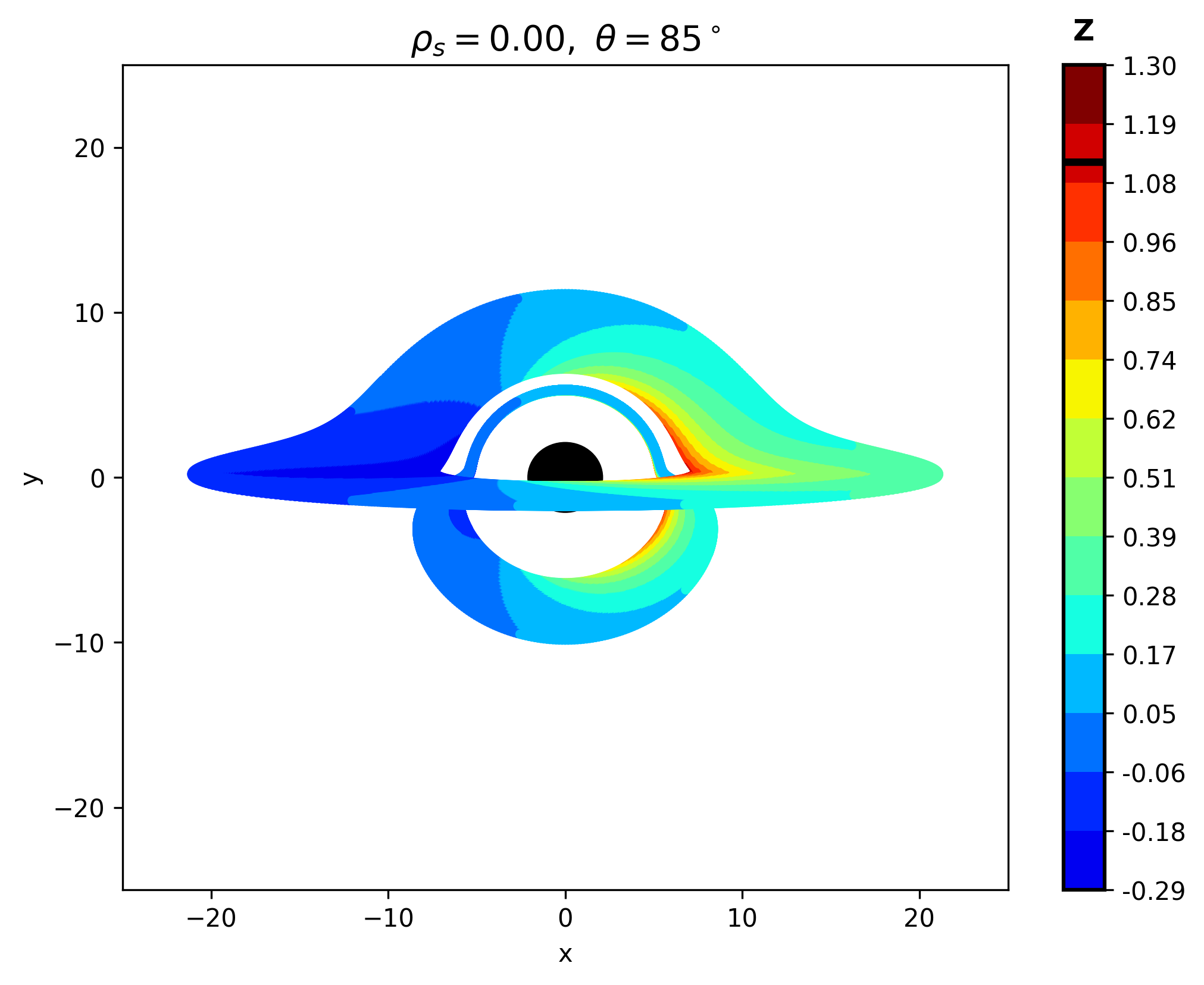}\hspace{-0.2cm}
  \includegraphics[scale=0.34]{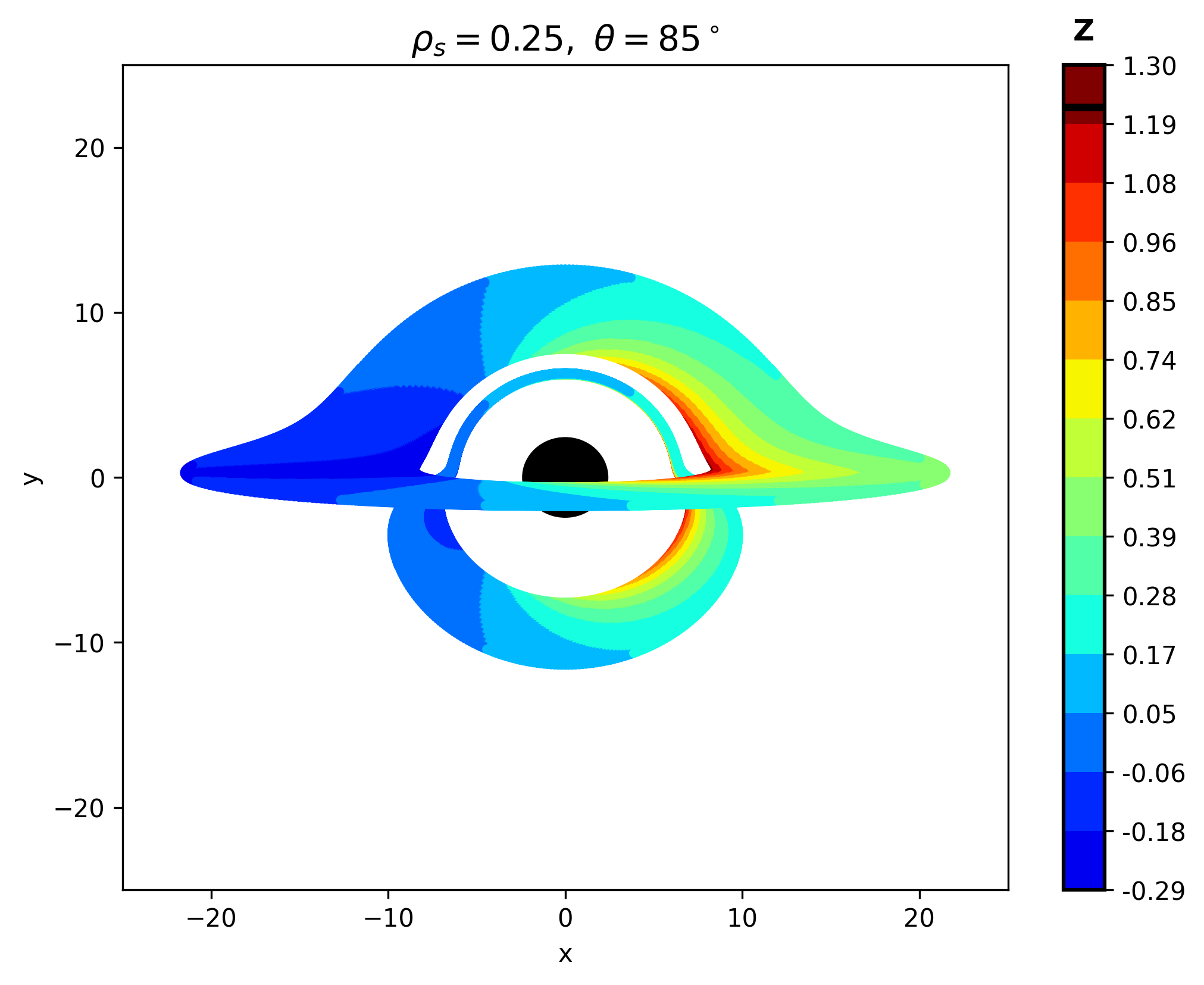}\hspace{-0.2cm}
  \includegraphics[scale=0.34]{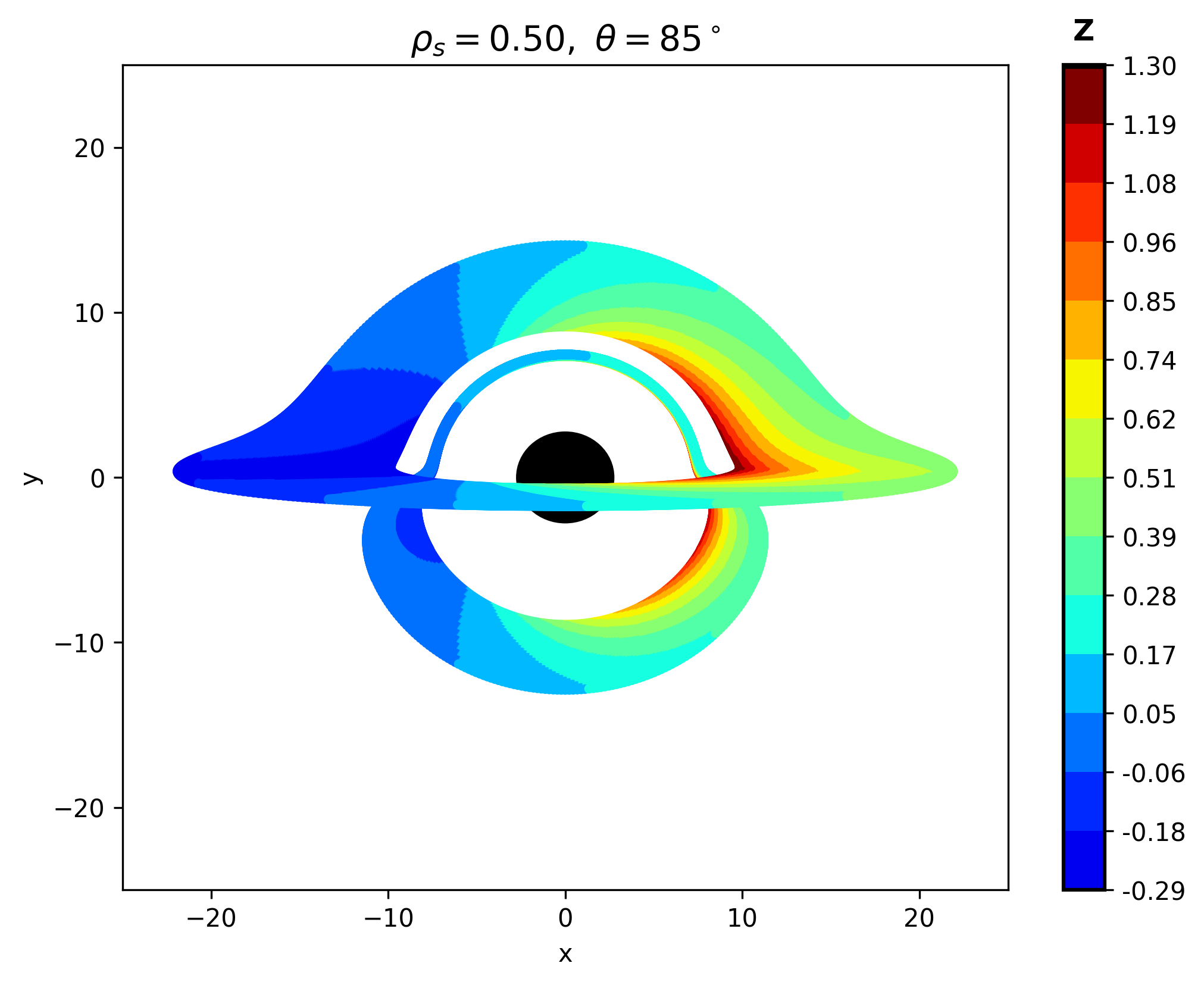}
  \end{tabular}
\caption{\label{fig:redeshiftdist}Distribution of the redshift in direct and secondary images of BHs surrounded by a King DM halo for inclination angles of $17^\circ$, $53^\circ$, and $85^\circ$, with the halo scale radius fixed at $r_s=0.3$.}
    \end{figure*}

Figure~\ref{fig:redeshiftdist} shows the redshift distribution across the accretion disk around a black hole surrounded by the DM halo. The combined effects of Doppler and gravitational redshifts cause the emitted light to be observed with a frequency shift. The rotation of the disk makes the image asymmetric, with one side appearing blueshifted and the other redshifted.
The black line on the color bar indicates the maximum value of the redshift factor $z$ for the accretion disk. The highest redshift value, $z_{\mathrm{max}} = 1.30$, is obtained for $\rho_s = 0.5$ and $\theta = 85^\circ$, whereas the lowest value, $z_{\mathrm{min}} = -0.29$, corresponds to $\rho_s = 0.0$ and $\theta = 85^\circ$. All panels use the same color scale limits for consistent comparison. From the distributions shown in the figure, it is evident that the redshift effect strengthens as the inclination angle increases. Moreover, the redshift ($z > 0$) is generally larger for the Schwarzschild BH surrounded by a King BH halo than for the Schwarzschild BH. These noticeable differences suggest that such systems could, in principle, be distinguished through optical observations of their accretion disks.
%
% \subsection{Observed flux and Redshift factor}

\section{CONCLUSIONS}\label{Sec:conclusion}

From an astrophysical viewpoint, a BH embedded in a DM halo provides a useful framework for studying BH–DM interactions and testing background gravity. In this work, we studied the properties of the Schwarzschild-like BH in the presence of a King DM halo and highlighted its differences from the classical Schwarzschild BH. We began by investigating the effect of the King DM halo on the geodesics of time-like particles. Based on the analysis of the bound orbits, we revealed that, as the King DM halo parameters $r_s$ and $\rho_s$ increase, the values of $R_{\mathrm{MBO}}$, $L_{\mathrm{MBO}}$, $R_{\mathrm{ISCO}}$, and $L_{\mathrm{ISCO}}$ also increase, while $E_{\mathrm{ISCO}}$ decreases (see Figs.~\ref{fig:mbo} and ~\ref{fig:isco}). These results indicate that the King DM halo has a gravitational nature that strengthens the overall gravitational field, thereby shifting the stable orbits to larger radii from the BH. 

To better understand the effect of the King DM halo on the geodesics of massive particles, we examined periodic orbits, which represent a special class of bound trajectories. For an orbit to be periodic, its fundamental orbital frequencies must satisfy a rational relation characterized by a rational number $q$. 
For that, any change in the particle’s energy $E$ and orbital angular momentum $L$ affects the configuration of the periodic orbits described by the integers $(z, w, v)$. This dependence becomes particularly sensitive near the maximum values of $E$ and $L$, where even small changes in these parameters lead to sharp variations in the rational number $q$ (as shown in Fig.~\ref{fig:q}). We evaluated the energies and orbital angular momenta corresponding to periodic orbits with different configurations $(z, w, v)$ for various combinations of the King DM parameters (see Table~\ref{table1}). Furthermore, we analyzed the differences between these orbits with higher zoom numbers $z$ that exhibit more intricate geometrical structures and with those with larger whirl numbers $w$ that perform a greater number of revolutions around the BH (see Figs.~\ref{fig:periodic} and \ref{fig:periodicL}). From the analysis of periodic orbits $(1, 1, 0)$, with and without the King DM halo profile, we found that, similarly to bound orbits, the presence of the King DM halo causes periodic orbits to shift outward, decreasing their energy and increasing their orbital angular momentum (see Fig.~\ref{fig:comparison}). This provides a clear distinction between the Schwarzschild-like BH spacetime within the King-type DM halo and the Schwarzschild BH. 

Furthermore, we investigated the optical properties of Schwarzschild BH surrounded by the King DM halo. By analyzing photon trajectories and accretion disk characteristics, we demonstrated that the presence of the King DM halo significantly alters the observable features of the BH compared to the Schwarzschild BH case. We found that the critical impact parameter, event horizon, and photon sphere radius increase as a consequence of the increase in DM halo parameters $\rho_s$ and $r_s$, indicating a strong dependence of light propagation on the surrounding King DM halo distribution (see Fig.~\ref{fig:ray1}). We found the critical impact parameter to be $b_c = 7.232$ for $r_s = 0.3$ and $\rho_s = 0.5$, whereas it is $b_c = 5.196$ for the Schwarzschild BH case. 

We provided a broader comparison between the Schwarzschild-like BH surrounded by the King DM halo and the standard Schwarzschild BH (see Table~\ref{tab:nb}). We found that, in the presence of the King DM halo, the photon ring range increases and extends to $7.213 < b < 7.298$, whereas in the pure Schwarzschild case it is $5.188 < b < 5.228$. We determined that this tendency is also observed in the lensed and direct ring regions, suggesting that the DM halo enhances gravitational lensing effects.

In addition, we showed that increasing $\rho_s$ and $r_s$ reduces the energy flux and radiation temperature of the accretion disk, causing the BH in the King DM halo to appear dimmer to distant observers (see Figs.~\ref{fig:Flux}-\ref{fig:fluxobs}). Based on our analysis of the redshift distribution $z$, we showed that higher inclination angles and larger DM halo parameters amplify redshift effects, with a maximum value of $z_{\max} = 1.30$ at $r_s = 0.3$, $\rho_s = 0.5$, and $\theta = 85^\circ$. These results confirmed that both gravitational and Doppler effects played a key role in shaping the observed image.

Our results highlight how the King DM halo significantly affects periodic geodesic orbits, photon motion, accretion disk properties, and image formation around BHs, providing insights into the spacetime structure and observable signatures of BH–DM systems. Moreover, the resulting analysis of disk radiation may serve as an alternative approach for detecting the presence of DM halos and constraining deviations from GR and its extensions.

\section*{Acknowledgements}

S.S. is supported by the National Natural Science Foundation of China under Grant No. W2433018. T.Z. is also supported by the National Natural Science Foundation of China under Grants No. 12275238 and 11675143, the National Key Research and Development Program under Grant No. 2020YFC2201503, and the Zhejiang Provincial Natural Science Foundation of China under Grants No. LR21A050001 and No. LY20A050002, and the Fundamental Research Funds for the Provincial Universities of Zhejiang in China under Grant No. RF-A2019015.

\bibliographystyle{apsrev4-1}
\bibliography{ref1,ref2}

\end{document}